\documentclass[12pt,oneside]{article}

\usepackage[latin1]{inputenc}
\usepackage{a4,german} 
\usepackage{amsmath,amssymb,amsxtra,exscale}
\usepackage{xspace}

\usepackage[square,comma,numbers,sort&compress]{natbib}

\newcommand{\eg}{e.\,g.\@\xspace}
\newcommand{\ie}{i.\,e.\@\xspace}
\newcommand{\cf}{cf.\@\xspace}
\newcommand{\etc}{etc.\@\xspace}
\newcommand{\rhs}{r.\,h.\,s.\@\xspace}

\newcommand{\eom}{e.\,o.\,m.\@\xspace}
\newcommand{\Sec}{Sec.~}
\newcommand{\App}{App.~}
\newcommand{\Tab}{Table~}

\newcommand{\1}{1\hspace{-0.243em}\text{l}} % Identity Matrix
\newcommand{\R}{\mathbb{R}}

\newcommand{\half}{\frac{1}{2}}
\newcommand{\const}{\operatorname{const}}

\newcommand{\cycl}{\operatorname{cycl}}
\newcommand{\gcycl}{\operatorname{gcycl}}
\newcommand{\sign}{\operatorname{sign}}
\newcommand{\id}{\operatorname{id}}

\newcommand{\BMf}{\mathcal{M}}
\newcommand{\TSp}{\mathcal{N}}
\newcommand{\Poisson}{\mathcal{P}}
\newcommand{\Casimir}{C}
\newcommand{\casimir}{c}

\newcommand{\Action}{L}
\newcommand{\N}{N}
\newcommand{\detv}{\Delta}
\newcommand{\derv}{\nabla}

\newcommand{\susy}[1]{\breve{#1}}

\DeclareMathAlphabet{\matheurm}{U}{eur}{m}{n} % \matheurm
\newcommand{\q}[1]{\matheurm{#1}}

\renewcommand{\^}{{}^}
\renewcommand{\_}{\!{}_}

\newcommand{\eqnsplit}{\notag \\ {} &\hspace{1.35em} {}}

\newcommand{\lvec}[1]{\overset{\raisebox{-0.5ex}{$\scriptscriptstyle
\leftarrow$}}{#1}{}}
\newcommand{\rvec}[1]{\overset{\,
\raisebox{-0.5ex}{$\scriptscriptstyle \rightarrow$}}{#1}{}}

\newcommand{\lpartial}{\lvec{\partial}}
\newcommand{\rpartial}{\rvec{\partial}}
\newcommand{\clmn}[2]{\left( \begin{array}{rr} #1 \\ #2
\end{array}\right) }

\newcommand{\mtrx}[4]{\left( \begin{array}{cc} #1 & #2 \\ #3 & #4
\end{array}\right) }

\let\a=\alpha \let\b=\beta \let\g=\gamma \let\d=\delta

\let\t=\tau \let\o=\omega  

\let\O=\Omega

\newcommand{\bibJBD}{%
  jordan55,Dicke:1957RM,Jordan:1959eg,Fierz:1956,Brans:1961sx}
\newcommand{\bibQuint}{%
  Wang:1999fa,Diaz-Rivera:1999wd,Matos:1999et,Coley:1999yq,%
  Bertolami:1999dp,Sokolowski:1995dk}
\newcommand{\bibQuintObserv}{%
  Perlmutter:1997hx,Perlmutter:1998np,Riess:1998cb,%
  Garnavich:1998nb,Garnavich:1998th,Schmidt:1998}
\newcommand{\bibTele}{%
  Hayashi:1977jd,Hayashi:1979qx,hehl79,kopczynski82,Kopczynski:1990af,%
  Muller-Hoissen:1983vc,mueller-hoissen85,Bakler:1988ub,%
  Mielke:1990my,mielke92,nester89}

\newcommand{\bibString}{green87,luest89,polchinski98}

\newcommand{\bibSAlg}{%
  Kac:1977qb,Scheunert:1976uf,Scheunert:1976ug,Frappat:1996pb}

\newcommand{\bibGrF}{Schmidt:1999wb,Obukhov:1997uc,Strobl:1999Habil}
\newcommand{\bibKV}{Katanaev:1986wk,Katanaev:1990qm}
\newcommand{\bibFOG}{Schaller:1994np,Schaller:1994es}
\newcommand{\bibDil}{%
  Banks:1991mk,Odintsov:1991qu,%
  Louis-Martinez:1994eh,Gegenberg:1995pv,%
  Klosch:1996fi,Klosch:1996qv,Kloesch:1997fi,Klosch:1998yh}
\newcommand{\bibJT}{%
  Barbashov:1979bm,Teitelboim:1983ux,teitelboim84,%
  jackiw84,Jackiw:1985je}
\newcommand{\bibDilKKL}{Katanaev:1996bh,Katanaev:1997ni}
\newcommand{\bibGrRed}{%
  Thomi:1984na,Hajicek:1984mz,Schmidt:1997mq,Schmidt:1998ih}
\newcommand{\bibSI}{Verlinde:1991rf,Cangemi:1992bj,Cangemi:1993sd}
\newcommand{\bibDBH}{Mandal:1991tz,Elitzur:1991cb,Witten:1991yr}
\newcommand{\bibDBHmatter}{%
  Callan:1992rs,%
  Mikovic:1992id,Mikovic:1993vy,Mikovic:1995ub,Mikovic:1997de,%
  Kuchar:1997zm,Varadarajan:1998qz,%
  Cangemi:1996yz,Benedict:1996qy}
\newcommand{\bibQGr}{Kummer:1997hy,Kummer:1997jj,Kummer:1998zs}
\newcommand{\bibPSM}{Schaller:1994es,Schaller:1994uj,Schaller:1995xk}
\newcommand{\bibNLGT}{Ikeda:1993qz,Ikeda:1993nk,Ikeda:1993aj}
\newcommand{\bibPSMstar}{Cattaneo:1999fm,Cattaneo:2000iw}
\newcommand{\bibPoisson}{Weinstein:1983MM,choquet-bruhat89}
\newcommand{\bibGrMatterC}{Kummer:1995qv,Grumiller:1999rz}

\newcommand{\bibSUGRA}{%
Freedman:1976xh,Freedman:1976py,Deser:1976eh,Deser:1976rb}
\newcommand{\bibSuperfield}{%
  berezin66,dewitt84,Vladimirov:1984zj,Vladimirov:1985ZJ,constantinescu94}
\newcommand{\bibGaugeSUGRA}{VanNieuwenhuizen:1981ae}

\newcommand{\bibSJTsuperfield}{Chamseddine:1991fg}
\newcommand{\bibSJTgauge}{Rivelles:1994xs,Cangemi:1994mj}
\newcommand{\bibNLSGT}{Ikeda:1994dr,Ikeda:1994fh}
\newcommand{\bibIzq}{%
  Ikeda:1994dr,Ikeda:1994fh,Izquierdo:1998hg,Strobl:1999zz}

\begin{document}

%********************************************************************
\begin{titlepage}
\renewcommand{\thefootnote}{\fnsymbol{footnote}}

\begin{flushright}
  TUW-00-35 \\
  FSU-TPI\ 12/00 \\
  hep-th/0012219
\end{flushright}

\bigskip

\begin{center}
  
  \begin{Large} \bfseries General Two-Dimensional Supergravity from
  Poisson Superalgebras \end{Large}

  \vspace{6ex}

  M.~Ertl\footnotemark[1], W.~Kummer\footnotemark[2],
  T.~Strobl\footnotemark[3],

  \vspace{6ex}

  {\footnotemark[1]\footnotemark[2]\footnotesize Institut f\"ur
    Theoretische Physik \\ Technische Universit\"at Wien \\ Wiedner
    Hauptstr.  8--10, A-1040 Wien, Austria}

  \medskip

  {\footnotemark[3]\footnotesize Institut f\"ur
    Theoretische Physik \\ Friedrich-Schiller-Universit\"at Jena \\ Max
    Wien Platz 1, D-07743 Jena, Germany}

  \footnotetext[1]{E-mail: \texttt{ertl@tph.tuwien.ac.at}}
  \footnotetext[2]{E-mail: \texttt{wkummer@tph.tuwien.ac.at}}
  \footnotetext[3]{E-mail: \texttt{Thomas.Strobl@tpi.uni-jena.de}}

\end{center}

\vspace{6ex}

\begin{abstract}
  
  We provide the geometric actions for most general $\N=1$
  supergravity in two spacetime dimensions. Our construction implies
  an extension to arbitrary $N$.  This provides a supersymmetrization
  of any generalized dilaton gravity theory or of any theory with an
  action being an (essentially) arbitrary function of curvature and
  torsion.
  
  Technically we proceed as follows: The bosonic part of any of these
  theories may be characterized by a generically nonlinear Poisson
  bracket on a three-dimensional target space. In analogy to a given
  ordinary Lie algebra, we derive all possible $\N=1$ extensions of
  any of the given Poisson (or $W$-) algebras. Using the concept of
  graded Poisson Sigma Models, any extension of the algebra yields a
  possible supergravity extension of the original theory, local
  Lorentz and super-diffeomorphism invariance follow by construction.
  Our procedure automatically restricts the fermionic extension to the
  minimal one; thus local supersymmetry is realized on-shell. By
  avoiding a superfield approach we are also able to circumvent in
  this way the introduction of constraints and their solution. For
  many well-known dilaton theories different supergravity extensions
  are derived.  In generic cases their field equations are solved
  explicitly.

\end{abstract}

\bigskip

December 2000

\end{titlepage}

%********************************************************************
%\tableofcontents

%********************************************************************
\section{Introduction}
\label{sec:intro}

The study of diffeomorphism invariant theories in $1+1$ dimensions for
quite some time has been a fertile ground for acquiring some insight
regarding the unsolved problems of quantum gravity in higher
dimensions. Indeed, the whole field of spherically symmetric gravity
belongs to this class, from $d$-dimensional Einstein theory to
extended theories like the Jordan-Brans-Dicke theory \cite{\bibJBD} or
`quintessence' \cite{\bibQuint} which may now seem to obtain
observational support \cite{\bibQuintObserv}.  Also equivalent
formulations for 4d Einstein theory with nonvanishing torsion
(`teleparallelism' \cite{\bibTele}) and alternative theories
including curvature and torsion \cite{Hehl:1995ue} are receiving
increasing attention.

On the other hand, supersymmetric extensions of gravity
\cite{\bibSUGRA} are believed to be a necessary ingredient for a
consistent solution of the problem to quantize gravity, especially
within the framework of string/brane theory \cite{\bibString}. These
extensions so far are based upon bosonic theories with vanishing
torsion.

In view of this situation it seems surprising that the following
problem so far has not been solved:

Given a general geometric action of pure gravity in two spacetime
dimensions of the form (\cf \cite{\bibGrF} and references therein)
\begin{equation}
  \Action^{\mathrm{gr}} = \int d^2\!x \sqrt{-g} F(R,\t^2), \label{grav}
\end{equation}
what are its possible supersymmetric generalizations?

Here $R$ and $\t^2$ denote the in two dimensions only two
algebraic-geometric invariants of curvature and torsion, respectively,
and $F$ is some sufficiently well-behaved function of them; for the
case that $F$ does not depend on its second argument, $R$ is
understood to be the Ricci scalar of the torsion-free Levi-Civita
connection.

As a prototype of a theory with dynamical torsion we may consider the
specification of (\ref{grav}) to the Katanaev-Volovich (KV) model
\cite{\bibKV}, quadratic in curvature and torsion.  Even for this
relatively simple particular case of (\ref{grav}), a supergravity
generalization has not been presented to this day.

The bosonic theory (\ref{grav}) may be reformulated as a first order
gravity action (FOG) by introducing auxiliary fields $\phi$ and $X^a$
(the standard momenta in a Hamiltonian reformulation of the model; \cf
\cite{\bibFOG} for particular cases and \cite{Strobl:1999Habil} for
the general discussion)
\begin{equation}
  \Action^\mathrm{FOG} = \int_\BMf \phi d\o + X_a De^a + \epsilon
  v(\phi,Y) \label{FOG}
\end{equation}
where $Y=X^2/2 \equiv X^a X_a/2$ and $v$ is some two-argument function
of the indicated variables.  In (\ref{FOG}) $e^a$ is the zweibein and
$\omega_{ab} = \omega\epsilon_{ab}$ the Lorentz or spin connection,
both 1-form valued, and $\epsilon = \frac{1}{2} e^a \wedge e^b
\epsilon_{ba} = e d^2\!x$ is the two-dimensional volume form ($e =
\det(e_m\^a)$). The torsion 2-form is $De^a = de^a + e^b \wedge \omega
\epsilon_b\^a$.

The function $v(\phi,Y)$ is the Legendre transform
\cite{Strobl:1999Habil} of $F(R,\t^2)$ with respect to the
\emph{three} arguments $R$ and $\t^a$, or, if $F$ depends on the
Levi-Civita curvature $R$ only, with respect to this single variable
($v$ depending only on $\phi$ then). In view of its close relation to
the corresponding quantity in generalized dilaton theories which we
recall below, we shall call $\phi$ the `dilaton' also within the
action (\ref{FOG}).\footnote{In the literature also $\Phi =
  -\frac{1}{2} \ln \, \phi$ carries this name. This definition is
  useful when, as it is often the case for specific models, $\phi$ is
  restricted to $\R_+$ only.}  The equivalence between (\ref{grav})
and the action (\ref{FOG}) holds at a global level, if there is a
globally well-defined Legendre transform of $F$. Prototypes are
provided by quadratic actions, \ie by $R^2$-gravity and the model of
\cite{\bibKV}. Otherwise generically the equivalence holds still
locally (patchwise). Only theories for which $v$ or $F$ even locally
do not have a Legendre transform, are not at all covered by the
respective other formulation. In any case, as far as the
supersymmetrization of 2d gravity theories is concerned, we will
henceforth focus on the family of actions given by (\ref{FOG}).

The FOG formulation (\ref{FOG}) also covers general dilaton theories
in two dimensions \cite{\bibDil} ($\widetilde{R}$ is the torsion free
curvature scalar),
\begin{equation}
  \label{dil}
  \Action^{\mathrm{dil}} = \int d^2\!x \sqrt{-g}
  \left[
    \frac{\widetilde{R}}{2} \phi - \half Z(\phi) (\partial^n \phi)
    (\partial_n \phi) + V(\phi)
  \right].
\end{equation}
Indeed, by eliminating $X^a$ and the torsion-dependent part of
$\omega$ in (\ref{FOG}) by their algebraic equations of motion, for
regular 2d space-times ($e = \sqrt{-g} \neq 0$) the theories
(\ref{dil}) and (\ref{FOG}) are locally and globally equivalent if in
(\ref{FOG}) the `potential' is chosen as \cite{\bibDilKKL} (\cf
  also \Sec\ref{sec:sdil} below for some details as well as
  \cite{Obukhov:1997uc} for a related approach)
\begin{equation}
  \label{vdil}
  v^{\mathrm{dil}}(\phi,Y) = Y Z(\phi) + V(\phi).
\end{equation}

There is also an alternative method for describing dilaton gravity by
means of an action of the form (\ref{FOG}), namely by using the
variables $e^a$ as a zweibein for a metric $\bar g$, related to $g$ in
(\ref{dil}) according to $g_{mn} = \Omega(\phi) \bar{g}_{mn}$ for a
suitable choice of the function $\Omega$ (it is chosen in such a way
that after transition from the Einstein-Cartan variables in
(\ref{FOG}) upon elimination of $X^a$ one is left with an action for
$\bar g$ of the form (\ref{dil}) with $\bar Z = 0$, \ie without
kinetic term for the dilaton, \cf \eg
\cite{Verlinde:1991rf,Banks:1991mk,Louis-Martinez:1994eh}).\footnote{Some
  details on the two approaches to general dilaton gravity may be
  found also in \cite{Strobl:1999Habil}.}  This formulation has the
advantage that the resulting potential $v$ depends on $\phi$ only. It
has to be noted, however, that due to a possibly singular behavior of
$\O$ (or $1/\O$) the global structures of the resulting spacetimes
(maximally extended with respect to $\bar g$ versus $g$) are in part
quite different.  Moreover, also the change of variables in a path
integral corresponding to the `torsion' description of dilaton
theories ($Y$-dependent potential (\ref{vdil})) seems advantageous over
the one in the `conformal' description. In this description even
interactions with (scalar) matter can be included in a systematic
perturbation theory, starting from the (trivially) exact path integral
for the geometric part (\ref{FOG}) \cite{\bibQGr}.  Therefore when
describing dilaton theories within the present paper we will primarily
focus on potentials (\ref{dil}), linear in $Y$.

In any case there is thus a huge number of 2d gravity theories
included in the present framework. We select only a few for
illustrative purposes, one of which is
spherically reduced Einstein gravity
(SRG) from $d$ dimensions \cite{\bibGrRed}
\begin{equation}
  \label{SRG}
  Z_{\mathrm{SRG}} = -\frac{d-3}{(d-2)\phi}, \qquad
  V_{\mathrm{SRG}} = -\lambda^2 \phi^{\frac{d-4}{d-2}} \, ,
\end{equation}
where $\lambda$ is some constant; in the `conformal approach' mentioned
above the respective potentials become 
\begin{equation}
  \label{EBH}
  Z_{\overline{\mathrm{SRG}}} = 0, \qquad
  V_{\overline{\mathrm{SRG}}} =
  -\frac{\lambda^2}{\phi^{\frac{1}{d-2}}}.
\end{equation}
The KV-model, already referred to above, results upon 
\begin{equation}
  \label{KV}
  Z_{\mathrm{KV}} = \alpha, \qquad
  V_{\mathrm{KV}} = \frac{\beta}{2} \phi^2 - \Lambda,
\end{equation}
where $\Lambda$, $\alpha$ and $\beta$ are constant.
 Two other particular examples are the
so-called Jackiw-Teitelboim (JT) model \cite{\bibJT} with vanishing
torsion in (\ref{FOG}) and no kinetic term of $\phi$ in (\ref{dil}),
\begin{equation}
  \label{JT}
  Z_{\mathrm{JT}} = 0, \qquad V_{\mathrm{JT}} = -\Lambda \phi,
\end{equation}
or the string inspired dilaton black hole (DBH) \cite{\bibDBH} (\cf
also \cite{\bibDBHmatter})
\begin{equation}
  \label{DBH}
  Z_{\mathrm{DBH}} = -\frac{1}{\phi}, \qquad
  V_{\mathrm{DBH}} = -\lambda^2 \phi,
\end{equation}
which, incidentally, may also be interpreted as the formal limit $d
\rightarrow \infty$ of (\ref{SRG}).

For our present purposes it will be crucial that (\ref{FOG}) may be
formulated as a Poisson Sigma Model (PSM) \cite{\bibPSM,Klosch:1996fi}
(\cf also \cite{\bibNLGT,\bibPSMstar}).  Collecting zero form and
one-form fields within (\ref{FOG}) as
\begin{equation}
  \label{ident1}
  (X^i) := (\phi,X^a), \qquad
  (A_i) = (dx^m A_{mi}(x)) := (\o,e_a),
\end{equation}
and after a partial integration, the action (\ref{FOG}) may be
rewritten identically as
\begin{equation}
  \label{PSM}
  \Action^{\mathrm{PSM}} = \int_\BMf dX^i \wedge A_i + \half
  \Poisson^{ij} A_j \wedge A_i,
\end{equation}
where the matrix $\Poisson^{ij}$ may be read off by direct comparison.
The basic observation in this framework is that this matrix defines a
Poisson bracket on the space spanned by coordinates $X^i$, which is
then identified with the target space of a Sigma Model. In the
present context this bracket $\{ X^i , X^j \} := \Poisson^{ij}$ has
the form
\begin{align}
  \{ X^a, \phi \} &= X^b \epsilon_b\^a, \label{LorentzB} \\
  \{ X^a, X^b \} &= v(\phi,Y) \epsilon^{ab},  \label{PB}
\end{align} 
where (throughout this paper) $Y \equiv \half X^a X_a$; this bracket
may be verified to obey the Jacobi identity.

The gravitational origin of the underlying model is reflected by the
first set of brackets: It shows that $\phi$ is the generator of
Lorentz transformations (with respect to the bracket) on the target
space $\R^3$. The form of the second set of brackets is already
completely determined by this: Indeed, antisymmetry of the bracket
leads to proportionality to the $\epsilon$-tensor, while the Jacobi
identity for the bracket requires $v$ to be a function of the Lorentz
invariant quantities $\phi$ and $X^2$ only.

Inspection of the local symmetries of a general PSM,
\begin{equation}
  \label{PSM-symms}
  \delta X^i = \Poisson^{ij} \epsilon_j, \qquad
  \delta A_i = -d\epsilon_i - (\partial_i \Poisson^{jk}) \epsilon_k
  A_j,
\end{equation}
shows that the Lorentz symmetry of the bracket gives rise to the
\emph{local} Lorentz symmetry of the gravity action (\ref{FOG})
(specialization of (\ref{PSM-symms}) to an $\epsilon$ with only
nonzero $\phi$ component, using the identification (\ref{ident1})).
The second necessary ingredient for the construction of a gravity
action, diffeomorphism invariance, on the other hand, is automatically
respected by an action of the form (\ref{PSM}). (It may be seen that
the diffeomorphism invariance is also encoded \emph{on-shell} by the
remaining two local symmetries (\ref{PSM-symms}), \cf \eg
\cite{Klosch:1996fi,Strobl:1994PhD}).

PSMs relevant for 2d gravity theories (without further gauge field
interactions) possess one `Casimir function' $\casimir(X)$ which is
characterized by vanishing of the Poisson brackets $\{ X^i,
\casimir\}$.  Different constant values of $\casimir$ characterize
symplectic leaves \cite{\bibPoisson}. In the language of gravity
theories, for models with asymptotic Minkowski behavior $\casimir$ is
proportional to the ADM mass of the system.\footnote{We
remark in parenthesis that an analogous conservation law may be
established also in the presence of additional matter fields 
\cite{\bibGrMatterC}.}

To summarize, the gravity models (\ref{PSM}), and thus implicitly also
any action of the form (\ref{FOG}) and hence generically of
(\ref{grav}), may be obtained from the construction of a Lorentz
invariant bracket on the two-dimensional Minkowski space $\R^2$
spanned by $X^a$, with $\phi$ entering as an additional
parameter.\footnote{Actually, this point of view was already used in
  \cite{Schaller:1994es} so as to arrive at (\ref{PSM}), without,
  however, fully realizing the relation to (\ref{grav}) at that time.
  Let us remark on this occasion that in principle one might also
  consider theories (\ref{PSM}) with $X^a$ replaced \eg by $X^a \cdot
  f(\phi,X^2)$. For a nonvanishing function $f$ after a suitable
  reparametrization of the target space this again yields a PSM. Also
  the identification of the gauge fields $A_i$ in (\ref{ident1}) could
  be modified in a similar manner. Hence we do not have to cover this
  possibility explicitly in what follows. Nevertheless, it could be
  advantageous to derive by this means a more complicated gravity
  model from a simpler PSM structure.}  The resulting bracket as well
as the corresponding models are seen to be parametrized by one
two-argument function $v$ in this way.

The Einstein-Cartan formulation of 2d gravities as in (\ref{FOG}) or,
even more so, in the PSM form (\ref{PSM}) will turn out to be
particularly convenient for obtaining the most general supergravities
in $d=2$.  Whereas the metrical formulation of gravity due to Einstein
in $d=4$ appeared very cumbersome for a supersymmetric generalization,
the Einstein-Cartan approach appeared to be best suited for the needs
of introducing additional fermionic degrees of freedom to pure gravity
\cite{\bibSUGRA}.

We now briefly digress to the corresponding strategy of constructing a
supersymmetric extension of a gravity theory in a spacetime of general
dimension $d$. By adding to vielbein $e_m\^a$ and Lorentz connection
$\omega_{ma}\^b$ appropriate terms containing a fermionic and spinor
valued 1-form $\psi_m\^\alpha$, the Rarita-Schwinger field, an action
invariant under local supersymmetry can be constructed, where
$\psi_m\^\alpha$ plays the role of the gauge field for that symmetry.
In this formulation the generic local infinitesimal supersymmetry
transformations are of the form
\begin{equation}
  \label{susytrafo}
  \delta e_m\^a = -2i (\epsilon \gamma^a \psi_m), \qquad
  \delta \psi_m\^\alpha = -D_m \epsilon^\alpha + \cdots
\end{equation}
with $\epsilon = \epsilon(x)$ arbitrary.

In the course of time various methods were developed to make the
construction of supergravity actions more systematical. One of these
approaches, relying on superfields \cite{\bibSuperfield}, extends the
Einstein-Cartan formalism by adding anticommuting coordinates to the
space-time manifold, thus making it a supermanifold, and,
simultaneously, by enlarging the structure group with a spinorial
representation of the Lorentz group. This method adds a huge number of
unphysical fields and unwanted symmetries to the theory, which can be
eliminated by choosing appropriate constraints on supertorsion and
supercurvature and by choosing a Wess-Zumino type gauge. The need for
constraints and their consistent embedding into the Bianchi identities
is the draw-back of this method, because the argumentation for a
particular set of them is mainly of technical nature and quite
involved.

The other systematic approach to construct supergravity models for
general $d$ is based on the similarity of gravity to a gauge theory.
The (inverse) vielbein and the Lorentz-connection are treated as gauge
fields on a similar footing as gauge fields of possibly additional
gauge groups.  Curvature and torsion appear as particular components
of the total field strength.\footnote{Note that nevertheless standard
  gravity theories cannot be just reformulated as YM gauge theories
  with all symmetries being incorporated in a principle fiber bundle
  description; there still is the infinite dimensional diffeomorphism
  group one has to deal with (\cf also \cite{Strobl:1993xt} for an
  illustration).}  By adding fermionic symmetries to the gravity gauge
group, usually taken as the Poincar\'e, de~Sitter, or conformal group,
one obtains the corresponding supergravity theories
\cite{\bibGaugeSUGRA}.

In the two-dimensional case, the supergravity multiplet was first
constructed using the superfield approach \cite{Howe:1979ia}. Based on
that formalism, it was straightforward to \emph{formulate} a
supersymmetric generalization of the dilaton theory (\ref{dil}), \cf
\cite{Park:1993sd}.  Before that the supersymmetric generalization of
the particular case of the Jackiw-Teitelboim or de~Sitter model
\cite{\bibJT} has been achieved within this framework in
\cite{\bibSJTsuperfield}.  Up to global issues, this solved implicitly
also the problem of a supersymmetrization of the theories (\ref{grav})
in the \emph{torsion-free} case.

Still, the supergravity multiplet obtained from the set of constraints
used in \cite{Howe:1979ia} consists of the vielbein, the
Rarita-Schwinger field and an auxiliary scalar field, but the
Lorentz-connection is lost as independent field. It is expressed in
terms of vielbein and Rarita-Schwinger field. Without a formalism
using an independent Lorentz-connection the construction of
supersymmetric versions of general theories of the $F(R, \t^2)$-type
is impossible.  Although there was some partial success in
constructing supergravity models of such theories by relaxing the
constraints of Howe, the result turned out to be too clumsy to allow
further development so far \cite{Ertl:1997ib}.

As in higher dimensional theories, the gauge theoretic approach
provides a much simpler method for supersymmetrization than the
superfield approach. However, it is restricted to relatively simple
Lagrangians such as the one of the Jackiw-Teitelboim model (\ref{JT})
\cite{\bibSJTgauge}. The generic model (\ref{FOG}) or also (\ref{dil})
cannot be treated in this fashion.

On the other hand, first attempts showed that super dilaton theories
may fit into the framework of `nonlinear' supergauge theories
\cite{\bibNLSGT}, and the action for a super dilaton theory was
obtained (without superfields) by a nonlinear deformation of the
graded de~Sitter group using free differential algebras in
\cite{Izquierdo:1998hg}.

Recently, it turned out \cite{Strobl:1999zz}
(but \cf also \cite{\bibNLSGT}) that the framework of
PSMs \cite{\bibPSM}, now with a graded target space, represents the
formalism of choice to deal with super dilaton theories. In
particular, it allowed for a simple derivation of the general solution
of the corresponding field equations, and in this process yielded the
somewhat surprising result that, \emph{in the absence of additional
matter fields}, the supersymmetrization of the dilaton theories
(\ref{dil}) is on-shell trivial. By this we mean that, up to the
choice of a gauge, in the general solution to the field equations all
fermionic fields can be made to vanish identically by an appropriate
choice of gauge while the bosonic fields satisfy the field equations
of the purely bosonic theory and are still subject to the symmetries
of the latter.  This local on-shell triviality of the supersymmetric
extension may be interpreted superficially to be yet another
consequence of the fact that, from the Hamiltonian point of view, the
`dynamics' of (\ref{PSM}) is described by just one variable (the
Casimir function) which does not change when fermionic fields are
added. This type of triviality will cease to prevail in the case of
additional matter fields (as is already obvious from a simple counting
of fields and local symmetries involved).  The supersymmetrization may
be used \cite{Park:1993sd}, furthermore, as a technical device to
prove positive energy theorems for supersymmetric \emph{and}
non-supersymmetric dilaton theories. Thus, the (local) on-shell
triviality of pure 2d supergravity theories by no means implicitly
demolishes all the possible interest in their supersymmetric
generalizations. This applies similarly to the $F(R,\t^2)$-theories
and to the FOG formulation (\ref{FOG}), we are primarily interested in
within the present paper.

Graded PSMs (gPSMs) turn out to also provide a unifying and most
efficient framework for the \emph{construction} of supersymmetric
extensions of a two-dimensional gravity theory, at least as far as
theories of the initially mentioned type (\ref{grav}) are considered.
This route, sketched already briefly in \cite{Strobl:1999zz}, will be
followed in detail within the present paper.

The main idea of this approach will be outlined in
\Sec\ref{sec:outline}. It will be seen that within this framework the
problem for a supersymmetric extension of a gravity theory
(\ref{grav}) is reduced to a finite dimensional problem: Given a
Lorentz invariant Poisson bracket on a two-dimensional Minkowski space
(which in addition depends also on the `dilaton' or, equivalently, on
the generator of Lorentz transformations $\phi$), one has to extend
this bracket consistently and in a Lorentz covariant manner to the
corresponding superspace.

In spirit this is closely related to the analogous extension of Lie
algebras to superalgebras \cite{\bibSAlg}. In fact, in the particular
case of a linear dependence of $v$ in (\ref{PB}) the original Poisson
bracket corresponds to a three-dimensional Lie algebra, and likewise
any \emph{linear} extension of this Poisson algebra to a superalgebra.
Here we are dealing with general nonlinear Poisson algebras,
particular cases of which can be interpreted as finite $W$-algebras
(\cf \cite{deBoer:1996nu}). Due to that nonlinearity the analysis
necessary for the fermionic extension is much more involved and there
is a much higher ambiguity in an extension (except if one considers
this only modulo arbitrary (super)diffeomorphisms). For this reason we
mainly focus on an $\N=1$ extension within this paper.

After recapitulating some material on gPSMs in \Sec\ref{sec:gPSM2},
setting also our notation and conventions, the solution of the
$\phi$-components of the Jacobi identities is given in
\Sec\ref{sec:ans-gP} simply by writing down the most general Lorentz
covariant ansatz for the Poisson tensor.  In \Sec\ref{sec:remJac} the
remaining Jacobi identities are solved in full generality for
nondegenerate and degenerate $\N=1$ fermionic extension.

The observation that a large degree of arbitrariness is present in
these extensions is underlined also by the study of target space
diffeomorphism in \Sec\ref{sec:diffeo}. We also point out the
advantages of this method in the quest for new algebras and
corresponding gravity theories.

In \Sec\ref{sec:Poisson} we shall consider particular examples of the
general result. This turns out to be much superior than to perform a
general abstract discussion of the results of \Sec\ref{sec:Jacobi}.
The more so, because fermionic extensions of specific bosonic 2d
gravity theories can be investigated which have been discussed already
in the literature. Supersymmetric extensions of the KV-model
(\ref{KV}) as compared to SRG (\ref{SRG}) will serve for illustrative
purposes. 

The corresponding actions and their relation to the initial problem
(\ref{grav}) are given in \Sec\ref{sec:models}.  Also the general
relation to the supersymmetric dilatonic theories (\ref{dil}) will be
made explicit using the results of \Sec\ref{sec:action}. Several
different supersymmetrizations (one of which is even parity violating)
for the example of SRG are compared to the one provided previously in
the literature \cite{Park:1993sd}.  For each model the corresponding
supersymmetry is given explicitly.

In \Sec\ref{sec:sdil-sol} the explicit solution for a supergravity
theory with the bosonic part corresponding to $v^{\mathrm{dil}}$ in
(\ref{vdil}) is given.

In the final \Sec\ref{sec:summary} we will summarize our findings and
comment on possible further investigations.

\App\ref{sec:notation} defines notations and summarizes some useful
identities.

%********************************************************************
\section{Graded Poisson Sigma Model}
\label{sec:gPSM}

%********************************************************************
\subsection{Outline of the approach}
\label{sec:outline}

The PSM formulation of gravity theories allows a direct
generalization, yielding possible supergravity theories. Indeed, from
this perspective it is suggestive to replace the Minkowski space with
its linear coordinates $X^a$ by its superspace analogue, spanned by
$X^a$ and (real, \ie Majorana) spinorial and (one or more)
Grassmann-valued coordinates $\chi^{\q i\alpha}$ (where $\q i =
1,\ldots,\N$).  In the purely bosonic case we required that $\phi$
generates Lorentz transformations on Minkowski space. We now extend
this so that $\phi$ is the generator of Lorentz transformations on
superspace. This implies in particular that beside (\ref{LorentzB})
now also
\begin{equation}
  \label{LorentzF}
  \{ \chi^{\q i\alpha}, \phi \} = 
  -\half\chi^{\q i\beta} (\gamma^3)_\beta\^\alpha,
\end{equation} 
has to hold, where $-\half (\gamma^3)_\beta\^\alpha$ is the generator
of Lorentz transformations in the spinorial representation. For the
choice of the $\gamma$-matrices and further details on notation and
suitable identities we refer to \App\ref{sec:notation}.

Within the present work we first focus merely on a consistent
extension of the original bosonic Poisson algebra to the total
superspace. This superspace can be built upon $\N$ pairs of
coordinates obeying (\ref{LorentzF}). Given such a graded Poisson
algebra, the corresponding Sigma Model provides a possible
$\N$-supergravity extension of the original gravity model
corresponding to the purely bosonic sigma model. We shall mainly focus
on the construction of a graded Poisson tensor $\Poisson^{IJ}$ for the
simplest supersymmetric extension $\N=1$, \ie on a (`warped') product
of the above superspace and the linear space spanned by the generator
$\phi$.  Upon restriction to the bosonic submanifold $\chi^\alpha=0$,
the bracket will be required to coincide with the bracket
(\ref{LorentzB}) and (\ref{PB}) corresponding to the bosonic theory
(\ref{PSM}). Just as the framework of PSMs turns out to provide a
fully satisfactory and consistent 2d gravity theory with all the
essential symmetries for any given (Lorentz invariant) Poisson bracket
(\ref{LorentzB}) and (\ref{PB}), the framework of graded Poisson Sigma
Models (gPSM) will provide possible generalizations for any of the
brackets $\Poisson^{ij}$ with a local `supersymmetry' of the generic
type (\ref{susytrafo}). In particular, by construction of the general
theory (cf \cite{Strobl:1999zz} or \Sec\ref{sec:Jacobi} below) and
upon an identification which is a straightforward extension of
(\ref{ident1}), the resulting gravity theory will be invariant
automatically with respect to local Lorentz transformations, spacetime
diffeomorphisms \emph{and} local supersymmetry transformations. In
particular, the Rarita-Schwinger field $\psi_\a$ (or $\psi_{\q
  i\alpha}$, $\q i = 1,\ldots,\N$ in the more general case) is seen to
enter naturally as the fermionic component of the one-form valued
multiplet $A_I$.  Likewise, specializing the local symmetries
(\ref{PSM-symms}) (or rather their generalization to the graded case
provided in (\ref{gPSM-symms}) below) to the spinorial part
$\epsilon_\a$, local supersymmetry transformations of the form
(\ref{susytrafo}) are found, which, by construction, are symmetries of
the action. (In fact, it is here where the graded Jacobi identity for
$\Poisson$ enters as an essential ingredient!) Finally, by
construction, the bosonic part of the action of the gPSM corresponding
to the bracket $\Poisson^{IJ}$ will coincide with (\ref{PSM}).  Thus,
for any such a bracket $\Poisson^{IJ}$, the resulting model should
allow the interpretation as a permissible supersymmetric
generalization of the original bosonic starting point.

The relations (\ref{LorentzB}), (\ref{LorentzF}) fix the $\phi$
components of the sought for (graded) Poisson tensor $\Poisson^{IJ}$.
We are thus left with determining the remaining components
$\Poisson^{AB}$, $A$ and $B$ being indices in the four-dimensional
superspace with $X^A = (X^a,\chi^\alpha)$. As will be recapitulated in
\Sec\ref{sec:gPSM2}, beside the graded symmetry of the tensor
$\Poisson^{IJ}$, the only other requirement it has to fulfil by
definition is the graded Jacobi identity. This is of the form of the
vanishing of a 3-tensor $J^{IJK}$ (\cf (\ref{gJacobi}) below),
which may be expressed also as the Schouten-Nijenhuis bracket $[ \cdot
, \cdot ]_{SN}$ of $(\Poisson^{IJ})$ with itself. In this formulation
$(\Poisson^{IJ})$ is meant to be the Poisson tensor itself and not its
components (abstract indices).  It is straightforward to verify (\cf
also \cite{Strobl:1999zz}) that the relations $J^{IJK}=0$ with at
least one of the indices coinciding with the one corresponding to
$\phi$ are satisfied, \emph{iff} $(\Poisson^{AB})$ is a Lorentz
covariant 2-tensor,
\begin{equation} \label{Pinv} \mathcal{L}_{(\Poisson^{A \phi})}
(\Poisson^{AB}) = 0,
\end{equation}
\ie depending on $X^a$, $\chi^\a$ and also on the Lorentz invariant
quantity $\phi$ in a covariant way as determined by its indices.  Thus
one is left with finding the general solution of $J^{ABC}=0$ starting
from a Lorentz covariant ansatz for $(\Poisson^{AB})$.

Let us note on this occasion that the above considerations do
\emph{not} imply that $(\Poisson^{AB})$ forms a bracket on the
Super-Minkowski space, a subspace of the target space
under discussion. The reason is that the
equations $J^{ABC}=0$ contain also derivatives of $\Poisson^{AB}$ with
respect to $\phi$: in terms of the Schouten-Nijenhuis bracket, the
remaining equations become
\begin{equation}
  \label{Schouten}
  [(\Poisson^{AB}),(\Poisson^{AB})]_{SN} = (\Poisson^{A
    \phi}) \wedge (\partial_\phi \Poisson^{AB}),
\end{equation} 
where the components of the supervector $(\Poisson^{A \phi})$ are
given implicitly by eqs.~(\ref{LorentzB}) and (\ref{LorentzF}) above.
So $(\Poisson^{AB})$ defines a graded Poisson bracket for the $X^A$
only if it is independent of $\phi$.  However, in the present context
$\phi$-independent Poisson tensors are uninteresting in view of our
discussion of actions of the form (\ref{FOG}).

It should be remarked that given a particular bosonic model and its
corresponding bracket, there is by no means a unique graded extension,
even for fixed $\N$. Clearly, any (super-)diffeomorphism leaving
invariant the bosonic sector as well as the brackets (\ref{LorentzF})
applied to a solution of the (graded) Jacobi identities yields another
solution.  This induces an ambiguity for the choice of a
superextension of a given gravity model (\ref{dil}) or also
(\ref{grav}). This is in contrast to the direct application of, say,
the superfield formalism of Howe \cite{Howe:1979ia}, which when
applied to the (necessarily torsionfree) theory (\ref{dil})
\cite{Park:1993sd}, yields one particular superextension. This now turns
out as just one possible choice within an infinite dimensional space
of admissible extensions.  {}From one or the other perspective,
however, different extensions (for a given $\N$) may be regarded also
as effectively equivalent. We shall come back to these issues below.

A final observation concerns the relation of our supersymmetric
extensions to `ordinary' supergravity. {}From the point of view of the
seminal work on the 2d analogue \cite{Deser:1976rb,Brink:1977vg} of 4d
supergravity our supergravity algebra is `deformed' by the presence of
a dilaton field. Such a feature is known also from the dimensional
reduction of supergravity theories in higher dimensions, where one or
more dilaton fields arise from the compactification.

%********************************************************************
\subsection{Details of the gPSM}
\label{sec:gPSM2}

In this Section we recollect for completeness some general and
elementary facts about graded Poisson brackets and the corresponding
Sigma Models. This Section (\cf also \App\ref{sec:notation}) also sets
the conventions about signs \etc used within the present paper, which
are adapted to those of \cite{Ertl:1997ib} and which differ on various
instances from those used in \cite{Strobl:1999zz}.

For the construction of the gPSM we take a 2-dimensional base manifold
$\BMf$, also called world sheet or space-time manifold, with purely
bosonic (commutative) coordinates $x^m$, and the target space $\TSp$
with coordinates $X^I = (\phi, X^A) = (\phi, X^a, \chi^\alpha)$,
$\phi$ and $X^a$ being bosonic and $\chi^\alpha$ fermionic
(anticommutative), promoting $\TSp$ to a supermanifold.  The
restriction to one Majorana spinor means that only the case $\N=1$ is
implied in what follows.  To the coordinate functions $X^I$ correspond
gauge fields $A_I$ which we identify with the usual Lorentz-connection
1-form $\omega$ and the vielbein 1-form $e_a$ of the Einstein-Cartan
formalism of gravity and with the Rarita-Schwinger 1-form
$\psi_\alpha$ of supergravity according to $A_I = (\omega, e_A) =
(\omega, e_a, \psi_\alpha)$.  They can be viewed as 1-forms on the
base manifold $\BMf$ with values in the cotangential space of $\TSp$,
and may be collected in the total 1-1-form
$A = dX^I A_I = dX^I dx^m A_{mI}$.  As the main structure of the model
we choose a Poisson tensor $\Poisson^{IJ} = \Poisson^{IJ}(X)$ on
$\TSp$, which encodes the desired symmetries and the dynamics of the
theory to be constructed.  Due to the grading of the coordinates of
$\TSp$ it is graded antisymmetric $\Poisson^{IJ} = -(-1)^{IJ}
\Poisson^{JI}$ and is assumed to fulfil the graded Jacobi identity
($\rpartial_I =\rpartial/\partial X^I$ is the right derivative,
\cf\App\ref{sec:notation}) of which we list also a convenient
alternative version
\begin{align}
  J^{IJK} &= \Poisson^{IL} \rpartial_L \Poisson^{JK} + \gcycl(IJK)
  \label{gJacobi} \\
  &= \Poisson^{IL} \rpartial_L \Poisson^{JK}
  + \Poisson^{JL} \rpartial_L \Poisson^{KI} (-1)^{I(J+K)}
  + \Poisson^{KL} \rpartial_L \Poisson^{IJ} (-1)^{K(I+J)} \\
  &= 3 \Poisson^{I]L} \rpartial_L \Poisson^{[JK} = 0.
\end{align}
The Poisson tensor defines the Poisson bracket of functions $f$, $g$
on $\TSp$,
\begin{equation}
  \label{gPB}
  \{f, g\} = (f \lpartial_J) \Poisson^{JI} (\rpartial_I g),
\end{equation}
implying for the coordinate functions $\{X^I, X^J\} = \Poisson^{IJ}$.
With (\ref{lr-partial}) the Poisson bracket (\ref{gPB}) may be written
also as
\begin{equation}
  \label{gPBv2}
  \{f, g\} = \Poisson^{JI} (\rpartial_I g) (\rpartial_J f) (-1)^{g(f+J)}.
\end{equation}
This bracket is graded anticommutative,
\begin{equation}
  \label{gPB-symm}
  \{f, g\} = -(-1)^{fg} \{g, f\},
\end{equation}
and fulfils the graded Jacobi identity
\begin{equation}
  \label{gPB-Jacobi}
  \{X^I, \{X^J, X^K\}\} (-1)^{IK} + \{X^J, \{X^K, X^I\}\} (-1)^{JI} +
  \{X^K, \{X^I, X^J\}\} (-1)^{KJ} = 0,
\end{equation}
which is equivalent to the graded derivation property
\begin{equation}
  \label{gPB-der}
  \{X^I, \{X^J, X^K\}\} = \{\{X^I, X^J\}, X^K\} + (-1)^{IJ} \{X^J, \{X^I,
  X^K\}\}.
\end{equation}

The PSM action (\ref{PSM}) generalizes to
\begin{equation}
  \label{gPSM}
  L^\mathrm{gPSM} = \int_\BMf dX^I A_I + \half \Poisson^{IJ} A_J A_I,
\end{equation}
where in the graded case the sequence of the indices is important.
The functions $X^I(x)$ represent a map from the base manifold to the
target space in the chosen coordinate systems of $\BMf$ and $\TSp$,
and $dX^I$ is the shorthand notation for the derivatives
$d^\BMf\!X^I(x) = dx^m \partial_m X^I(x)$ of these functions. The
reader may notice the overloading of the symbols $X^I$ which sometimes
are used to denote the map from the base manifold to the target space
and sometimes, as in the paragraph above, stand for target space
coordinates. This carries over to other expressions like $dX^I$ which
denote the coordinate differentials $d^\TSp\!X^I$ on $\TSp$ and, on
other occasions, as in the action (\ref{gPSM}), the derivative of the
map from $\BMf$ to $\TSp$.

The variation of $A_I$ and $X^I$ in (\ref{gPSM}) yields the gPSM field
equations
\begin{gather}
  dX^I + \Poisson^{IJ} A_J = 0, \label{gPSM-eomX} \\
  dA_I + \half (\rpartial_I \Poisson^{JK}) A_K A_J = 0.
  \label{gPSM-eomA}
\end{gather}
These are first order differential equations of the fields $X^I(x)$
and $A_{mI}(x)$ and the Jacobi identity (\ref{gJacobi}) of the Poisson
tensor ensures the closure of (\ref{gPSM-eomX}) and (\ref{gPSM-eomA}).
As a consequence of (\ref{gJacobi}) the action exhibits the symmetries
\begin{equation}
  \label{gPSM-symms}
  \delta X^I = \Poisson^{IJ} \epsilon_J, \qquad
  \delta A_I = -d\epsilon_I - (\rpartial_I \Poisson^{JK}) \epsilon_K A_J,
\end{equation}
where corresponding to each gauge field $A_I$ we have a symmetry
parameter $\epsilon_I(x)$ with the same grading which is a function of
$x$ only. When calculating the commutator of these symmetries, in
general parameters depending on both $x$ and $X$ are obtained.  For
two parameters $\epsilon_{1I}(x,X)$ and $\epsilon_{2I}(x,X)$
\begin{align}
  \left( \delta_1 \delta_2 - \delta_2 \delta_1 \right) X^I &= \delta_3
  X^I, \label{gCommX} \\
  \left( \delta_1 \delta_2 - \delta_2 \delta_1 \right) A_I &= \delta_3
  A_I + \left( dX^J + \Poisson^{JK} A_K \right) (\rpartial_J
  \rpartial_I \Poisson^{RS}) \epsilon_{1S} \epsilon_{2R}
  \label{gCommA}
\end{align}
follows, where $\epsilon_{3I}(x,X)$ of the resulting variation
$\delta_3$ are given by the Poisson bracket (or Koszul-Lie bracket) of
the 1-forms $\epsilon_1 = dX^I \epsilon_{1I}$ and $\epsilon_2 = dX^I
\epsilon_{2I}$, defined according to
\begin{equation}
  \label{gCommEp3}
  \epsilon_{3I} = \{ \epsilon_2, \epsilon_1 \}_I
  := (\rpartial_I \Poisson^{JK}) \epsilon_{1K} \epsilon_{2J} +
  \Poisson^{JK} \left( \epsilon_{1K} \rpartial_J \epsilon_{2I} -
    \epsilon_{2K} \rpartial_J \epsilon_{1I} \right).
\end{equation}
Note, that the commutator of the PSM symmetries closes if the Poisson
tensor is linear, for non-linear Poisson tensors the algebra closes
only on-shell (\ref{gCommA}).

Right and left Hamiltonian vector fields are defined by $\rvec{T}^I =
\{X^I, \cdot\}$ and $\lvec{T}^I = \{\cdot, X^I\}$, respectively, \ie
by
\begin{equation}
  \label{gHVF}
  \rvec{T}^I \cdot f = \{X^I, f\} = \Poisson^{IJ} (\rpartial_J f),
  \qquad
  f \cdot \lvec{T}^I = \{f, X^I\} = (f \lpartial_J) \Poisson^{JI}.
\end{equation}
The vector fields $\lvec{T}^I$ are the generators of the symmetries,
$\delta X^I = X^I \cdot \lvec{T}^J \epsilon_J$. From their commutator
the algebra
\begin{equation}
  \label{gStrucFunc}
  [\lvec{T}^I, \lvec{T}^J] = \lvec{T}^K f_K\^{IJ}(X)
\end{equation}
follows with the structure functions $f_K\^{IJ} = (\rpartial_K
\Poisson^{IJ})$.  Structure constants and therefore Lie algebras are
obtained when the Poisson tensor depends only linearly on the
coordinates, which is true for Yang-Mills gauge theory and simple
gravity models like (anti-)de~Sitter gravity.

As in the purely bosonic case the kernel of the graded Poisson
algebra determines the so-called Casimir functions $\Casimir$ obeying
$\{ \Casimir, X^I\} = 0$. When the co-rank of the bosonic
theory---with one Casimir function---is not changed we shall call this
case non-degenerate. Then $\Poisson^{\alpha\beta}|$, the bosonic part
of the fermionic extension, must be of full rank. For $\N=1$
supergravity and thus one target space Majorana spinor $\chi^\alpha$,
the expansion of $\Casimir$ in $\chi^\alpha$ reads ($\chi^2 =
\chi^\alpha \chi_\alpha$, \cf\App\ref{sec:notation})
\begin{equation}
  \label{Casimir}
  \Casimir = \casimir + \half \chi^2 \casimir_2,
\end{equation}
where $\casimir$ and $\casimir_2$ are functions of $\phi$ and $Y
\equiv \half X^a X_a$ only.  This assures that the Poisson bracket
$\{\phi, \Casimir\}$ is zero.  From the bracket $\{X^a, \Casimir\} =
0$, to zeroth order in $\chi^\alpha$, the defining equation of the
Casimir function for pure bosonic gravity PSMs becomes
\begin{equation}
  \label{c}
  \derv \casimir :=
  \left( \partial_\phi - v \partial_Y \right) \casimir = 0.
\end{equation}
This is the well-known partial differential equation of that quantity
\cite{Klosch:1996fi,Strobl:1999Habil}. The solution of (\ref{c}) for
bosonic potentials relevant for kinetic dilaton theories (\ref{vdil})
can be given by ordinary integration,
\begin{alignat}{2}
  \casimir(\phi,Y) &= Y e^{Q(\phi)} + W(\phi), \label{c-sol} \\
  Q(\phi) &= \int_{\phi_1}^{\phi} Z(\varphi) d\varphi, &\qquad
  W(\phi) &= \int_{\phi_0}^{\phi} e^{Q(\varphi)} V(\varphi)
  d\varphi. \label{QW}
\end{alignat}
The new component $\casimir_2$ is derived by considering the terms
proportional to $\chi^\beta$ in the bracket $\{\chi^\alpha, \Casimir\}
= 0$. Thus $\casimir_2$ will depend on the specific fermionic
extension. In the degenerate case, when $\Poisson^{\alpha\beta}$ is
not of full rank, there will be more than one Casimir function,
including purely Grassmann valued ones (see \Sec\ref{sec:DFS-P} and
\ref{sec:CPA-P}).

%********************************************************************
\section{Solution of the Jacobi-Identities}
\label{sec:Jacobi}

As mentioned above, in order to obtain the general solution of the
graded Jacobi identities a suitable starting point is the use of
Lorentz symmetry in a most general ansatz for $\Poisson^{IJ}$.
Alternatively, one could use a simple $\Poisson^{IJ}_{(0)}$ which
trivially fulfils (\ref{gJacobi}). Then the most general
$\Poisson^{IJ}$ may be obtained by a general diffeomorphism in target
space. The first route will be followed within this section. We will
comment upon the second one in \Sec\ref{sec:diffeo}.

%********************************************************************
\subsection{Lorentz-Covariant Ansatz for the Poisson-Tensor}
\label{sec:ans-gP}

Lorentz symmetry determines the mixed components $\Poisson^{A\phi}$ of
$\Poisson^{IJ}$,
\begin{equation}
  \label{P-Lorentz}
  \Poisson^{a\phi} = X^b \epsilon_b\^a, \qquad
  \Poisson^{\alpha\phi} = - \half \chi^\beta 
  (\gamma^3)_\beta\^\alpha.
\end{equation}
All other components of the Poisson tensor must be Lorentz-covariant
(\cf the discussion around (\ref{Pinv})).  Expanding them in
terms of invariant tensors $\eta^{ab}$, $\epsilon^{ab}$,
$\epsilon^{\alpha\beta}$ and $\gamma$-matrices yields
\begin{align}
  \Poisson^{ab} &= V \epsilon^{ab}, \label{ans-P-bb} \\
  \Poisson^{\alpha b} &= \chi^\beta (F^b)_\beta\^\alpha,
  \label{ans-P-fb} \\
  \Poisson^{\alpha\beta} &= U (\gamma^3)^{\alpha\beta} + i \widetilde U X^c
  (\gamma_c)^{\alpha\beta} + i \widehat U X^c \epsilon_c\^d
  (\gamma_d)^{\alpha\beta}. \label{ans-P-ff}
\end{align}
The quantities $V$, $U$, $\widetilde U$ and $\widehat U$ are functions
of $\phi$, $Y$ and $\chi^2$.  Due to the anticommutativity of
$\chi^\alpha$ the dependence on $\chi^2$ is at most linear. Therefore
\begin{equation}
  \label{ans-V}
  V = v(\phi, Y) + \half \chi^2\, v_2(\phi, Y)
\end{equation}
depends on two Lorentz-invariant functions $v$ and $v_2$ of $\phi$ and
$Y$. An analogous notation will be implied for $U, \widetilde U$ and
$\widehat{U}$, using the respective small case letter for the
$\chi$-independent component of the superfield and an additional index
2 for the respective $\chi^2$-component.  The component
(\ref{ans-P-fb}) contains the spinor matrix $(F^a)_\beta\^\gamma$,
which may be first expanded in terms of the linearly independent
$\gamma$-matrices,
\begin{equation}
  \label{ans-f-bff}
  (F^a)_\beta\^\gamma = f_{(1)}^a \delta_\beta\^\gamma + i f^{ab}
  (\gamma_b)_\beta\^\gamma + f_{(5)}^a (\gamma^3)_\beta\^\gamma \; .
\end{equation}
The Lorentz-covariant coefficient functions in (\ref{ans-f-bff}) are
further decomposed according to
\begin{align}
  f_{(1)}^a &= f_{(11)} X^a - f_{(12)} X^b \epsilon_b\^a,
  \label{ans-f1-b} \\
  f_{(5)}^a &= f_{(51)} X^a - f_{(52)} X^b \epsilon_b\^a,
  \label{ans-f5-b} \\
  f^{ab} &= f_{(s)} \eta^{ab} + f_{(t)} X^a X^b - f_{(h)} X^c
  \epsilon_c\^a X^b + f_{(a)} \epsilon^{ab}. \label{ans-f-bb}
\end{align}
The eight Lorentz-invariant coefficients $f_{(11)}$, $f_{(12)}$,
$f_{(51)}$, $f_{(52)}$, $f_{(s)}$, $f_{(t)}$, $f_{(h)}$ and $f_{(a)}$
are functions of $\phi$ and $Y$ only. The linearity in $\chi^\alpha$
of (\ref{ans-P-fb}) precludes any $\chi^2$ term in (\ref{ans-f-bff}).

Below it will turn out to be convenient to use a combined notation for
the bosonic and the $\chi^2$-dependent part of
$\Poisson^{\alpha\beta}$,
\begin{equation}
  \label{v-ff}
  \Poisson^{\alpha\beta} = v^{\alpha\beta} + \half \chi^2
  v_2^{\alpha\beta},
\end{equation}
where $v^{\alpha\beta}$ and $v_2^{\alpha\beta}$ are particular
matrix-valued functions of $\phi$ and $X^a$, namely, in the notation
above (\cf also \App\ref{sec:notation} for the definition of $X^{++}$
and $X^{--}$),
\begin{equation}
  \label{matrix}
  v^{\alpha\beta} = \mtrx{\sqrt{2}X^{++}(\tilde u - \hat u)}{-u}
  {-u}{\sqrt{2}X^{--}(\tilde u + \hat u)},
\end{equation}
and likewise with suffix 2. Note that the symmetric $2 \times 2$
matrix $v^{\alpha\beta}$ still depends on three arbitrary real
functions; as a consequence of Lorentz invariance, however, they are
functions of $\phi$ and $Y$ only. A similar explicit matrix
representation may be given also for $F^{\pm \pm}\_\a\^\b$.

%********************************************************************
\subsection{Remaining Jacobi Identities}
\label{sec:remJac}

The Jacobi identities $J^{\phi BC}=0$ have been taken care of
automatically by the Lorentz covariant parametrization introduced in
\Sec\ref{sec:ans-gP}. In terms of these functions we write the
remaining identities as
\begin{align}
  J^{\alpha\beta\gamma} &= \rvec T^\a(\Poisson^{\b\g}) + \cycl(\a\b\g)
  = 0, \label{J-fff} \\
  J^{\alpha\beta c} &= \rvec T^c(\Poisson^{\alpha\beta}) + \rvec
  T^\alpha(\chi F^c)^\beta + \rvec T^\beta(\chi F^c)^\alpha = 0,
  \label{J-ffb} \\
  \half J^{\alpha bc} \epsilon_{cb} &= \rvec T^\alpha(V) - \rvec
  T^b(\chi F^c)^\alpha \epsilon_{cb} = 0. \label{J-fbb}
\end{align}
Here $\rvec T^a$ and $\rvec T^\a$ are Hamiltonian vector fields
introduced in (\ref{gHVF}), yielding ($\partial_\phi =
\frac{\partial}{\partial\phi}$, $\partial_a = \frac{\partial}{\partial
  X^a}$, $\partial_\alpha = \frac{\partial}{\partial \chi^\alpha}$)
\begin{align}
  \rvec T^a &= X^b \epsilon_b\^a \partial_\phi + \left( v +
    \half\chi^2 v_2 \right) \epsilon^{ab}
  \partial_b - (\chi F^a)^\beta \partial_\beta, \label{T-b} \\
  \rvec T^\alpha &= -\half (\chi\gamma^3)^\alpha \partial_\phi + (\chi
  F^b)^\alpha \partial_b + \left( v^{\alpha\beta} + \half\chi^2
    v_2\^{\alpha\beta} \right) \partial_\beta.
  \label{T-f}
\end{align}
To find the solution of (\ref{J-fff})--(\ref{J-fbb}) it is necessary
to expand in terms of the anticommutative coordinate $\chi^\alpha$.
Therefore, it is convenient to split off any dependence on
$\chi^\alpha$ and its derivative also in (\ref{T-b}) and (\ref{T-f}),
using instead the special Lorentz vector and spinor matrix valued
derivatives\footnote{When (\ref{derv-b}) acts on an invariant function
  of $\phi$ and $Y$, $\derv^c$ essentially reduces to the `scalar'
  derivative, introduced in (\ref{c}).}
\begin{align}
  \derv^c &:= X^d \epsilon_d\^c \partial_\phi + v \epsilon^{cd}
  \partial_d, \label{derv-b} \\
  \derv_\delta\^\alpha &:= -\half (\gamma^3)_\delta\^\alpha
  \partial_\phi + F^d\_\delta\^\alpha \partial_d. \label{derv-ff}
\end{align}
Then the Jacobi identities, arranged in the order $J^{\alpha\beta
  c}|$, $J^{\alpha\beta\gamma}|_{\chi}$, $J^{\alpha bc}|_{\chi}$ and
$J^{\alpha\beta c}|_{\chi^2}$, that is the order of increasing
complexity best adapted for our further analysis, read
\begin{gather}
  v^{\alpha)\gamma} F^c\_\gamma\^{(\beta} + \half \derv^c v^{\alpha\beta}
  = 0, \label{J-ffb-0} \\
  v_\delta\^\alpha v_2^{\beta\gamma} - \derv_\delta\^\alpha
  v^{\beta\gamma} + \cycl(\alpha\beta\gamma) = 0, \label{J-fff-1}
  \\
  v_\delta\^\alpha v_2 - \derv_\delta\^\alpha v + \derv^c
  F^b\_\delta\^\alpha \epsilon_{bc} - (F^c F^b)_\delta\^\alpha
  \epsilon_{bc} = 0 , \label{J-fbb-1} \\
  \derv^c v_2^{\alpha\beta} - F^c\_\delta\^\delta v_2^{\alpha\beta} +
  v_2 \epsilon^{cd} \partial_d v^{\alpha\beta} + 2
  \derv^{\delta(\alpha|} F^c\_\delta\^{|\beta)} + 2
  v_2^{\alpha)\delta} F^c\_\delta\^{(\beta} = 0. \label{J-ffb-2}
\end{gather}

All known solutions for $d=2$ supergravity models found in the
literature have the remarkable property that the Poisson tensor has
(almost everywhere, \ie except for isolated points) constant rank
four, implying exactly one conserved Casimir function $\Casimir$
\cite{Strobl:1999zz}. Since the purely bosonic Poisson tensor has
(almost everywhere) maximal rank two, this implies that the respective
fermionic bracket $\Poisson^{\alpha\beta}$ (or, equivalently, its
$\chi$-independent part $v^{\alpha\beta}$) must be of full rank if
only one Casimir function is present in the fermionic extension. In
the following subsection we will consider this case, \ie we will
restrict our attention to (regions in the target space with)
invertible $\Poisson^{\alpha\beta}$. For describing the rank we
introduce the notation $(B|F)$. Here $B$ denotes the rank of the
bosonic body of the algebra, $F$ the one of the extension. In this
language the nondegenerate case has rank $(2|2)$. The remaining
degenerate cases with rank $(2|0)$ and $(2|1)$ will be analyzed in a
second step (\Sec\ref{sec:DFS-P} and \Sec\ref{sec:CPA-P}).

%********************************************************************
\subsubsection{Nondegenerate Fermionic Sector}
\label{sec:sol-gP}

When the matrix $v^{\alpha\beta}$ in (\ref{v-ff}) is nondegenerate,
\ie when its determinant
\begin{equation}
  \label{detv}
  \detv := \det(v^{\alpha\beta}) = \half v^{\alpha\beta}
  v_{\beta\alpha}
\end{equation}
is nonzero, for a given bosonic bracket this yields all supersymmetric
extensions of maximal total rank. We note in parenthesis that due to
the two-dimensionality of the spinor space (and the symmetry of
$v^{\alpha\beta}$) the inverse matrix to $v^{\alpha\beta}$ is nothing
else but $v_{\a\b}/\detv$, which is used in several intermediary steps
below.

The starting point of our analysis of the remaining Jacobi identities
$J^{ABC}=0$ will always be a certain ansatz, usually for
$v^{\alpha\beta}$. Therefore, it will be essential to proceed in a
convenient sequence so as to obtain the restrictions on the remaining
coefficient functions in the Poisson tensor with the least effort.
This is also important because it turns out that several of these
equations are redundant. This sequence has been anticipated in
(\ref{J-ffb-0})--(\ref{J-ffb-2}). There are already redundancies
contained in the second and third step (eqs.\ (\ref{J-fff-1}) and
(\ref{J-fbb-1})), while the $\chi^2$-part of $J^{\alpha\beta c} = 0$
(eq.\ (\ref{J-ffb-2})) turns out to be satisfied identically because
of the other equations.  It should be noted, though, that this
peculiar property of the Jacobi identities is \emph{not} a general
feature, resulting \eg from some hidden symmetry, it holds true only
in the case of a nondegenerate $\Poisson^{\a\b}$ (\cf the discussion
of the degenerate cases below).

For fixed (nondegenerate) $v^{\alpha\beta}$, all solutions of
(\ref{J-ffb-0}) are parametrized by a Lorentz vector field $f^a$ on
the coordinate space $(\phi, X^a)$:
\begin{equation}
  \label{sol-f-bff}
  F^c\_\alpha\^\beta =  \left[ f^c \epsilon^{\g\beta} - \derv^c
    v^{\g\b} \right] \frac{v_{\g\a}}{2 \detv}
\end{equation}
Eq.\ (\ref{J-fff-1}) can be solved to determine $v_2^{\alpha\beta}$
in terms of $v^{\alpha\beta}$:
\begin{align}
  v_2\^{\alpha\beta} &= -\frac{1}{4\detv} v_\gamma\^\delta \left[
    \derv_\delta\^\gamma v^{\alpha\beta} + \cycl(\alpha\beta\gamma)
  \right] \label{sol-v2-ff}
\end{align}
Multiplying (\ref{J-fbb-1}) by $v_\beta\^\gamma$ yields
\begin{equation}
  \detv \delta_\beta\^\alpha v_2 =  v_\beta\^\delta \left[ -
    \derv_\delta\^\alpha v + \derv^c F^b\_\delta\^\alpha \epsilon_{bc}
    - (F^c F^b)_\delta\^\alpha \epsilon_{bc} \right]. \label{sol-v2}
\end{equation}
The trace of (\ref{sol-v2}) determines $v_2$, which is thus seen to
depend also on the original bosonic potential $v$ of (\ref{PB}).

Neither the vanishing
traces of (\ref{sol-v2}) multiplied with $\gamma^3$ or with
$\gamma_a$, nor the identity (\ref{J-ffb-2}) provide new restrictions
in the present case. This has been checked by extensive computer
calculations \cite{Ertl:Index-0.14.2}, based upon the explicit
parametrization (\ref{P-Lorentz})--(\ref{ans-f-bb}), which were
necessary because of the extreme algebraic complexity of this problem.
It is a remarkable feature of (\ref{sol-f-bff}), (\ref{sol-v2-ff}) and
(\ref{sol-v2}) that the solution of the Jacobi identities for the
nondegenerate case can be obtained from algebraic equations only.

As explained at the end of \Sec\ref{sec:gPSM2} the fermionic extension
of the bosonic Casimir function $\casimir$ can be derived from $\{
\chi^\alpha, \Casimir \} = 0$. The general result for the
nondegenerate case we note here for later reference
\begin{equation}
  \label{c2}
  \casimir_2 = -\frac{1}{2\detv} v_\alpha\^\beta \left( -\half
    (\gamma^3)_\beta\^\alpha \partial_\phi + F^d\_\beta\^\alpha
    \partial_d \right) \casimir.
\end{equation}

The algebra of full rank $(2|2)$ with the above solution for
$F^c\_\alpha\^\beta$, $v_2\^{\alpha\beta}$ and $v_2$ depends on 6
independent functions $v$, $v^{\alpha\beta}$ and $f^a$ and their
derivatives.  The original bosonic model determines the `potential'
$v$ in (\ref{FOG}) or (\ref{PB}). Thus the arbitrariness of
$v^{\alpha\beta}$ and $f^a$ indicates that the supersymmetric
extensions, obtained by fermionic extension from the PSM, are far from
unique. This has been mentioned already in the previous section and we
will further illuminate it in the following one.

%*********************************************************************
\subsubsection[Degenerate Fermionic Sector, Rank $(2|0)$]{Degenerate
  Fermionic Sector, Rank \mathversion{bold}$(2|0)$}
\label{sec:DFS-P}

For vanishing rank of $\Poisson^{\alpha\beta}|$, \ie $v^{\alpha\beta}
= 0$, the identities (\ref{J-ffb-0}) and (\ref{J-fff-1}) hold
trivially whereas the other Jacobi identities become complicated
differential equations relating $F^a$, $v$ and $v_2^{\alpha\beta}$.
However, these equations can again be reduced to algebraic ones for
these functions when the information on additional Casimir functions
is employed, which appear in this case. These have to be of fermionic
type with the general ans\"a{}tze
\begin{align}
  \Casimir^{(+)} &= \chi^+ \left| \frac{X^{--}}{X^{++}}
  \right|^{\frac{1}{4}} \casimir_{(+)}, \label{DFS-Cf+} \\
  \Casimir^{(-)} &= \chi^- \left| \frac{X^{--}}{X^{++}}
  \right|^{-\frac{1}{4}} \casimir_{(-)}. \label{DFS-Cf-}
\end{align}
The quotients $X^{--}/X^{++}$ assure that $\casimir_{(\pm)}$ are
Lorentz invariant functions of $\phi$ and $Y$.  This is possible
because the Lorentz boosts in two dimensions do not mix chiral
components and the light cone coordinates $X^{\pm\pm}$.

Taking a Lorentz covariant ansatz for the Poisson tensor as specified
in \Sec\ref{sec:ans-gP}, $\Casimir^{(+)}$ and $\Casimir^{(-)}$ must
obey $\{ X^a, \Casimir^{(+)} \} = \{ X^a, \Casimir^{(-)} \} = 0$. Both
expressions are linear in $\chi^\alpha$, therefore, the coefficients
of $\chi^\alpha$ have to vanish separately. This leads to $(F^a)_-\^+
= 0$ and $(F^a)_+\^- = 0$ immediately. With the chosen representation
of the $\gamma$-matrices (\cf \App\ref{sec:notation}) it is seen that
(\ref{ans-f-bff}) is restricted to $f^{ab}=0$, \ie the potentials
$f_{(s)}$, $f_{(t)}$, $f_{(h)}$ and $f_{(a)}$ have to vanish. A
further reduction of the system of equations reveals the further
conditions $f_{(11)} = 0$\footnote{In fact $f_{(11)}$ vanishes in all
  cases, \ie also for rank $(2|2)$ and $(2|1)$.} and
\begin{equation}
  \label{DFS-v}
  v = 4Y f_{(51)}.
\end{equation}
This leaves the differential equations for $\casimir_{(+)}$ and
$\casimir_{(-)}$
\begin{align}
  \left( \derv + f_{(12)} + f_{(52)} \right) \casimir_{(+)} &= 0,
  \label{DFS-c+} \\
  \left( \derv + f_{(12)} - f_{(52)} \right) \casimir_{(-)} &=
  0. \label{DFS-c-}
\end{align}
The brackets $\{ \chi^+, \Casimir^{(+)} \}$ and $\{ \chi^-,
\Casimir^{(-)} \}$ are proportional to $\chi^2$; the resulting
equations require $\tilde{u}_2 = \hat{u}_2 = 0$. The only surviving
term $u_2$ of $\Poisson^{\alpha\beta}$ is related to $F^a$ via $u_2 =
-f_{(12)}$ as can be derived from $\{ \chi^-, \Casimir^{(+)} \} = 0$
as well as from $\{ \chi^+, \Casimir^{(-)} \} = 0$, which are
equations of order $\chi^2$ too.

Thus the existence of the fermionic Casimir functions (\ref{DFS-Cf+})
and (\ref{DFS-Cf-}) has lead us to a set of \emph{algebraic} equations
among the potentials of the Lorentz covariant ansatz for the Poisson
tensor, and the number of independent potentials has been reduced
drastically.  The final question, whether the Jacobi identities are
already fulfilled with the relations found so far finds a positive
answer, and the general Poisson tensor with degenerate fermionic
sector, depending on four parameter functions $v(\phi,Y)$,
$v_2(\phi,Y)$, $f_{(12)}(\phi,Y)$ and $f_{(52)}(\phi,Y)$ reads
\begin{align}
  \Poisson^{ab} &= \left( v + \half\chi^2 v_2 \right)
  \epsilon^{ab}, \label{DFS-P-bb} \\
  \Poisson^{\alpha b} &= \frac{v}{4Y} X^b (\chi\gamma^3)^\alpha -
  f_{(52)} X^c \epsilon_c\^b (\chi\gamma^3)^\alpha - f_{(12)} X^c
  \epsilon_c\^b \chi^\alpha, \label{DFS-P-fb} \\
  \Poisson^{\alpha\beta} &= -\half\chi^2 f_{(12)} \gamma^3\^{\alpha\beta}.
  \label{DFS-P-ff}
\end{align}
This Poisson tensor possesses three Casimir functions: two fermionic
ones defined in regions $Y \neq 0$ according to (\ref{DFS-Cf+}) and
(\ref{DFS-Cf-}), where $\casimir_{(+)}$ and $\casimir_{(-)}$ have to
fulfil the first order differential equations (\ref{DFS-c+}) and
(\ref{DFS-c-}), respectively, and one bosonic Casimir function
$\Casimir$ of the form (\ref{Casimir}), where $\casimir$ is a solution
of the bosonic differential equation (\ref{c})---note the definition
of $\derv$ therein---and where $\casimir_2$ has to obey
\begin{equation}
  \left( \derv + 2 f_{(12)} \right) \casimir_2 = v_2 \partial_Y
  \casimir.
\end{equation}

Let us finally emphasize that it was decisive within this subsection
to use the information on the \emph{existence} of Casimir functions.
This follows from the property of the bivector $\Poisson^{IJ}$ to be
surface-forming, which in turn is a consequence of the (graded) Jacobi
identity satisfied by the bivector. However, the inverse does not hold
in general: Not any surface-forming bivector satisfies the Jacobi
identities.  Therefore, it was necessary to check their validity in a
final step.

%*********************************************************************
\subsubsection[Degenerate Fermionic Sector, Rank $(2|1)$]{Degenerate
  Fermionic Sector, Rank \mathversion{bold}$(2|1)$}
\label{sec:CPA-P}

When the fermionic sector has maximal rank one, again the existence of
a fermionic Casimir function is very convenient. We start with
`positive chirality'\footnote{`Positive chirality' refers to the
  structure of (\ref{CPA-v-ff}). It does not preclude the coupling to
  the negative chirality component $\chi^-$ in other terms. A genuine
  chiral algebra (similar to $\N=(1,0)$ supergravity) is a special
  case to be discussed below in \Sec\ref{sec:CPA}}. We choose
  the ansatz (\cf \App\ref{sec:notation})
\begin{equation}
  \label{CPA-v-ff}
  \Poisson^{\alpha\beta}| = v^{\alpha\beta} = i\tilde{u} X^c (\gamma_c
  P_{+})^{\alpha\beta} = \mtrx{\sqrt{2}\tilde{u} X^{++}}{0}{0}{0}.
\end{equation}
The most general case of rank $(2|1)$ can be reduced to
  (\ref{CPA-v-ff}) by a (target space) transformation of the spinors.
Negative chirality where $P_{+}$ is replaced with $P_{-}$ is
considered below. Testing the ans\"a{}tze (\ref{DFS-Cf+}) and
(\ref{DFS-Cf-}) reveals that $\Casimir^{(-)}$ now again is a Casimir
function, but $\Casimir^{(+)}$ is not.  Indeed $\{ \chi^+,
\Casimir^{(+)} \}| \propto \tilde{u} \casimir_{(+)} \neq 0$ in
general, whereas $\{ \chi^+, \Casimir^{(-)} \}| \equiv 0$ shows that
the fermionic Casimir function for positive chirality is
$\Casimir^{(-)}$, where $\casimir_{(-)} = \casimir_{(-)}(\phi,Y)$ has
to fulfil a certain differential equation, to be determined below.

The existence of $\Casimir^{(-)}$ can be used to obtain information
about the unknown components of $\Poisson^{AB}$. Indeed an
investigation of $\{ X^A, \Casimir^{(-)} \} = 0$ turns out to be much
simpler than trying to get that information directly from the Jacobi
identities. The bracket $\{ X^a, \Casimir^{(-)} \} = 0$ results in
$(F^a)_+\^- = 0$ and from $\{ \chi^\alpha, \Casimir^{(-)} \} = 0$ the
relation $v_2\^{--} = 0$ can be derived. This is the reason why the
ansatz (\ref{ans-f-bff}) and (\ref{ans-f-bb}), retaining
(\ref{ans-f1-b}) and (\ref{ans-f5-b}), attains the simpler form
\begin{align}
  F^a &= f_{(1)}^a \1 + i f^{ab} (\gamma_b P_{+}) + f_{(5)}^a \gamma^3,
  \label{CPA-F-bff} \\
  f^{ab} &= f_{(s)} \eta^{ab} + f_{(t)} X^a X^b. \label{CPA-f-bb}
\end{align}
Likewise for the $\chi^2$-component of $\Poisson^{\alpha\beta}$ we set
\begin{equation}
  \label{CPA-v2-ff}
  v_2\^{\alpha\beta} = i \tilde{u}_2 X^c (\gamma_c P_{+})^{\alpha\beta}
  + u_2 (\gamma^3)^{\alpha\beta}.
\end{equation}

Not all information provided by the existence of $\Casimir^{(-)}$ has
been introduced at this point. Indeed using the chiral ansatz
(\ref{CPA-v-ff}) together with (\ref{CPA-F-bff})--(\ref{CPA-v2-ff})
the calculation of $\{ X^a, \Casimir^{(-)} \} = 0$ in conjunction with
the Jacobi identities $J^{\alpha\beta c}| = 0$ (\cf
(\ref{J-ffb-0})) requires $f_{(11)} = 0$ and
\begin{equation}
  \label{CPA-v}
  v = 4Y f_{(51)}.
\end{equation}
It should be noted that the results $f_{(11)} = 0$ and (\ref{CPA-v})
follow from $\{ \chi^\alpha, \Casimir \} = 0$ too, where $\Casimir$ is
a bosonic Casimir function. The remaining equation in $\{ X^a,
\Casimir^{(-)} \} = 0$ together with $\{ \chi^\alpha,\Casimir^{(-)} \}
= 0$ yields $u_2 = -f_{(12)}$. With the solution obtained so far any
calculation of $\{ X^A, \Casimir^{(-)} \} = 0$ leads to one and only
one differential equation (\ref{DFS-c-}) which must be satisfied in
order that (\ref{DFS-Cf-}) is a Casimir function.

We now turn our attention to the Jacobi identities.  The inspection of
$J^{++c}|=0$ (\ref{J-ffb-0}), $J^{+++}|_{\chi}=0$ (\ref{J-fff-1}) and
$J^{+bc}|_{\chi}=0$ (\ref{J-fbb-1}) leads to the conditions
\begin{align}
  f_{(52)} &= \half (\derv\ln|\tilde{u}|) - f_{(12)} - \frac{v}{4Y},
  \label{CPA-f52+} \\
  \tilde{u}_2 &= f (\partial_Y \ln|\tilde{u}|) + f_{(t)},
  \label{CPA-uT2} \\
  v_2 &= \left( \derv + 2 f_{(12)} + (\partial_Y v) \right)
  \frac{f}{\tilde{u}}, \label{CPA-v2}
\end{align}
respectively. In order to simplify the notation we introduced 
\begin{equation}
  \label{CPA-f}
  f = f_{(s)} + 2 Y f_{(t)}.
\end{equation}
All other components of the Jacobi tensor are found to vanish
identically.

The construction of graded Poisson tensors with `negative
chirality', \ie with fermionic sector of the form
\begin{equation}
  \Poisson^{\alpha\beta}| = v^{\alpha\beta} = i\tilde{u} X^c (\gamma_c
  P_{-})^{\alpha\beta} = \mtrx{0}{0}{0}{\sqrt{2}\tilde{u} X^{--}},
\end{equation}
proceeds by the same steps as for positive chirality. Of course, the
relevant fermionic Casimir function is now $\Casimir^{(+)}$ of
(\ref{DFS-Cf+}) and $P_{+}$ in (\ref{CPA-F-bff}) and (\ref{CPA-v2-ff})
has to be replaced by $P_{-}$.  The results $f_{(11)}=0$,
(\ref{CPA-v}), (\ref{CPA-uT2}) and (\ref{CPA-v2}) remain the same,
only $f_{(52)}$ acquires an overall minus sign,
\begin{equation}
  \label{CPA-f52-}
  f_{(52)} = -\half (\derv\ln|\tilde{u}|) + f_{(12)} + \frac{v}{4Y},
\end{equation}
to be inserted in the differential equation (\ref{DFS-c+}) for
  $\casimir_{(+)}$.

The results for graded Poisson tensors of both chiralities can be
summarized as (\cf (\ref{CPA-f}))
\begin{align}
  \Poisson^{ab} &= \left( v + \half\chi^2 \left[ \derv + 2 f_{(12)} +
      (\partial_Y v) \right] \frac{f}{\tilde{u}} \right) \epsilon^{ab},
  \label{CPA-P-bb} \\
  \Poisson^{\alpha b} &= (\chi F^b)^\alpha \label{CPA-P-fb} \\
  \Poisson^{\alpha\beta} &= i \left( \tilde{u} + \half\chi^2 \left[ f
      (\partial_Y \ln|\tilde{u}|) + f_{(t)} \right] \right) X^c
  (\gamma_c P_{\pm})^{\alpha\beta} - \half\chi^2 f_{(12)}
  \gamma^3\^{\alpha\beta}.
  \label{CPA-P-ff}
\end{align}
Eq.\ (\ref{CPA-P-fb}) reads explicitly
\begin{multline}
  \label{CPA-F-b}
  F^b = \frac{v}{4Y} (X^b \pm X^c \epsilon_c\^b) \gamma^3 - 2 f_{(12)}
  X^c \epsilon_c\^b P_{\mp} \\
  + i f_{(s)} (\gamma^b P_{\pm}) + i f_{(t)} X^b X^c (\gamma_c P_{\pm})
  \mp \half (\derv\ln|\tilde{u}|) X^c \epsilon_c\^b \gamma^3.
\end{multline}
Eqs.\ (\ref{CPA-P-bb})--(\ref{CPA-F-b}) represent the generic solution
of the graded $\N=1$ Poisson algebra of rank $(2|1)$. It depends
beside $v(\gamma,Y)$ on four parameter functions $\tilde{u}$,
$f_{(12)}$, $f_{(s)}$ and $f_{(t)}$, all depending on $\phi$ and $Y$.

Each chiral type possesses a bosonic Casimir function $\Casimir =
\casimir + \half\chi^2 \casimir_2$, where $\casimir(\phi,Y)$ and
$\casimir_2(\phi,Y)$ are determined by $\derv \casimir = 0$ and
\begin{equation}
  \label{CPA-c2}
  \casimir_2 = \frac{f \partial_Y \casimir}{\tilde{u}}.
\end{equation}
The fermionic Casimir function for positive chirality is
$\Casimir^{(-)}$ and for negative chirality $\Casimir^{(+)}$ (\cf
(\ref{DFS-Cf-}) and (\ref{DFS-Cf+})), where $\casimir_{(\mp)}(\phi,Y)$
are bosonic scalar functions solving the same differential equation in
both cases when eliminating $f_{(52)}$,
\begin{equation}
  \label{CPA-c1}
  \left( \derv + 2 f_{(12)} + \frac{v}{4Y} - \half(\derv\ln|\tilde{u}|)
  \right) \casimir_{(\mp)} = 0,
\end{equation}
derived from (\ref{DFS-c-}) with (\ref{CPA-f52+}) and from
(\ref{DFS-c+}) with (\ref{CPA-f52-}).

%*********************************************************************
\section{Target space diffeomorphisms}
\label{sec:diffeo}

When subjecting the Poisson tensor of the action (\ref{gPSM}) to a
diffeomorphism
\begin{equation}
  \label{diffeo}
  X^I \to \bar{X}^I = \bar{X}^I(X)
\end{equation}
on the target space $\TSp$, another action of gPSM form is generated
with the new Poisson tensor
\begin{equation}
  \label{tr-P}
  \bar{\Poisson}^{IJ} = (\bar{X}^I \lpartial_K ) \Poisson^{KL}
  (\rpartial_L \bar{X}^J).
\end{equation}
It must be emphasized that in this manner a \emph{different} model is
created with---in the case of 2d gravity theories and their fermionic
extensions---in general different bosonic `body' (and global
topology). Therefore, such transformations are a powerful tool to
create new models from available ones. This is important, because---as
shown in \Sec\ref{sec:Jacobi} above---the solution of the Jacobi
identities as a rule represents a formidable computational problem.
This problem could be circumvented by starting from a simple
$\bar{\Poisson}^{IJ}(\bar{X})$, whose Jacobi identities have been
solved rather trivially. As a next step a transformation
(\ref{diffeo}) is applied. The most general Poisson tensor can be
generated by calculating the inverse of the Jacobi matrices
\begin{alignat}{2}
  J_I\^{\bar J}(X) &= \rpartial_I \bar{X}^J, &\qquad
  J_I\^{\bar K} (J^{-1})_{\bar K}\^J &= \delta_I\^J, \\
  I^{\bar I}\_J(X) &= \bar{X}^I \lpartial_J, &\qquad
  (I^{-1})^I\_{\bar K} I^{\bar K}\_J &= \delta_I\^J.
\end{alignat}
According to
\begin{equation}
  \label{generalP}
  P^{IJ}(X) = (I^{-1})^I\_{\bar K} \bar{P}^{KL}|_{\bar{X}(X)}
  (J^{-1})_{\bar L}\^J
\end{equation}
the components $\Poisson^{IJ}$ of the transformed Poisson tensor are
expressed in terms of the coordinates $X^I$ without the need to invert
(\ref{diffeo}).

The drawback of this argument comes from the fact that in our problem
the (bosonic) part of the `final' algebra is given, and the inverted
version of the procedure described here turns out to be very difficult
to implement.

Nevertheless, we construct explicitly the diffeomorphisms connecting
the dilaton prepotential superalgebra given in \Sec\ref{sec:Izq-P}
with a prototype Poisson tensor in its simplest form, \ie with a
Poisson tensor with constant components.  Coordinates where the
nonzero components take the values $\pm 1$ are called Casimir-Darboux
coordinates. This immediately provides the explicit solution of the
corresponding gPSM too; for details \cf \Sec\ref{sec:sdil-sol}.

In addition, we have found target space diffeomorphisms very useful to
incorporate \eg bosonic models related by conformal transformations.
An example of that will be given in \Sec\ref{sec:cIzq-P} where an
algebra referring to models without bosonic torsion---the just
mentioned dilaton prepotential algebra---can be transformed quite
simply to one depending quadratically on torsion and thus representing
a dilaton theory with kinetic term ($Z\neq 0$ in (\ref{dil})) in its
dilaton version.  There the identification $A_I=(\omega, e_a,
\psi_\alpha)$ with `physical' Cartan variables is used to determine
the solution of the latter theory ($Z \neq 0$) from the simpler model
($\bar{Z}=0$) with PSM variables $(\bar{X}^I, \bar{A}_I)$ by
\begin{equation}
  \label{tr-A}
  A_I = (\rpartial_I \bar{X}^J) \bar{A}_J.
\end{equation}

Also from target space diffeomorphisms interesting information can be
collected, regarding the arbitrariness to obtain supersymmetric
extensions of bosonic models as found in the general solutions of
\Sec\ref{sec:Jacobi}.  Imagine that a certain gPSM has been found,
solving the Jacobi identities with a particular ansatz. A natural
question would be to find out which other models have the same bosonic
body. For this purpose we single out at first all transformations
(\ref{diffeo}) which leave the components $\Poisson^{A\phi}$ form
invariant as given by (\ref{LorentzB}) and (\ref{LorentzF}):
\begin{equation}
  \bar{\phi} = \phi, \qquad
  \bar{X}^a = X^b C_b\^a, \qquad
  \bar{\chi}^\alpha = \chi^\beta h_\beta\^\alpha.
\end{equation}
Here $C_b\^a$ and $h_\beta\^\alpha$ are Lorentz covariant functions
(resp.\ spinor matrices)
\begin{align}
  C_b\^a &= L \delta_b\^a + M \epsilon_b\^a = c_b\^a + \half\chi^2
  (c_2)_b\^a, \label{tr-bb} \\
  h_\beta\^\alpha &= \left[ h_{(1)} \1 + h_{(2)} \gamma^3 + i h_{(3)}
    X^c \gamma_c + i h_{(4)} X^d \epsilon_d\^c \gamma_c
  \right]_\beta\^\alpha, \label{tr-ff}
\end{align}
when expressed in terms of $\chi^2$ ($L = l + \half\chi^2 l_2$ and
similar for $M$) and in terms of $\phi$ and $Y$ ($l$, $l_2$, $m$,
$m_2$, $h_{(i)}$).

The `stabilisator' ($\bar{\Poisson}^{ab} = \Poisson^{ab}$) of the
bosonic component $v(\phi,Y) = v(\bar{\phi},\bar{Y})$ of a graded
Poisson tensor will be given by the restriction of $c_b\^a$ to a
Lorentz transformation on the target space $\TSp$ with $l^2 - m^2 = 1$
in (\ref{tr-bb}). Furthermore from the two parameters $h_{(1)}$ and
$h_{(2)}$ a Lorentz transformation can be used to reduce them to one
independent parameter. Thus no less than five arbitrary two argument
functions are found to keep the bosonic part of $\Poisson^{ab}$
unchanged, but produce different fermionic extensions with
supersymmetries different from the algebra we started from. This
number for rank $(2|2)$ exactly coincides with the number of arbitrary
invariant functions found in \Sec\ref{sec:sol-gP}. For rank $(2|1)$
in the degenerate case a certain `chiral' combination of $h_{(3)}$
and $h_{(4)}$ in (\ref{tr-ff}) must be kept fixed, reducing that
number to four---again in agreement with \Sec\ref{sec:CPA-P}. In a
similar way also the appearance of just three arbitrary functions in
\Sec\ref{sec:DFS-P} for rank $(2|0)$ can be understood.

%*********************************************************************
\section{Particular Poisson Superalgebras}
\label{sec:Poisson}

The compact formulas of the last sections do not seem suitable for a
general discussion, especially in view of the large arbitrariness of
gPSMs. We, therefore, elucidate the main features in special models of
increasing complexity. The corresponding actions and their relations
to the alternative formulations (\ref{grav}) and, or the dilaton
theory form (\ref{dil}) will be discussed in \Sec\ref{sec:models}.

%*********************************************************************
\subsection{Block Diagonal Algebra}
\label{sec:BDA}

The most simple ansatz which, nevertheless, already shows the generic
features appearing in fermionic extensions, consists in setting the
mixed components $\Poisson^{\alpha b}=0$ so that the nontrivial
fermionic brackets are restricted to the block
$\Poisson^{\alpha\beta}$. Then (\ref{J-ffb-0})--(\ref{J-ffb-2}) reduce
to
\begin{gather}
  \derv^c v^{\alpha\beta} = 0, \label{BDA-J-ffb-0} \\
  v_\delta\^\alpha v_2\^{\beta\gamma} + \half \gamma^3\_\delta\^\alpha
  \partial_\phi v^{\beta\gamma} + \cycl(\alpha\beta\gamma) = 0,
  \label{BDA-J-fff-1} \\
  v_\delta\^\alpha v_2 + \half \gamma^3\_\delta\^\alpha \partial_\phi
  v = 0, \label{BDA-J-fbb-1} \\
  v_2 \epsilon^{cd} \partial_d v^{\alpha\beta} + \derv^c
  v_2^{\alpha\beta} = 0. \label{BDA-J-ffb-2}
\end{gather}
Eq.\ (\ref{BDA-J-fbb-1}) implies the spinorial structure
\begin{equation}
  v^{\alpha\beta} = u(\gamma^3)^{\alpha\beta}.
\end{equation}
The trace of (\ref{BDA-J-ffb-0}) with $\gamma^3$ leads to the
condition (\ref{c}) for $u$, \ie $u = u(\casimir(\phi, Y))$
depends on the combination of $\phi$ and $Y$ as determined by the
bosonic Casimir function.

For $u \neq 0$ the remaining equations (\ref{BDA-J-fff-1}),
(\ref{BDA-J-fbb-1}) and (\ref{BDA-J-ffb-2}) are fulfilled by
\begin{equation}
  \label{BDA-v2}
  v_2^{\alpha\beta} = -\frac{\partial_\phi u}{2u}
  (\gamma^3)^{\alpha\beta}, \qquad
  v_2 = -\frac{\partial_\phi v}{2u}.
\end{equation}
For the present case according to (\ref{c2}) the Casimir function is
\begin{equation}
  \label{BDA-C}
  \Casimir = \casimir - \half\chi^2 \frac{\partial_\phi\casimir}{2
    u(\casimir)}.
\end{equation}
It is verified easily that
\begin{equation}
  \label{BDA-U}
  U = u(\Casimir) = u(\casimir) + \half\chi^2 u_2, \qquad u_2 =
  -\frac{\partial_\phi\casimir}{2u(\casimir)} \frac{du}{d\casimir}.
\end{equation}

Already in this case we observe that in the fermionic extension
$\detv^{-1}$, $u^{-1}$ from the inverse of $v^{\alpha\beta}$ may
introduce singularities. It should be emphasized that $u=u(c)$ is an
arbitrary function of $c(\phi,Y)$. Except for $u=u_0=\const$ (see
below) any generic choice of the arbitrary function $u(\casimir)$ by
the factors $u^{-1}$ in (\ref{BDA-v2}), thus may introduce
restrictions on the allowed range of $\phi$ and $Y$ or new
singularities on a certain surface where $u(\casimir(\phi,Y))$
vanishes, not present in the purely bosonic bracket. Indeed, these
obstructions in certain fermionic extensions are a generic feature of
gPSMs. The singularities are seen to be caused here by $\detv^{-1}$,
the inverse of the determinant (\ref{detv}), except for cases with
$\detv = \const$ or when special cancellation mechanisms are invoked.
Another source for the same phenomenon will appear below in connection
with the appearance of a `prepotential' for $v$.  Still, such
`obstructions' can be argued to be rather harmless. We will come back
to these issues in several examples below, especially when discussing
an explicit solution in \Sec\ref{sec:sdil-sol}.

This complication can be made to disappear by choosing $u = u_0 =
\const \neq 0$. Then the fermionic extension ($v'=\partial_\phi v$)
\begin{align}
  \Poisson^{ab} &= \left( v - \frac{1}{4u_0} \chi^2 v' \right)
  \epsilon^{ab}, \label{BDA-P-bb} \\
  \Poisson^{\alpha b} &= 0, \label{BDA-P-fb} \\
  \Poisson^{\alpha\beta} &= u_0 (\gamma^3)^{\alpha\beta}
  \label{BDA-P-ff}
\end{align}
does not lead to restrictions on the purely bosonic part $v(\phi, Y)$
of the Poisson tensor, nor does it introduce additional singularities,
beside the ones which may already be present in the potential
$v(\phi,Y)$. However, then no genuine supersymmetry survives (see
\Sec\ref{sec:BDS} below).

It should be noted that all dilaton models mentioned in the
introduction can be accommodated in a nontrivial version $u \neq
\const$ of this gPSM. We shall call the corresponding supergravity
actions their `diagonal extensions'.

%*********************************************************************
\subsection{Nondegenerate Chiral Algebra}
\label{sec:SUSY/2-P}

Two further models follow by setting $u=u_0=\const \neq 0$ and
$\hat{u}=\pm \tilde{u}_0=\const$.  In this way a generalization with
full rank of the chiral $\N=(1,0)$ and $\N=(0,1)$ algebras is obtained
(\cf \App\ref{sec:notation})
\begin{equation}
  v^{\alpha\beta} = i \tilde{u}_0 X^c (\gamma_c P_\pm)^{\alpha\beta}
  + u_0 (\gamma^3)^{\alpha\beta}.
\end{equation}
This particular choice for the coefficients of $X^c$ in
$v^{\alpha\beta}$ also has the advantage that $X^c$ drops out from
$\detv = -u_0^2/4$, thus its inverse exists everywhere.  Restricting
furthermore $f^c=0$ we arrive at
\begin{equation}
  F^c\_\alpha\^\beta = -\frac{i\tilde{u}_0 v}{2u_0}
  (\gamma^c P_\pm)_\alpha\^\beta, \qquad
  v_2\^{\alpha\beta} = 0, \qquad
  v_2 = -\frac{v'}{2u_0}.
\end{equation}
This yields another graded Poisson tensor for the arbitrary bosonic
potential $v(\phi,Y)$
\begin{align}
  \Poisson^{ab} &= \left( v - \frac{1}{4u_0} \chi^2 v' \right)
  \epsilon^{ab},
  \label{SUSY/2-P-bb} \\
  \Poisson^{\alpha b} &= -\frac{i\tilde{u}_0v}{2u_0}
  (\chi\gamma^b P_\pm)^\alpha, \label{SUSY/2-P-fb} \\
  \Poisson^{\alpha\beta} &= i \tilde{u}_0 X^c
  (\gamma_c P_\pm)^{\alpha\beta} + u_0 (\gamma^3)^{\alpha\beta}.
  \label{SUSY/2-P-ff}
\end{align}
Again there are no obstructions for such models corresponding to any
bosonic gravity model, given by a particular choice of $v(\phi,Y)$.

The Casimir function (\cf (\ref{c2})) reads
\begin{equation}
  \label{SUSY/2-C}
  \Casimir = \casimir - \frac{1}{4 u_0} \chi^2 c',
\end{equation}
where $\casimir$ must obey (\ref{c}).

%*********************************************************************
\subsection{Deformed Rigid Supersymmetry}
\label{sec:SUSY-P}

The structure of rigid supersymmetry is encoded within the Poisson
tensor by means of the components $v=0$ and (\cf (\ref{matrix}))
\begin{equation}
  \label{SUSY-v-ff}
  v^{\alpha\beta} = i \tilde{u}_0 X^c \gamma_c\^{\alpha\beta} =
  \mtrx{\sqrt{2} \tilde{u}_0 X^{++}}{0}{0}{\sqrt{2} \tilde{u}_0 X^{--}},
\end{equation}
where again $\tilde{u}=\tilde{u}_0=\const \neq 0$. Here $\detv = 2Y
\tilde{u}_0^2$ and
\begin{equation}
  \frac{1}{\detv} v_{\alpha\beta} = \frac{i}{2Y \tilde{u}_0} X^c
  \gamma_{c\alpha\beta} = \mtrx{\frac{1}{\sqrt{2} \tilde{u}_0
      X^{++}}}{0}{0}{\frac{1}{\sqrt{2} \tilde{u}_0 X^{--}}}.
\end{equation}

Generalizing this ansatz to $v \neq 0$, the simplest choice $f^c = 0$
with an arbitrary function $v(\phi,Y)$ (deformed rigid supersymmetry,
DRS) yields
\begin{equation}
  F^c\_\alpha\^\beta = \frac{v}{4Y} X^a (\gamma_a \gamma^c
  \gamma^3)_\alpha\^\beta, \qquad
  v_2\^{\alpha\beta} = \frac{v}{4Y} \gamma^3\^{\alpha\beta},
  \qquad 
  v_2 = 0,
\end{equation}
and thus for $\Poisson^{IJ}$
\begin{align}
  \Poisson^{ab} &= v \epsilon^{ab}, \label{SUSY-P-bb} \\
  \Poisson^{\alpha b} &= \frac{v}{4Y} X^c
  (\chi\gamma_c\gamma^b\gamma^3)^\alpha, \label{SUSY-P-fb} \\
  \Poisson^{\alpha\beta} &= i \tilde{u}_0 X^c (\gamma_c)^{\alpha\beta}
  + \frac{1}{2} \chi^2 \frac{v}{4Y} (\gamma^3)^{\alpha\beta},
  \label{SUSY-P-ff}
\end{align}
and for the Casimir function $\Casimir = \casimir$ with (\ref{c}).

{}From (\ref{SUSY-P-bb})--(\ref{SUSY-P-ff}) it is clear---in contrast
to the algebras \ref{sec:BDA} and \ref{sec:SUSY/2-P}---that this
fermionic extension for a generic $v \neq 0$ introduces a possible
further singularity at $Y=0$, which cannot be cured by further
assumptions on functions which are still arbitrary.

Of course, in order to describe flat space-time, corresponding to the
Poisson tensor of rigid supersymmetry, one has to set $v(\phi,Y)=0$.
Then the singularity at $Y=0$ in the extended Poisson tensor
disappears.

We remark already here that despite the fact that for $v \neq 0$ the
corresponding supersymmetrically extended action functional (in
contrast to its purely bosonic part) becomes singular at field values
$Y\equiv \half X^aX_a =0$, we expect that if solutions of the field
equations are singular there as well, such singularities will not be
relevant if suitable `physical' observables are considered. We have in
mind the analogy to curvature invariants which are not affected by
`coordinate singularities'. We do, however, not intend to prove this
statement in detail within the present paper; in
\Sec\ref{sec:sdil-sol} below we shall only shortly discuss the similar
singularities, caused by the prepotential in an explicit solution of
the related field-theoretical model.

%*********************************************************************
\subsection{Dilaton Prepotential Algebra}
\label{sec:Izq-P}

We now assume that the bosonic potential $v$ is restricted to be a
function of the dilaton $\phi$ only, $\dot{v} = \partial_Y v = 0$.
Many models of 2d supergravity, already known in the literature, are
contained within algebras of this type, one of which was described in
ref.~\cite{\bibIzq}. Let deformed rigid supersymmetry of
\Sec\ref{sec:SUSY-P} again be the key component of the Poisson tensor
(\ref{SUSY-v-ff}). Our attempt in \Sec\ref{sec:SUSY-P} to provide a
Poisson tensor for arbitrary $v$ built around that component produced
a new singularity at $Y=0$ in the fermionic extension. However, the
Poisson tensor underlying the model considered in \cite{\bibIzq} was
not singular in $Y$.  Indeed there exists a mechanism by which this
singularity can be cancelled in the general solution
(\ref{detv})--(\ref{sol-v2}), provided the arbitrary functions are
chosen in a specific manner.

For this purpose we add to (\ref{SUSY-v-ff}), keeping
$\tilde{u}=\tilde{u}_0=\const$, the fermionic potential $u(\phi)$,
\begin{equation}
  \label{Izq-v-ff}
  v^{\alpha\beta} = i \tilde{u}_0 X^c (\gamma_c)^{\alpha\beta} + u
  (\gamma^3)^{\alpha\beta} =
  \mtrx{\sqrt{2} \tilde{u}_0 X^{++}}{-u}{-u}{\sqrt{2} \tilde{u}_0 X^{--}},
\end{equation}
with determinant 
\begin{equation}
  \label{Izq-detv}
  \detv = 2Y \tilde{u}_0^2 - u^2.
\end{equation}

The Hamiltonian vector field $\rvec T^c$ in the solution
(\ref{sol-f-bff}) generates a factor $f_{(t)} \neq 0$ in
(\ref{ans-f-bb}). The independent vector field $f^c$ can be used to
cancel that factor provided one chooses
\begin{equation}
  \label{Izq-f-b}
  f^c = \half u' X^c.
\end{equation}
Then the disappearance of $f_{(t)}$ is in agreement with the solution
given in ref.~\cite{Izquierdo:1998hg}.  The remaining coefficient
functions then follow as
\begin{align}
  F^c\_\alpha\^\beta &= \frac{1}{2\detv} \left( \tilde{u}_0^2 v + u u'
  \right) X^a (\gamma_a \gamma^c \gamma^3)_\alpha\^\beta +
  \frac{i\tilde{u}_0}{2\detv} \left( u v + 2Y u' \right)
  (\gamma^c)_\alpha\^\beta, \label{Izq-f-bff} \\
  v_2\^{\alpha\beta} &= \frac{1}{2\detv} \left( \tilde{u}_0^2 v + u u'
  \right) \gamma^3\^{\alpha\beta}, \label{Izq-v2-ff} \\
  v_2 &= \frac{uv}{2\detv^2} \left( \tilde{u}_0^2 v + u u' \right) +
  \frac{uu'}{2\detv^2} \left( u v + 2Y u' \right) + \frac{1}{2\detv} \left(u
    v' + 2Y u' \dot{v} + 2Y u'' \right). \label{Izq-v2}
\end{align}

Up to this point the bosonic potential $v$ and the potential $u$ have
been arbitrary functions of $\phi$. Demanding now that
\begin{equation}
  \label{Izq-v}
  \tilde{u}_0^2 v + u u' = 0,
\end{equation}
the singularity at $\detv=0$ is found to be cancelled not only in the
respective first terms of (\ref{Izq-f-bff})--(\ref{Izq-v2}), but also
in the rest:
\begin{alignat}{2}
  v &= -\frac{(u^2)'}{2\tilde{u}_0^2}, &\qquad
  F^c\_\alpha\^\beta &= \frac{iu'}{2\tilde{u}_0}
  \gamma^c\_\alpha\^\beta, \label{Izq-pot1} \\
  v_2\^{\alpha\beta} &= 0, &\qquad
  v_2 &= \frac{u''}{2\tilde{u}_0^2}. \label{Izq-pot2}
\end{alignat}
Furthermore the fermionic potential $u(\phi)$ is seen to be promoted
to a `prepotential' for $v(\phi)$.  A closer look at (\ref{Izq-v})
with (\ref{Izq-detv}) shows that this relation is equivalent to $\derv
\detv = 0$ which happens to be precisely the defining equation
(\ref{c}) of the Casimir function $\casimir(\phi,Y)$ of the bosonic
model. The complete Casimir function follows from (\ref{c2}):
\begin{equation}
  \label{Izq-c2}
  \casimir_2 = \frac{1}{2\detv} \left( u
    \partial_\phi + 2Y u' \partial_Y \right) \casimir
\end{equation}
so that
\begin{equation}
  \label{Izq-C}
  \Casimir = \detv + \half\chi^2 u'.
\end{equation}

Thus the Poisson tensor for $v=v(\phi)$, related to $u(\phi)$ by
(\ref{Izq-v}), becomes
\begin{align}
  \Poisson^{ab} &= \frac{1}{2\tilde{u}_0^2} \left( -(u^2)'
    + \half \chi^2 u'' \right) \epsilon^{ab}, \label{Izq-P-bb} \\
  \Poisson^{\alpha b} &= \frac{iu'}{2\tilde{u}_0} (\chi\gamma^b)^\alpha,
  \label{Izq-P-fb} \\
  \Poisson^{\alpha\beta} &= i \tilde{u}_0 X^c (\gamma_c)^{\alpha\beta}
  + u (\gamma^3)^{\alpha\beta}, \label{Izq-P-ff}
\end{align}
which is indeed free from singularities produced by the supersymmetric
extension. However, this does not eliminate all pitfalls: Given a
bosonic model described by a particular potential $v (\phi)$ where
$\phi$ is assumed to take values in the interval $I \subseteq \R$, we
have to solve (\ref{Izq-v}) for the prepotential $u(\phi)$, \ie the
quadratic equation
\begin{equation}
  \label{Izq-u}
  u^2 = -2 \tilde{u}_0^2 \int_{\phi_0}^{\phi} v(\varphi) d\varphi,
\end{equation}
which may possess a solution within the real numbers only for a
restricted range $\phi \in J$.  The interval $J$ may have a nontrivial
intersection with $I$ or even none at all. Clearly no restrictions
occur if $v$ contains a potential for the dilaton which happens to
lead to a negative definite integral on the \rhs of (\ref{Izq-u}) for
\emph{all} values of $\phi$ in $I$ (this happens \eg if $v$ contains
only odd powers of $\phi$ with negative prefactors).  On the other
hand, the domain of $\phi$ is always restricted if $v$ contains even
potentials, as becomes immediately clear when viewing the special
solutions given in \Tab\ref{tab:IzqMdls}.
\begin{table}[ht]
  \begin{center}
    \begin{tabular}{|l||l|l|} \hline
      Model & $v(\phi) = -\frac{(u^2)'}{2\tilde{u}_0^2}$ & $u(\phi)$
      \\
      \hline\hline
      & $0$ & $\tilde{u}_0 \lambda$ \\
      String & $-\Lambda$ & $\pm \tilde{u}_0 \sqrt{2\Lambda
        (\phi-\phi_0)}$ \\
      JT & $-\lambda^2 \phi$ & $\tilde{u}_0 \lambda \phi$ \\
      $R^2$ & $-\frac{\alpha}{2} \phi^2$ & $\pm \tilde{u}_0
      \sqrt{\frac{\alpha}{3} (\phi^3-\phi_0^3)}$ \\
      Howe & $-2 \lambda^2 \phi^3$ & $\tilde{u}_0 \lambda \phi^2$ \\
      \hline
      $\bar{\mathrm{SRG}}$ & $-\frac{\lambda^2}{\sqrt{\phi}}$ & $2
      \tilde{u}_0 \lambda \sqrt[4]{\phi}$ \\
      \hline
    \end{tabular}
  \end{center}
  \caption{Special Dilaton Prepotential Algebras} \label{tab:IzqMdls}
\end{table}
There the different potentials $v(\phi)$ are labelled according to the
models: The string model with $\Lambda=\const$ of
\cite{\bibSI,\bibDBHmatter}, JT is the Jackiw-Teitelboim model
(\ref{JT}), $\overline{\mathrm{SRG}}$ the spherically reduced black
hole (\ref{EBH}) in the conformal description (\cf
\Sec\ref{sec:intro}); the cubic potential appeared in
\cite{Howe:1979ia}, $R^2$ gravity is self-explaining. Note that in the
case of $\overline{\mathrm{SRG}}$ $I=J=\R_+$ ($\phi > 0$), there is
already a (harmless) restriction on allowed values of $\phi$ at the
purely bosonic level, \cf (\ref{EBH}).

So, as argued above, one may get rid of the singularities at $Y=0$ of
supersymmetric extensions obtained in the previous section. In some
cases, however, this leads to a restricted range for allowed values of
the dilaton, or, alternatively, to complex valued Poisson tensors.
Similarly to our expectation of the harmlessness of the above
mentioned $1/Y$-singularities on the level of the solutions (\cf also
\cite{Strobl:1999zz}), we expect that also complex-valued Poisson
tensors are no serious obstacle (both of these remarks apply to the
classical analysis only!). In fact, a similar scenario was seen to be
harmless (classically) also in the Poisson Sigma formulation of the
$G/G$ model for compact gauge groups like $SU(2)$, \cf
\cite{Schaller:1995xk,Alekseev:1995py}. We further illustrate these
remarks for the class of supergravity models considered in
\cite{\bibIzq} at the end of \Sec\ref{sec:sdil-sol}.

%*********************************************************************
\subsection[Bosonic Potential Linear in $Y$]{Bosonic Potential Linear
in \mathversion{bold}$Y$}
\label{sec:cIzq-P}

In order to retain the $Y$-dependence and thus an algebra with bosonic
torsion, we take solution (\ref{Izq-v-ff})--(\ref{Izq-v2}) but instead
of (\ref{Izq-v}) we may also choose
\begin{equation} \label{cIzq-v} v = -\frac{(u^2)'}{2\tilde{u}_0^2} -
\frac{\detv}{2} f,
\end{equation} where $f$ is an arbitrary function of $\phi$ and
$Y$. Thanks to the factor $\detv$ also in this case the fermionic
extension does not introduce new singularities at
$\detv=0$.\footnote{Clearly also in (\ref{cIzq-v}) the replacement
$\detv f \rightarrow G(\detv,\phi,Y)$ with $G(\detv,\phi,Y)/\detv$
regular at $\detv=0$ has a similar effect. But linearity in $\detv$ is
sufficient for our purposes.}  Even if $f$ is a function of $\phi$
only ($\dot f = 0$), this model is quadratic in (bosonic) torsion,
because of (\ref{Izq-detv}).  A straightforward calculation using
(\ref{cIzq-v}) gives
\begin{align} F^c\_\alpha\^\beta &= -\frac{\tilde{u}_0^2 f}{4} X^a
(\gamma_a \gamma^c \gamma^3)_\alpha\^\beta + i \left( \frac{u'}{2
\tilde{u}_0} - \frac{\tilde{u}_0 u f}{4} \right)
(\gamma^c)_\alpha\^\beta, \label{cIzq-f-bff} \\ v_2\^{\alpha\beta} &=
-\frac{\tilde{u}_0^2 f}{4} \gamma^3\^{\alpha\beta}, \label{cIzq-v2-ff}
\\ v_2 &= \half \left( \frac{u''}{\tilde{u}_0^2} - u'f - \frac{uf'}{2}
+ \frac{\tilde{u}_0^2 u f^2}{4} - \frac{2Y u' \dot{f}}{2}
\right). \label{cIzq-v2}
\end{align}

It is worthwhile to note that the present algebra, where the bosonic
potential $v$ is of the type (\ref{vdil}), can be reached from the
algebra of \Sec\ref{sec:Izq-P} with $\bar{v}=\bar{v}(\bar\phi)$ by a
conformal transformation, \ie a target space diffeomorphism in the
sense of \Sec\ref{sec:diffeo}. We use bars to denote quantities and
potentials of the algebra of \Sec\ref{sec:Izq-P}, but not for
$\tilde{u}_0$ because it remains unchanged, \ie $\bar{v} =
-\frac{(\bar{u}^2)'}{2\tilde{u}_0^2}$. By
\begin{equation} \label{conf-tr-X} \phi = \bar{\phi}, \qquad X^a =
e^{\varphi(\phi)} \bar{X}^a, \qquad \chi^\alpha = e^{\half
\varphi(\phi)} \bar{\chi}^\alpha,
\end{equation} the transformed Poisson tensor, expanded in terms of
unbarred coefficient functions (\cf \Sec\ref{sec:ans-gP}) becomes
\begin{alignat}{3} \tilde{u} &= \tilde{u}_0, &\qquad \tilde{u}_2 &= 0,
\\ u &= e^\varphi \bar{u}, &\qquad u_2 &= -\half \varphi', \\ v &=
e^{2 \varphi} \bar{v} - 2 Y \varphi', &\qquad v_2 &= e^\varphi
\frac{\bar{u}''}{2\tilde{u}_0^2}, \\ f_{(12)} &= \half \varphi',
&\qquad f_{(51)} &= -\half \varphi', \\ f_{(s)} &= e^\varphi
\frac{\bar{u}'}{2\tilde{u}_0}
\end{alignat} and $f_{(11)} = f_{(52)} = f_{(t)} = f_{(h)} = 0$. When
$u(\phi)$ and $\varphi(\phi)$ are taken as basic independent
potentials we arrive at
\begin{align} v &= -\frac{1}{2\tilde{u}_0^2} e^{2\varphi} \left(
e^{-2\varphi} u^2 \right)' - 2 Y \varphi', \label{cIzq-v-alt} \\ v_2
&= \frac{1}{2\tilde{u}_0^2} e^{\varphi} \left( e^{-\varphi} u
\right)'', \\ f_{(s)} &= \frac{1}{2\tilde{u}_0} e^{\varphi} \left(
e^{-\varphi} u \right)'.
\end{align} If we set $\varphi' = \tilde{u}_0^2 f / 2$ we again obtain
solution (\ref{cIzq-v})--(\ref{cIzq-v2}) for $Y$-independent $f$.  The
components $\bar{\Poisson}^{a\phi}$ and $\bar{\Poisson}^{\alpha \phi}$
remain form invariant,
\begin{equation} \Poisson^{a\phi} = X^b \epsilon_b\^a, \qquad
\Poisson^{\alpha \phi} = -\half \chi^\beta (\gamma^3)_\beta\^\alpha,
\end{equation} in agreement with the requirement determined for this
case in
\Sec\ref{sec:ans-gP}. For completeness we also list the transformation
of the 1-forms $A_I = (\omega, e_a, \psi_\alpha)$ according to
(\ref{tr-A})
\begin{equation} \label{conf-tr-A} \omega = \bar{\omega} - \varphi'
\left( \bar{X}^b \bar{e}_b + \half \bar{\chi}^\beta \bar{\psi}_\beta
\right), \qquad e_a = e^{-\varphi} \bar{e}_a, \qquad \psi_\alpha =
e^{-\half \varphi} \bar{\psi}_\alpha.
\end{equation} The second equation in (\ref{conf-tr-A}) provide the
justification for the name `conformal transformation'.

With the help of the scaling parameter $\varphi$ we can write
(\ref{cIzq-v-alt}), and also (\ref{cIzq-v}), in its equivalent form
$\derv(e^{-2\varphi} \detv) = 0$, thus exposing the Casimir function
to be $\casimir(\phi,Y) = e^{-2\varphi} \detv$.  Now $u(\phi)$ and
$\varphi(\phi)$ are to be viewed as two independent parameter
functions labelling specific types of Poisson tensors.  The solution
\begin{align} \Poisson^{ab} &= \left( -\frac{1}{2\tilde{u}_0^2}
e^{2\varphi} \left( e^{-2\varphi} u^2 \right)' - 2 Y \varphi' +
\frac{1}{4\tilde{u}_0^2} \chi^2 e^{\varphi} \left( e^{-\varphi} u
\right)'' \right) \epsilon^{ab}, \label{cIzq-P-bb} \\ \Poisson^{\alpha
b} &= -\half \varphi' X^a (\chi\gamma_a \gamma^b \gamma^3)^\alpha +
\frac{i}{2 \tilde{u}_0} e^{\varphi} \left( e^{-\varphi} u \right)'
(\chi\gamma^b)^\alpha, \label{cIzq-P-fb} \\ \Poisson^{\alpha\beta} &=
i \tilde{u}_0 X^c (\gamma_c)^{\alpha\beta} + \left( u -
\frac{1}{4}\chi^2 \varphi' \right) (\gamma^3)^{\alpha\beta}
\label{cIzq-P-ff}
\end{align} does not introduce a new singularity at $Y=0$, but in
order to provide the extension of the bosonic potential (\ref{vdil})
we have to solve (\ref{cIzq-v-alt}) for the scaling parameter
$\varphi(\phi)$ and the fermionic potential $u(\phi)$, which may again
lead to obstructions similar to the ones described at the end of
\Sec\ref{sec:Izq-P}. With the integrals over $Z(\phi)$ and $V(\phi)$
introduced in (\ref{QW}) we find
\begin{align} \varphi &= -\half Q(\phi), \label{cIzq-tr} \\ u &= \pm
\sqrt{-2 \tilde{u}_0^2 e^{-Q(\phi)} W(\phi)}. \label{cIzq-u}
\end{align} Now we can read off the restriction to be $W(\phi) < 0$,
yielding singularities at the boundary $W(\phi) = 0$. The ansatz
(\ref{cIzq-v}) can be rewritten in the equivalent form
\begin{equation} \label{cIzq-c} \derv(e^Q \detv) = 0 \Leftrightarrow
\casimir(\phi,Y) = e^Q \detv = 2\tilde{u}_0^2 (Y e^Q + W).
\end{equation} The complete Casimir from (\ref{c2}), which again
exhibits the simpler form (\ref{Izq-c2}), reads
\begin{equation} \label{cIzq-C} \Casimir = e^Q \left( \detv +
\half\chi^2 e^{-\half Q} \left( e^{\half Q} u \right)' \right).
\end{equation} As expected from ordinary 2d gravity $\Casimir$ is
conformally invariant.

Expressing the Poisson tensor in terms of the potentials $V(\phi)$ and
$Z(\phi)$ of the original bosonic theory, and with $u(\phi)$ as in
(\ref{cIzq-u}) we arrive at
\begin{align} \Poisson^{ab} &= \left( V + Y Z - \half \chi^2 \left[
\frac{V Z + V'}{2u} + \frac{\tilde{u}_0^2 V^2}{2u^3} \right] \right)
\epsilon^{ab}, \label{cIzq-P-bb-alt} \\ \Poisson^{\alpha b} &=
\frac{Z}{4} X^a (\chi\gamma_a \gamma^b \gamma^3)^\alpha -
\frac{i\tilde{u}_0V}{2u} (\chi\gamma^b)^\alpha, \label{cIzq-P-fb-alt}
\\ \Poisson^{\alpha\beta} &= i \tilde{u}_0 X^c
(\gamma_c)^{\alpha\beta} + \left( u + \frac{Z}{8}\chi^2 \right)
(\gamma^3)^{\alpha\beta}.  \label{cIzq-P-ff-alt}
\end{align} As will be shown in \Sec\ref{sec:sdil} this provides a
supersymmetrization for all the dilaton theories (\ref{dil}), because
it covers all theories (\ref{FOG}) with $v$ linear in $Y$. Among these
two explicit examples, namely SRG and the KV model, will be treated in
more detail now.

%*********************************************************************
\subsubsection[SRG, Nondiagonal Extension I]{Spherically Reduced
Gravity (SRG), Nondiagonal Extension I}
\label{sec:SRG1}

In contrast to the KV-model below, no obstructions are found when
(\ref{cIzq-f-bff})--(\ref{cIzq-v2}) with $v$ given by (\ref{cIzq-v})
is used for SRG.  For simplicity we take in (\ref{SRG}) the case
$d=4$ and obtain $Q(\phi) = -\half \ln(\phi)$, $W(\phi) = -2 \lambda^2
\sqrt{\phi}$ and
\begin{equation} \label{SRG1-u-and-f} u = 2 \tilde{u}_0 \lambda
\sqrt{\phi}, \qquad \varphi = \frac{1}{4} \ln(\phi),
\end{equation} where $u_0$ is a constant. Here already the bosonic
theory is defined for $\phi > 0$ only. From
(\ref{cIzq-P-bb})--(\ref{cIzq-P-ff}) in the Poisson tensor of SRG
\begin{align} P^{ab} &= \left( -\lambda^2 - \frac{Y}{2\phi} -
\frac{3\lambda}{32 \tilde{u}_0 \phi^{3/2}} \chi^2 \right)
\epsilon^{ab}, \label{SRG1-P-bb} \\ P^{\alpha b} &= -\frac{1}{8\phi}
X^c (\chi \gamma_c \gamma^b \gamma^3)^\alpha +
\frac{i\lambda}{4\sqrt{\phi}} \, (\chi\gamma^b)^\alpha,
\label{SRG1-P-fb} \\ P^{\alpha\beta} &= i \tilde{u}_0 X^c
(\gamma_c)^{\alpha\beta} + \left( 2 \tilde{u}_0 \lambda \sqrt{\phi} -
\frac{1}{16\phi} \chi^2 \right) (\gamma^3)^{\alpha\beta}
\label{SRG1-P-ff}
\end{align} the singularity of the bosonic part simply carries over to
the extension, without introducing any new restriction for $\phi > 0$.

The bosonic part of the Casimir function (\ref{cIzq-C}) is
proportional to the ADM mass for SRG.

%*********************************************************************
\subsubsection[KV, Nondiagonal Extension I]{Katanaev-Volovich Model
(KV), Nondiagonal Extension I}
\label{sec:SKV1}

The bosonic potential (\ref{KV}) leads to $Q(\phi) = \alpha \phi$,
thus $\varphi = -\frac{\alpha}{2} \phi$, and
\begin{equation}
  W(\phi) = \int_{\phi_0}^{\phi} e^{\alpha\eta} \left(
    \frac{\beta}{2} \eta^2 - \Lambda \right) d\eta = \left. e^{\alpha\eta}
    \left[ \frac{\beta}{2} \left( \frac{2}{\alpha^3} -
        \frac{2\eta}{\alpha^2} + \frac{\eta^2}{\alpha} \right) -
      \frac{\Lambda}{\alpha} \right]
  \right|_{\phi_0}^{\phi}. \label{SKV1-W}
\end{equation}
With $u(\phi)$ calculated according to (\ref{cIzq-u}) the Poisson
tensor is
\begin{align}
  \Poisson^{ab} &= \left( \frac{\beta}{2} \phi^2 - \Lambda
    + \alpha Y + \half \chi^2 v_2 \right) \epsilon^{ab},
  \label{SKV1-P-bb} \\
  \Poisson^{\alpha b} &= \frac{\alpha}{4} X^a (\chi\gamma_a \gamma^b
  \gamma^3)_\alpha\^\beta - \frac{i\tilde{u}_0}{2u} \left(
    \frac{\beta}{2} \phi^2 - \Lambda \right) (\chi\gamma^b)_\alpha\^\beta,
  \label{SKV1-P-fb} \\
  \Poisson^{\alpha\beta} &= i \tilde{u}_0 X^c
  (\gamma_c)^{\alpha\beta} + \left( u + \frac{\alpha}{8}\chi^2 \right)
  (\gamma^3)^{\alpha\beta}, \label{SKV1-P-ff}
\end{align}
with
\begin{equation}
  \label{SKV1-v2}
  v_2 = -\frac{\alpha \left( \frac{\beta}{2} \phi^2 - \Lambda \right)
    + \beta \phi}{2u} - \frac{\tilde{u}_0^2 \left( \frac{\beta}{2}
      \phi^2 - \Lambda \right)^2}{2u^3}.
\end{equation}

For general parameters $\alpha$, $\beta$, $\Lambda$ from
(\ref{cIzq-u}) restrictions upon the range of $\phi$ will emerge in
general, if we do not allow singular and complex Poisson
tensors---even no allowed interval for $\phi$ may be found to exist.
In fact, as we see from (\ref{SKV1-v2}), in the present case, the
`problem' of complex-valued Poisson tensors comes together with the
`singularity-problem'.

On the other hand, for $\beta \le 0$ and $\Lambda \ge 0$, where at
least one of these parameter does not vanish, the integrand in
(\ref{SKV1-W}) becomes negative definite, leading to the restriction
$\phi > \phi_0$ with singularities at $\phi = \phi_0$. If we further
assume $\alpha > 0$ we can set $\phi_0 = -\infty$. In contrast to the
torsionless $R^2$ model (see \Tab\ref{tab:IzqMdls}) the restriction
for this particular case disappears and the fermionic potential
becomes
\begin{equation}
  \label{SKV1-u}
  u = \pm \tilde{u}_0 \sqrt{-\frac{\beta}{\alpha^3} \left( \left(
        1-\alpha\phi \right)^2 + 1 \right) +
    \frac{2\Lambda}{\alpha}}.
\end{equation}

%*********************************************************************
\subsection{General Prepotential Algebra}
\label{sec:GPA-P}

This algebra represents the immediate generalization of the
torsionless one of \Sec\ref{sec:Izq-P}, when (\ref{Izq-v-ff}) is taken
for $v^{\alpha\beta}$, but now with $u$ depending on both $\phi$ and
$Y$.  Here also $v=v(\phi,Y)$.  Again we have $\detv = 2Y
\tilde{u}_0^2 - u^2$. By analogy to the step in \Sec\ref{sec:Izq-P} we
again cancel the $f_{(t)}$ term (\cf (\ref{ans-f-bb})) by the choice
\begin{equation}
  f^c = \half (\derv u) X^c.
\end{equation}
This yields
\begin{align}
  F^c\_\alpha\^\beta &= \frac{1}{2\detv} \left( \tilde{u}_0^2 v + u
    \derv u \right) X^a (\gamma_a \gamma^c \gamma^3)_\alpha\^\beta +
  \frac{i\tilde{u}_0}{2\detv} \left( u v + 2Y \derv u \right)
  (\gamma^c)_\alpha\^\beta, \label{GPA-f-bff} \\
  v_2\^{\alpha\beta} &= \frac{1}{2\detv} \left( \left( \tilde{u}_0^2 v
      + u \derv u \right) + \dot{u} \left( u v + 2Y \derv u \right)
  \right) \gamma^3\^{\alpha\beta}, \label{GPA-v2-ff} \\
  v_2 &= \frac{uv}{2\detv^2} \left( \tilde{u}_0^2 v + u \derv u
  \right) + \frac{u \derv u}{2\detv^2} \left( u v + 2Y \derv u \right)
  + \frac{1}{2\detv} \left(u v' + 2Y \dot{v} \derv u + 2Y \derv^2 u
  \right). \label{GPA-v2}
\end{align}

The factors $\detv^{-1}$, $\detv^{-2}$ indicate the appearance of
action functional singularities, at values of the fields where $\detv$
vanishes. Again we have to this point kept $u$ independent of $v$.  In
this case, even when we relate $v$ and $u$ by imposing \eg
\begin{equation}
  \tilde{u}_0^2 v + u \derv u = 0 \Leftrightarrow
\derv\detv = 0 \Leftrightarrow \casimir(\phi,Y) = \detv,
\end{equation}
in order to cancel the first terms in
(\ref{GPA-f-bff})--(\ref{GPA-v2}), a generic singularity obstruction
is seen to persist (previous remarks on similar occasions should apply
here, too, however).

%*********************************************************************
\subsection[Algebra with $u(\phi,Y)$ and $\tilde{u}(\phi,Y)$]{Algebra
with \mathversion{bold}$u(\phi,Y)$ and $\tilde{u}(\phi,Y)$}
\label{sec:uTu-P}

For $v^{\alpha\beta}$ we retain (\ref{Izq-v-ff}), but now with both
$u$ and $\tilde{u}$ depending on $\phi$ and $Y$.  Again the
determinant
\begin{equation}
  \label{uTu-detv}
  \detv = 2Y \tilde{u}^2 - u^2
\end{equation}
will introduce singularities. If we want to cancel the $f_{(t)}$ term
(\cf (\ref{ans-f-bb})) as we did in \Sec\ref{sec:Izq-P} and
\Sec\ref{sec:GPA-P}, we have to set here
\begin{equation}
  f^c = \frac{1}{2\tilde{u}} \left( \tilde{u} \derv u - u
    \derv\tilde{u} \right) X^c.
\end{equation}
This leads to
\begin{equation}
  \label{uTu-f-bff}
  F^c = -\frac{1}{4} (\derv\ln\detv) X^a \gamma_a \gamma^c \gamma^3 +
  \half (\derv\ln\tilde{u}) X^c \gamma^3 + \frac{i \tilde{u}
    u}{2\detv} \left[ v + 2Y \left( \derv\ln\frac{u}{\tilde{u}}
    \right) \right] \gamma^c.
\end{equation}

Again we could try to fix $u$ and $\tilde{u}$ suitably so as to cancel
\eg the first term in (\ref{uTu-f-bff}):
\begin{equation}
  \label{uTu-v}
  -\half \derv \detv = \tilde{u}^2 v + u \derv u - 2Y \tilde{u} \derv
  \tilde{u} = 0.
\end{equation}
But then the singularity obstruction resurfaces in (\cf (\ref{uTu-v}))
\begin{equation}
  v = \frac{\detv'}{\dot{\detv}} = \frac{-uu' + 2Y
    \tilde{u}\tilde{u}'}{\tilde{u}^2 - u\dot{u} + 2Y
    \tilde{u}\Dot{\Tilde{u}}}\; .
\end{equation}
Eq.\ (\ref{uTu-f-bff}) becomes
\begin{equation}
  F^c = \frac{1}{\dot{\detv}} \left( \tilde{u}
    \tilde{u}' - \frac{u}{\tilde{u}} \left( \tilde{u}' \dot{u} -
      \Dot{\Tilde{u}} u' \right) \right) X^c \gamma^3 +
  \frac{i}{\dot{\detv}} \left( \tilde{u} u' - 2Y \left( \tilde{u}'
      \dot{u} - \Dot{\Tilde{u}} u' \right) \right) \gamma^c.
\end{equation}

The general formulas for the Poisson tensor are not very illuminating.
Instead, we consider two special cases.

%*********************************************************************
\subsubsection[SRG, Nondiagonal Extension II]{Spherically Reduced
Gravity (SRG), Nondiagonal Extension II}
\label{sec:SRG2}

For SRG also \eg the alternative
\begin{equation}
  \label{SRG2-v} v^{\mathrm{SRG}}(\phi,Y) = \frac{\detv'}{\dot{\detv}}
\end{equation}
exists, where $\tilde{u}$ and $u$ are given by
\begin{equation}
  \label{SRG2-uTu}
  \tilde{u} = \frac{\tilde{u}_0}{\sqrt[4]{\phi}}, \qquad u = 2
  \tilde{u}_0\lambda \sqrt[4]{\phi}\,,
\end{equation}
and $\tilde{u}_0 = \const$. The Poisson tensor is
\begin{align}
  P^{ab} &= \left( -\lambda^2 - \frac{Y}{2\phi} -
    \frac{3\lambda}{32\tilde{u}_0 \phi^{5/4}} \chi^2 \right)
  \epsilon^{ab}, \label{SRG2-P-bb} \\
  P^{\alpha b} &= -\frac{1}{8\phi} X^b (\chi\gamma^3)^\alpha +
  \frac{i\lambda}{4\sqrt{\phi}} (\chi\gamma^b)^\alpha,
  \label{SRG2-P-fb} \\
  P^{\alpha\beta} &= \frac{i\tilde{u}_0}{\sqrt[4]{\phi}} X^c
  (\gamma_c)^{\alpha\beta} + 2\tilde{u}_0\lambda \sqrt[4]{\phi}\,
  (\gamma^3)^{\alpha\beta}. \label{SRG2-P-ff}
\end{align}
Regarding the absence of obstructions this solution is as acceptable
as (and quite similar to) (\ref{SRG1-P-bb})--(\ref{SRG1-P-ff}).
Together with the diagonal extension implied by \Sec\ref{sec:BDA} and
the nondegenerate chiral extension of \Sec\ref{sec:SUSY/2-P}, these
four solutions for the extension of the physically motivated 2d
gravity theory in themselves represent a counterexample to the
eventual hope that the requirement for nonsingular, real extensions
might yield a unique answer, especially also for a supersymmetric
$\N=1$ extension of SRG.

%*********************************************************************
\subsubsection[KV, Nondiagonal Extension II]{Katanaev-Volovich Model
(KV), Nondiagonal Extension II}
\label{sec:SKV2}

Within the fermionic extension treated now also another alternative
version of the KV case may be formulated. As for SRG in
\Sec\ref{sec:SRG2} we may identify the bosonic potential (\ref{KV})
with
\begin{equation}
  \label{SKV2-v} v^{\mathrm{KV}}(\phi,Y) = \frac{\detv'}{\dot{\detv}}.
\end{equation}
Then $\tilde{u}$ and $u$ must be chosen as
\begin{align}
  \tilde{u} &= \tilde{u}_0 e^{\frac{\alpha}{2} \phi},
  \label{SKV2-uT} \\
  u &= \pm \sqrt{-2\tilde{u}_0^2 W(\phi)}, \label{SKV2-u}
\end{align}
where $\tilde{u}_0 = \const$ and $W(\phi)$ has been defined in
(\ref{SKV1-W}). Instead of (\ref{SKV1-P-bb})--(\ref{SKV1-P-ff}) we
then obtain
\begin{align}
  \Poisson^{ab} &= \left( \frac{\beta}{2} \phi^2 - \Lambda
    + \alpha Y + \half \chi^2 v_2 \right) \epsilon^{ab},
  \label{SKV2-P-bb} \\
  \Poisson^{\alpha b} &= \frac{\alpha}{4} X^b (\chi
  \gamma^3)_\alpha\^\beta - \frac{i\tilde{u}}{2u} \left( \frac{\beta}{2}
    \phi^2 - \Lambda \right) (\chi\gamma^b)_\alpha\^\beta,
  \label{SKV2-P-fb} \\
  \Poisson^{\alpha\beta} &= i \tilde{u} X^c (\gamma_c)^{\alpha\beta} +
  u (\gamma^3)^{\alpha\beta}. \label{SKV2-P-ff}
\end{align}
with
\begin{equation}
  \label{SKV2-v2} v_2 = -\frac{\alpha \left(
      \frac{\beta}{2} \phi^2 - \Lambda \right) + \beta \phi}{2u} -
  \frac{\tilde{u}^2 \left( \frac{\beta}{2} \phi^2 - \Lambda
    \right)^2}{2u^3},
\end{equation}
which, however, is beset with the same obstruction problems as the
nondiagonal extension I.

%********************************************************************
\section{Supergravity Actions}
\label{sec:action}

The algebras discussed in the last section have been selected in view
of their application in specific gravitational actions.

%********************************************************************
\subsection{First Order Formulation}
\label{sec:gFOG}

With the notation introduced in \Sec\ref{sec:ans-gP}, the
identification (\ref{ident1}), and after a partial integration, the
action (\ref{gPSM}) takes the explicit form (remember $e_A = (e_a,
\psi_\alpha)$)
\begin{equation}
  \label{gFOG}
  \Action^{\mathrm{gFOG}} = \int_\BMf \phi d\omega + X^a De_a +
  \chi^\alpha D\psi_\alpha + \half \Poisson^{AB} e_B e_A,
\end{equation}
where the elements of the Poisson structure by expansion in Lorentz
covariant components in the notation of \Sec\ref{sec:ans-gP} can be
expressed explicitly as (\cf (\ref{ans-P-bb})--(\ref{ans-f-bb}))
\begin{align}
  \label{ans-L} \half \Poisson^{AB} e_B e_A &= - \half U
  (\psi \gamma^3 \psi) - \frac{i}{2} \widetilde U X^a (\psi \gamma_a
  \psi) - \frac{i}{2} \widehat U X^a \epsilon_a\^b (\psi \gamma_b \psi)
  \eqnsplit + (\chi F^a e_a \psi) + \half V \epsilon^{ba} e_a e_b.
\end{align}
Here $F^a = (F^a)_\beta\^\gamma$, the quantity of (\ref{ans-f-bff}),
provides the direct coupling of $\psi$ and $\chi$, and $D$ is the
Lorentz covariant exterior derivative,
\begin{equation}
  \label{CovExtDer} DX^a = dX^a + X^b \omega
  \epsilon_b\^a, \qquad D\chi^\alpha = d\chi^\alpha - \half \chi^\beta
  \omega (\gamma^3)_\beta\^\alpha.
\end{equation}

Of course, at this point the Jacobi identity had not been used as yet
to relate the arbitrary functions; hence the action functional
(\ref{ans-L}) is not invariant under a local supersymmetry.  On the
other hand, when the Jacobi identities restrict those arbitrary
functions, the action (\ref{gFOG}) possesses the local symmetries
(\ref{gPSM-symms}), where the parameters $\epsilon_I = (l, \epsilon_a,
\epsilon_\alpha)$ correspond to Lorentz symmetry, diffeomorphism
invariance and, in addition, to supersymmetry, respectively. We,
nevertheless, already at this point may list the explicit
supersymmetry transformations with parameter $\epsilon_I = (0, 0,
\epsilon_\alpha)$ for the scalar fields,
\begin{align}
  \delta \phi &= \half (\chi \gamma^3 \epsilon), \label{symm-ph} \\
  \delta X^a &= - (\chi F^a \epsilon), \label{symm-X-b} \\
  \delta \chi^\alpha &= U (\gamma^3 \epsilon)^\alpha
  + i \widetilde{U} X^c (\gamma_c \epsilon)^\alpha + i \widehat{U} X^d
  \epsilon_d\^c (\gamma_c \epsilon)^\alpha, \label{symm-X-f}
\end{align}
and also for the gauge fields,
\begin{align}
  \delta \omega &= U' (\epsilon \gamma^3 \psi) + i \widetilde{U}' X^b
  (\epsilon \gamma_b \psi) + i \widehat{U}' X^a \epsilon_a\^b
  (\epsilon \gamma_b \psi) + (\chi \partial_\phi F^b
  \epsilon) e_b, \label{symm-om} \\
  \delta e_a &= i \widetilde{U} (\epsilon \gamma_a \psi) + i
  \widehat{U} \epsilon_a\^b (\epsilon \gamma_b \psi) + (\chi
  \partial_a F^b \epsilon) e_b \eqnsplit + X_a \left[ \Dot{U}
    (\epsilon \gamma^3 \psi) + i \Dot{\widetilde{U}} X^b (\epsilon
    \gamma_b \psi) + i \Dot{\widehat{U}} X^b \epsilon_b\^c (\epsilon
    \gamma_c \psi) \right], \label{symm-e-b} \\
  \delta \psi_\alpha &= -D\epsilon_\alpha + (F^b \epsilon)_\alpha e_b
  \eqnsplit + \chi_\alpha \left[ u_2 (\epsilon \gamma^3 \psi) + i
    \tilde{u}_2 X^b (\epsilon \gamma_b \psi) + i \hat{u}_2 X^a
    \epsilon_a\^b (\epsilon \gamma_b \psi) \right], \label{symm-ps-f}
\end{align}
with the understanding that they represent symmetries of the action
(\ref{gFOG}) only after the relations between the still arbitrary
functions for some specific algebra are implied. The only
transformation independent of those functions is (\ref{symm-ph}).

%********************************************************************
\subsection[Elimination of the Auxiliary Fields $X^I$]{Elimination of
the Auxiliary Fields \mathversion{bold}$X^I$}
\label{sec:Legendre}

We can eliminate the fields $X^I$ by a Legendre transformation.  To
sketch the procedure, we rewrite the action (\ref{gPSM}) in a
suggestive form as Hamiltonian action principle ($d^2\!x = dx^1 \wedge
dx^0$)
\begin{equation}
  \label{calHamilton-S}
  \Action = \int_{\BMf} d^2\!x
  \left( X^I \dot{\mathcal{A}}_I - \mathcal{H}(X,A) \right),
\end{equation}
where $X^I$ should be viewed as the `momenta' conjugate to the
`velocities' $\dot{\mathcal{A}}_I = \partial_0 A_{1I} - \partial_1
A_{0I}$ and $A_{mI}$ as the `coordinates'.  Velocities
$\dot{\mathcal{A}}_I$ and the `Hamiltonian' $\mathcal{H}(X,A) =
\Poisson^{JK} A_{0K} A_{1J}$ are densities in the present definition.
The second PSM field equation (\ref{gPSM-eomA}), in the form obtained
when varying $X^I$ in (\ref{calHamilton-S}), acts as a Legendre
transformation of $\mathcal{H}(X,A)$ with respect to the variables
$X^I$,
\begin{equation}
  \label{Legendre-calAdot}
  \dot{\mathcal{A}}_I = \frac{\rpartial \mathcal{H}(X,A)}{\partial
    X^I},
\end{equation}
also justifying the interpretation of $\dot{\mathcal{A}}_I$ as
conjugate to $X^I$. When (\ref{Legendre-calAdot}) can be solved for
all $X^I$, we get $X^I = X^I(\dot{A},A)$.  Otherwise, not all of the
$X^I$ can be eliminated and additional constraints
$\Phi(\dot{\mathcal{A}},A) = 0$ emerge. In the latter situation the
constraints may be used to eliminate some of the gauge fields $A_I$ in
favour of others. When all $X^I$ can be eliminated the Legendre
transformed density
\begin{equation}
  \label{Legendre-calF}
  \mathcal{F}(\dot{\mathcal{A}},A) = X^I(\dot{\mathcal{A}},A)
  \dot{\mathcal{A}}_I - \mathcal{H}(X(\dot{\mathcal{A}},A),A)
\end{equation}
follows, as well as the second order Lagrangian action principle
\begin{equation}
  \label{Lagrange-calS} \Action = \int_{\BMf}
  d^2\!x\,\mathcal{F}(\dot{\mathcal{A}},A),
\end{equation}
where the coordinates $A_{mI}$ must be varied
independently.

The formalism already presented applies to any (graded) PSM. If there
is an additional volume form $\epsilon$ on the base manifold $\BMf$ it
may be desirable to work with functions instead of densities. This is
also possible if the volume is dynamical as in gravity theories,
$\epsilon = \epsilon(A)$, because a redefinition of the velocities
$\dot\mathcal{A}_I$ containing coordinates $A_{mI}$ but not momenta
$X^I$ is always possible, as long as we can interpret the field
equations from varying $X^I$ as Legendre transformation. In particular
we use $\dot{A}_I = \star d A_I = \epsilon^{mn} \partial_n A_{mI}$ as
velocities and $H(X,A) = \star \left( -\half \Poisson^{JK} A_K A_J
\right) = \half P^{JK} \epsilon^{mn} A_{nK} A_{mJ}$ as Hamiltonian
function, yielding
\begin{equation}
  \label{Hamilton-S} \Action = \int_{\BMf} \epsilon
  \left( X^I \dot{A}_I - H(X,A) \right).
\end{equation}
Variation of $X^I$ leads to
\begin{equation}
  \label{Legendre-Adot} \dot{A}_I = \frac{\rpartial
    H(X,A)}{\partial X^I}.
\end{equation}
Solving this equation for $X^I = X^I(\dot{A},A)$ the Legendre
transformed function $F(\dot{A},A) = X^I(\dot{A},A) \dot{A}_I -
H(X(\dot{A},A),A)$ constitutes the action
\begin{equation}
  \label{Lagrange-S} \Action = \int_{\BMf} \epsilon
  F(\dot{A},A).
\end{equation}
If the Poisson tensor is linear in the coordinates $\Poisson^{JK} =
X^I f_I\^{JK}$, where $f_I\^{JK}$ are structure constants,
(\ref{Legendre-Adot}) cannot be used to solve for $X^I$, instead the
constraint $\dot{A}_I - \half f_I\^{JK} \epsilon^{mn} A_{nK} A_{mJ} =
0$ appears, implying that the field strength of ordinary gauge theory
is zero.  For nonlinear Poisson tensor we have always the freedom to
move $X^I$-independent terms of the rhs of (\ref{Legendre-Adot}) to
the lhs, thus using this particular type of covariant derivatives as
velocities conjugate to the momenta $X^I$ in the Legendre
transformation. This redefinition can already be done in the initial
action (\ref{Hamilton-S}) leading to a redefinition of the Hamiltonian
$H(X,A)$.

In order to bring 2d gravity theories into the form
(\ref{Hamilton-S}), but with covariant derivatives, it is desirable to
split off $\phi$-components of the Poisson tensor and to define the
`velocities' (\cf (\ref{ep-tensor}) and (\ref{HodgeDual}) in
\App\ref{sec:forms}) as
\begin{align}
  \rho &:= \star d\omega = \epsilon^{mn} (\partial_n
  \omega_{m}), \label{Ldre-ro} \\
  \tau_a &:= \star De_a = \epsilon^{mn}
  (\partial_n e_{ma}) - \omega_a, \label{Ldre-tau} \\
  \sigma_\alpha &:= \star D\psi_\alpha = \epsilon^{mn} \left(
    \partial_n \psi_{m\alpha} + \half \omega_n (\gamma^3\psi_m)_\alpha
  \right). \label{Ldre-si}
\end{align}
Here $\rho = R/2$ is proportional to the Ricci scalar; $\tau_a$ and
$\sigma_\alpha$ are, respectively, the torsion vector and the spinor
built from the derivative of the Rarita-Schwinger field. As a
consequence the Lorentz connection $\omega_m$ is absent in the
Hamiltonian,
\begin{equation}
  V(\phi,X^A;e_{mA}) = \half \Poisson^{BC}
  \epsilon^{mn} e_{nC} e_{mB} = \frac{1}{e} \Poisson^{BC} e_{0C} e_{1B},
\end{equation}
of the supergravity action
\begin{equation}
  \label{Hamilton-sugra-S}
  \Action = \int_{\BMf} \epsilon \left( \phi \rho + X^a \tau_a +
    \chi^\alpha \sigma_\alpha - V(\phi,X^A;e_{mA}) \right).
\end{equation}

%********************************************************************
\subsection{Superdilaton Theory}
\label{sec:sdil}

As remarked already in \Sec\ref{sec:intro}, first order formulations
of (bosonic) 2d gravity (and hence PSMs) allow for an at least on the
classical level globally equivalent description of general dilaton
theories (\ref{dil}). Here we show that this statement remains valid
also in the case of additional supersymmetric partners (\ie for
gPSMs).  We simply have to eliminate the Lorentz connection $\omega_a$
and the auxiliary field $X^a$. Of course, also the validity of an
algebraic elimination procedure in the most general case should (and
can) be checked by verifying that the correct \eom{}s also follow from
the final action (\ref{sdil-S2}) or (\ref{sdil-S3}).  (Alternatively
to the procedure applied below one may also proceed as in
\cite{Strobl:1999Habil}, performing two `Gaussian integrals' to
eliminate $X^a$ and $\t^a$ from the action).  In fact, in the present
section we will allow also for Poisson structures characterized by a
bosonic potential $v$ not necessarily linear in $Y \equiv \half X^a
X_a$ as in (\ref{vdil}).

Variation of $X^a$ in (\ref{gFOG}) yields the torsion equation
\begin{equation}
  \label{sdil-tau} \tau_a = \half (\partial_a
  \Poisson^{AB}) \epsilon^{mn} e_{nB} e_{mA}.
\end{equation}
{}From (\ref{sdil-tau}) using $\tilde{\omega}_a := \star de_a =
\epsilon^{mn} (\partial_n e_{ma})$ and $\tau_a = \tilde{\omega}_a -
\omega_a$ (\cf (\ref{Ldre-tau}))\footnote{For specific supergravities
  it may be useful to base this separation upon a different
  SUSY-covariant Lorentz connection $\susy{\omega}_a$ (\cf
  \Sec\ref{sec:Izq} and \ref{sec:cIzq}).} we get
\begin{equation}
  \label{sdil-om}
  \omega_a = \tilde{\omega}_a - \half
  (\partial_a \Poisson^{AB}) \epsilon^{mn} e_{nB} e_{mA}.
\end{equation}
Using (\ref{sdil-om}) to eliminate $\omega_a$ the separate terms of
(\ref{gFOG}) read
\begin{align}
  \phi d\omega &= \phi d\tilde{\omega} + \epsilon
  \epsilon^{an} (\partial_n \phi) \tau_a + \text{total div.},
  \label{sdil-phdom} \\
  X^a De_a &= \epsilon X^a \tau_a, \\
  \chi^\alpha D\psi_\alpha &= \chi^\alpha \tilde{D}\psi_\alpha + \half
  \epsilon \epsilon^{an} (\chi\gamma^3\psi_n) \tau_a,
\end{align}
where $\tau_a = \tau_a(X^I,e_{mA})$ is determined by (\ref{sdil-tau}).
The action, discarding the boundary term in (\ref{sdil-phdom}),
becomes
\begin{multline}
  \label{sdil-S1} \Action = \int_\BMf d^2\!x e \biggl[
  \phi \tilde{\rho} + \chi^\alpha \tilde{\sigma}_\alpha - \half
  \Poisson^{AB} \epsilon^{mn} e_{nB} e_{mA} \\
  + \left( X^a + \epsilon^{ar} \partial_r\phi + \half \epsilon^{ar}
    (\chi\gamma^3\psi_r) \right) \half (\partial_a \Poisson^{AB})
  \epsilon^{mn} e_{nB} e_{mA} \biggr].
\end{multline}
Here $\tilde{\rho}$ and $\tilde{\sigma}$ are defined in analogy to
(\ref{Ldre-ro}) and (\ref{Ldre-si}), but calculated with
$\tilde{\omega}_a$ instead of $\omega_a$.

To eliminate $X^a$ we vary once more with respect to $\delta X^b$:
\begin{equation}
  \left[ X^a + \epsilon^{an} (\partial_n \phi) + \half
    \epsilon^{an} (\chi\gamma^3\psi_n) \right] (\partial_b\partial_a
  \Poisson^{AB}) \epsilon^{mn} e_{nB} e_{mA} = 0.
\end{equation}
For $(\partial_b\partial_a \Poisson^{AB}) \epsilon^{mn} e_{nB} e_{mA}
\neq 0$ this leads to the (again \emph{algebraic}) equation
\begin{equation}
  \label{sdil-X-b} X^a = -\epsilon^{an} \left[
    (\partial_n \phi) + \half (\chi\gamma^3\psi_n) \right]
\end{equation}
for $X^a$. It determines $X^a$ in a way which does not depend on the
specific gPSM, because (\ref{sdil-X-b}) is nothing else than the \eom
for $\delta\omega$ in (\ref{gFOG}).

We thus arrive at the superdilaton action for an arbitrary gPSM
\begin{equation}
  \label{sdil-S2}
  \Action = \int_\BMf \phi d\tilde{\omega} + \chi^\alpha
  \tilde{D}\psi_\alpha + \half \Poisson^{AB} \bigg|_{X^a} e_B e_A,
\end{equation}
where $\vert_{X^a}$ means that $X^a$ has to be replaced by
(\ref{sdil-X-b}). The action (\ref{sdil-S2}) expressed in component
fields
\begin{equation}
  \label{sdil-S3} \Action = \int_\BMf d^2\!x e \left[
    \phi \frac{\tilde{R}}{2} + \chi^\alpha \tilde{\sigma}_\alpha - \half
    \Poisson^{AB}\bigg|_{X^a} \epsilon^{mn} e_{nB} e_{mA} \right]
\end{equation}
explicitly shows the fermionic generalization of the bosonic dilaton
theory (\ref{dil}) for any gPSM. A quadratic term $X^2=X^a X_a$ in
$\Poisson^{AB}$ because of the first term in (\ref{sdil-X-b}) provides
the usual kinetic term of the $\phi$-field in (\ref{dil}) if we take
the special case (\ref{vdil}),
\begin{align}
  \Action &= \Action^{\mathrm{dil}} + \Action^{\mathrm{f}}
  \\
  \Action^{\mathrm{f}} &= \int_\BMf d^2\!x e \biggl[ \chi^\alpha
  \tilde{\sigma}_\alpha - \frac{Z}{2} (\partial^n \phi)
  (\chi\gamma^3\psi_n) + \frac{Z}{16} \chi^2 (\psi^n \psi_n) \eqnsplit
  +\half\chi^2 v_2\bigg|_{X^a} + (\chi F^a\bigg|_{X^a} \epsilon_a\^m
  \psi_m) - \half \Poisson^{\alpha\beta}\bigg|_{X^a} \epsilon^{mn}
  \psi_{n\beta} \psi_{m\alpha} \biggr].
\end{align}
However, (\ref{sdil-S3}) even allows an arbitrary dependence on $X^2$
and a corresponding dependence on higher powers of $(\partial^n \phi)
(\partial_n \phi)$. For the special case where $\Poisson^{AB}$ is
linear in $X^a$ (\ref{sdil-S1}) shows that $X^a$ drops out of that
action without further elimination. However, the final results
(\ref{sdil-S2}) and (\ref{sdil-S3}) are the same.

%*********************************************************************
\section{Actions for Particular Models}
\label{sec:models}

Whereas in \Sec\ref{sec:Poisson} a broad range of solutions of graded
Poisson algebras has been constructed, we now discuss the related
actions and their (eventual) relation to a supersymmetrization of
(\ref{grav}) or (\ref{dil}). It will be found that, in contrast to the
transition from (\ref{FOG}) to (\ref{dil}) the form (\ref{grav}) of
the supersymmetric action requires that the different functions in the
gPSM solution obey certain conditions which are not always fulfilled.

For example, in order to obtain the supersymmetrization of
(\ref{grav}), $\phi$ and $X^a$ should be eliminated by a Legendre
transformation.  This is possible only if the Hessian determinant of
$v(\phi,Y)$ with respect to $X^i=(\phi,X^a)$ is regular,
\begin{equation}
  \label{HesseDet}
  \det\left( \frac{\partial^2 v}{\partial X^i \partial X^j} \right)
  \neq 0.
\end{equation}
Even in that case the generic situation will be that no closed
expression of the form (\ref{grav}) can be given.

In the following subsections for each algebra of \Sec\ref{sec:Poisson}
the corresponding FOG action (\ref{gFOG}), the related supersymmetry
and examples of the superdilaton version (\ref{sdil-S3}) will be
presented. In the formulas for the local supersymmetry we always drop
the (common) transformation law for $\delta\phi$ (\ref{symm-ph}).

%*********************************************************************
\subsection{Block Diagonal Supergravity}
\label{sec:BDS}

The action functional can be read off from (\ref{gFOG}) and
(\ref{ans-L}) for the Poisson tensor of \Sec\ref{sec:BDA} (\cf
(\ref{BDA-U}) for $U$ and (\ref{BDA-v2}) together with (\ref{ans-V})
to determine $V$). It reads
\begin{equation}
  \label{BDS-S}
  \Action^{\mathrm{BDS}} = \int_\BMf \phi d\omega + X^a De_a +
  \chi^\alpha D\psi_\alpha - \half U (\psi \gamma^3 \psi) + \half V
  \epsilon^{ba} e_a e_b,
\end{equation}
and according to (\ref{symm-ph})--(\ref{symm-ps-f}) possesses the
local supersymmetry
\begin{alignat}{3}
  & &\qquad
  \delta X^a &= 0, &\qquad
  \delta \chi^\alpha &= U (\gamma^3 \epsilon)^\alpha,
  \label{BDS-symm-X} \\
  \delta \omega &= U' (\epsilon \gamma^3 \psi), &\qquad
  \delta e_a &= X_a \Dot{U} (\epsilon \gamma^3 \psi), &\qquad
  \delta \psi_\alpha &= -D\epsilon_\alpha + \chi_\alpha u_2 (\epsilon
  \gamma^3 \psi). \label{BDS-symm-A}
\end{alignat}
This transformation leads to a translation of $\chi^\alpha$ on the
hypersurface $\Casimir = \const$ if $u \neq 0$. Comparing with the
usual supergravity type symmetry (\ref{susytrafo}) we observe that the
first term in $\delta\psi_\alpha$ has the required form
(\ref{susytrafo}), but the variation $\delta e_a$ is quite different.

The fermionic extension (\ref{BDA-C}) of the Casimir function
(\ref{Casimir}) for this class of theories implies an absolute
conservation law $\Casimir = \casimir_0 = \const$.

Whether the supersymmetric extension of the action of type
(\ref{grav}) can be reached depends on the particular choice of the
bosonic potential $v$. An example where the elimination of all target
space coordinates $\phi$, $X^a$ and $\chi^\alpha$ is feasible and
actually quite simple is $R^2$-supergravity with $v =
-\frac{\alpha}{2} \phi^2$ and $U = u_0 = \const$.\footnote{Clearly
  then no genuine supersymmetry is implied by (\ref{BDS-symm-A}). But
  we use this example as an illustration for a complete elimination of
  different combinations of $\phi$, $X^a$ and $\chi^\alpha$.} The
result in this case is (\cf \Sec\ref{sec:Legendre} and especially
(\ref{Hamilton-sugra-S}))
\begin{equation}
  \Action^{R^2} = \int_\BMf d^2\!x e \left[ \frac{1}{8\alpha}
    \tilde{R}^2 - \frac{2 u_0}{\tilde{R}} \tilde{\sigma}^\alpha
    \tilde{\sigma}_\alpha + \frac{u_0}{2} \epsilon^{nm} (\psi_m
    \gamma^3 \psi_n) \right].
\end{equation}
Here the tilde in $\tilde{R}$ and $\tilde{\sigma}^\alpha$ indicates
that the torsion-free connection $\tilde{\omega}_a = \epsilon^{mn}
\partial_n e_{ma}$ is used to calculate the field strengths. In
supergravity it is not convenient to eliminate the field $\phi$.
Instead it should be viewed as the `auxiliary field' of supergravity
and therefore remain in the action. Thus eliminating only $X^a$ and
$\chi^\alpha$ yields
\begin{equation}
  \Action^{R^2} = \int_\BMf d^2\!x e \left[ \phi \frac{\tilde{R}}{2} -
    \frac{u_0}{\alpha \phi} \tilde{\sigma}^\alpha
    \tilde{\sigma}_\alpha - \frac{\alpha}{2} \phi^2 + \frac{u_0}{2}
    \epsilon^{nm} (\psi_m \gamma^3 \psi_n) \right].
\end{equation}

Also for SRG in $d$-dimensions (\ref{SRG}) such an elimination is
possible if $d \neq 4$, $d < \infty$. However, interestingly enough,
the Hessian determinant vanishes just for the physically most relevant
dimension four (SRG) and for the DBH (\ref{DBH}), preventing just
there a transition to the form (\ref{grav}).

The formula for the equivalent superdilaton theories (\ref{sdil-S3})
is presented for the restriction (\ref{vdil}) to quadratic torsion
only, in order to have a direct comparison with (\ref{dil}). The
choice $U = u_0 = \const$ yields
\begin{align}
  \Action^{\mathrm{BDA}} &= \Action^{\mathrm{dil}} +
  \Action^{\mathrm{f}} \\
  \intertext{with the fermionic extension}
  \Action^{\mathrm{f}} &= \int_\BMf d^2\!x e \biggl[ \chi^\alpha
  \tilde{\sigma}_\alpha - \frac{Z}{2} (\partial^n \phi)
  (\chi\gamma^3\psi_n) + \frac{Z}{16} \chi^2 (\psi^n \psi_n) \eqnsplit
  -\frac{1}{4u_0} \chi^2 \left( V' - \frac{Z'}{2} (\partial^n \phi)
    (\partial_n \phi) \right) + \frac{u_0}{2} \epsilon^{mn} (\psi_n
  \gamma^3 \psi_m) \biggr].
\end{align}
It should be noted that this model represents a superdilaton theory
for arbitrary functions $V(\phi)$ and $Z(\phi)$ in (\ref{dil}).

%*********************************************************************
\subsection{Parity Violating Supergravity}
\label{sec:PVS}

The action corresponding to the algebra of \Sec\ref{sec:SUSY/2-P}
inserted into (\ref{gFOG}) and (\ref{ans-L}) becomes
\begin{multline}
  \label{PVS-S}
  \Action^{\mathrm{PVS}} = \int_\BMf \phi d\omega + X^a De_a +
  \chi^\alpha D\psi_\alpha + \epsilon \left( v - \frac{1}{4 u_0}
    \chi^2 v' \right) \\
   - \frac{i \tilde{u}_0 v}{2 u_0} (\chi \gamma^a P_{\pm} e_a \psi)
  - \frac{i \tilde{u}_0}{2} X^a (\psi \gamma_a P_{\pm} \psi) -
  \frac{u_0}{2} (\psi \gamma^3 \psi),
\end{multline}
with local supersymmetry
\begin{equation}
  \delta X^a = \frac{i \tilde{u}_0 v}{2 u_0} (\chi \gamma^a P_{\pm}
  \epsilon), \qquad
  \delta \chi^\alpha = i \tilde{u}_0 X^b (\gamma_b P_{\pm}
  \epsilon)^\alpha + u_0 (\gamma^3 \epsilon)^\alpha,
\end{equation}
as well as
\begin{align}
  \delta \omega &= -\frac{i \tilde{u}_0 v'}{2 u_0} (\chi \gamma^b
  P_{\pm} \epsilon) e_b, \\
  \delta e_a &= i \tilde{u}_0 (\epsilon \gamma_a P_{\pm} \psi) -
  \frac{i \tilde{u}_0}{2 u_0} X_a \dot{v} (\chi \gamma^b P_{\pm}
  \epsilon) e_b, \\
  \delta \psi_\alpha &= -D\epsilon_\alpha - \frac{i \tilde{u}_0 v}{2
    u_0} (\gamma^b P_{\pm} \epsilon)_\alpha e_b,
\end{align}
and the absolute conservation law (\ref{SUSY/2-C}). Here, in contrast
to the model of \Sec\ref{sec:BDS}, the transformation law of $e_a$
essentially has the `canonical' form (\ref{susytrafo}).

As seen from the action (\ref{PVS-S}) and the symmetry transformations
the two chiralities are treated differently, but we do not have the
case of a genuine chiral supergravity (\cf \Sec\ref{sec:CPA} below).

%*********************************************************************
\subsection{Deformed Rigid Supersymmetry}
\label{sec:SUSY}

In this case with the algebra (\ref{SUSY-P-bb})--(\ref{SUSY-P-ff}) we
obtain
\begin{multline}
  \label{SUSY-S}
  \Action^\mathrm{DRS} = \int_\BMf \phi d\omega + X^a De_a +
  \chi^\alpha D\psi_\alpha + \epsilon v \\
  + \frac{v}{4Y} X^a (\chi \gamma^3 \gamma_a \gamma^b e_b \psi) -
  \frac{i \tilde{u}_0}{2} X^a (\psi \gamma_a \psi) - \frac{v}{16Y}
  \chi^2 (\psi \gamma^3 \psi)
\end{multline}
with local supersymmetry (\ref{symm-ph}),
\begin{equation}
  \delta X^a = -\frac{v}{4Y} X^b (\chi \gamma_b \gamma^a \gamma^3
  \epsilon), \qquad
  \delta \chi^\alpha = i \tilde{u}_0 X^b (\gamma_b \epsilon)^\alpha +
  \half\chi^2 \frac{v}{4Y} (\gamma^3 \epsilon)^\alpha,
\end{equation}
and
\begin{align}
  \delta \omega &= \left( \frac{v}{4Y} \right)' \left[ X^c
    (\chi\gamma_c\gamma^b\gamma^3\epsilon) e_b + \half\chi^2
    (\epsilon\gamma^3\psi) \right], \\
  \delta e_a &= i \tilde{u}_0 (\epsilon \gamma_a \psi) + \frac{v}{4Y}
  (\chi\gamma_a\gamma^b\gamma^3\epsilon) e_b \eqnsplit
  + X_a \left( \frac{v}{4Y} \right)\spdot \left[ X^c
    (\chi\gamma_c\gamma^b\gamma^3\epsilon) e_b + \half\chi^2
    (\epsilon\gamma^3\psi) \right], \\
  \delta \psi_\alpha &= -D\epsilon_\alpha + \frac{v}{4Y} \left[ X^c
    (\gamma_c\gamma^b\gamma^3\epsilon)_\alpha e_b + \chi_\alpha
    (\epsilon\gamma^3\psi) \right].
\end{align}

Clearly, this model exhibits a `genuine' supergravity symmetry
(\ref{susytrafo}). As pointed out already in \Sec\ref{sec:SUSY-P} the
bosonic potential $v(\phi,Y)$ is not restricted in any way by the
super-extension. However, a new singularity of the action functional
occurs at $Y = 0$. The corresponding superdilaton theory can be
derived along the lines of \Sec\ref{sec:sdil}.

%*********************************************************************
\subsection{Dilaton Prepotential Supergravities}
\label{sec:Izq}

In its FOG version the action from (\ref{gFOG}) with
(\ref{Izq-P-bb})--(\ref{Izq-P-ff}) reads
\begin{multline}
  \label{Izq-S}
  \Action^{\mathrm{DPA}} = \int_\BMf \phi d\omega + X^a De_a +
  \chi^\alpha D\psi_\alpha - \frac{1}{2\tilde{u}_0^2} \epsilon \left(
    (u^2)' - \half \chi^2 u'' \right) \\
   + \frac{iu'}{2\tilde{u}_0} (\chi \gamma^a e_a \psi) - \frac{i
     \tilde{u}_0}{2} X^a (\psi \gamma_a \psi) - \frac{u}{2} (\psi
   \gamma^3 \psi),
\end{multline}
where $\tilde{u}_0=\const$ and the `prepotential' $u$ depends on
$\phi$ only. The corresponding supersymmetry becomes (\ref{symm-ph}),
\begin{equation}
  \label{Izq-symm-X}
  \delta X^a = -\frac{iu'}{2\tilde{u}_0} (\chi\gamma^a\epsilon),
  \qquad
  \delta \chi^\alpha = i \tilde{u}_0 X^b (\gamma_b \epsilon)^\alpha +
  u (\gamma^3 \epsilon)^\alpha,
\end{equation}
and further
\begin{align}
  \delta \omega &= u' (\epsilon \gamma^3 \psi) +
  \frac{iu''}{2\tilde{u}_0} (\chi\gamma^b\epsilon) e_b,
  \label{Izq-symm-om} \\
  \delta e_a &= i \tilde{u}_0 (\epsilon \gamma_a \psi),
  \label{Izq-symm-e-b} \\
  \delta \psi_\alpha &= -D\epsilon_\alpha + \frac{iu'}{2\tilde{u}_0}
  (\gamma^b\epsilon)_\alpha e_b. \label{Izq-symm-ps-f}
\end{align}

Here we have the special situation of an action linear in $X^a$,
described at the end of \Sec\ref{sec:sdil}. Variation of $X^a$ in
(\ref{Izq-S}) leads to the constraint $De_a - \frac{i\tilde{u}_0}{2}
(\psi\gamma_a\psi) = 0$.  It can be used to eliminate the Lorentz
connection, \ie $\omega_a = \susy{\omega}_a$, where we introduced the
SUSY-covariant connection
\begin{equation}
  \label{Izq-om}
  \susy{\omega}_a := \epsilon^{mn} (\partial_n e_{ma}) +
  \frac{i\tilde{u}_0}{2} \epsilon^{mn} (\psi_n\gamma_a\psi_m).
\end{equation}
The action reads
\begin{multline}
  \label{Izq-S1}
  \Action^{\mathrm{DPA}} = \int_\BMf d^2\!x e \biggl[ \phi
  \frac{\susy{R}}{2} + \chi^\alpha \susy{\sigma}_\alpha -
  \frac{1}{2\tilde{u}_0^2} \left( (u^2)' - \half \chi^2 u'' \right) \\
  + \frac{iu'}{2\tilde{u}_0} (\chi \gamma^m \epsilon_m\^n \psi_n) +
  \frac{u}{2} \epsilon^{mn}(\psi_n \gamma^3 \psi_m) \biggr],
\end{multline}
where $\susy{R}$ and $\susy{\sigma}$ indicate that these covariant
quantities are built with the spinor dependent Lorentz connection
(\ref{Izq-om}).

The present model is precisely the one studied in \cite{\bibIzq}.  In
\Sec\ref{sec:sdil-sol} we give the explicit solution of the PSM field
equations for this model. The $R^2$ model and the model of Howe will
be treated in a little more detail now.

%*********************************************************************
\subsubsection[$R^2$ Model]{\mathversion{bold}$R^2$ Model}
\label{sec:R2}

The supersymmetric extension of $R^2$-gravity with bosonic potential
$v = -\frac{\alpha}{2} \phi^2$ is obtained with the general solution
$u = \pm \tilde{u}_0 \sqrt{\frac{\alpha}{3} (\phi^3 - \phi_0^3)}$ (\cf
(\ref{Izq-u})). In order to simplify the analysis we choose $u =
\tilde{u}_0 \sqrt{\frac{\alpha}{3} \phi^3}$.  The parameter $\alpha$
can have both signs, implying the restriction on the range of the
dilaton field such that $\alpha \phi > 0$. Thus the superdilaton
action (\ref{Izq-S1}) becomes
%\begin{multline}
%  \label{R2-S}
%  \Action^{R^2} = \int_\BMf d^2\!x e \left[ \phi \rho + X^a \tau_a +
%    \chi^\alpha \sigma_\alpha - \frac{\alpha}{2} \phi^2 +
%    \frac{1}{16\tilde{u}_0} \chi^2 \sqrt{\frac{3\alpha}{\phi}}
%  \right. \\
%  \left. + \frac{3i\alpha}{4} \sqrt{\frac{\phi}{3\alpha}}
%    (\chi\gamma^m\epsilon_m\^n\psi_n) + \frac{i\tilde{u}_0}{2} X^a
%    \epsilon^{mn} (\psi_n\gamma_a\psi_m) + \frac{\tilde{u}_0}{2}
%    \sqrt{\frac{\alpha}{3} \phi^3} \epsilon^{mn}
%    (\psi_n\gamma^3\psi_m) \right]
%\end{multline}
\begin{multline}
  \label{R2-S}
  \Action^{R^2} = \int_\BMf d^2\!x e \biggl[ \phi \frac{\susy{R}}{2} +
  \chi^\alpha \susy{\sigma}_\alpha - \frac{\alpha}{2} \phi^2 +
  \frac{1}{16\tilde{u}_0} \chi^2 \sqrt{\frac{3\alpha}{\phi}} \\
  + \frac{3i\alpha}{4} \sqrt{\frac{\phi}{3\alpha}}\,
  (\chi\gamma^m\epsilon_m\^n\psi_n) + \frac{\tilde{u}_0}{2}
  \sqrt{\frac{\alpha}{3} \phi^3}\, \epsilon^{mn}
  (\psi_n\gamma^3\psi_m) \biggr].
\end{multline}

The equation obtained when
varying $\chi^\alpha$ yields
\begin{equation}
  \label{R2-ch}
  \chi^\alpha = -8 \tilde{u}_0 \sqrt{\frac{\phi}{3\alpha}}\,
  \susy{\sigma}^\alpha - 2i \tilde{u}_0 (\psi^n \epsilon_n\^m
  \gamma_m)^\alpha.
\end{equation}
Eliminating the $\chi^\alpha$ field gives
\begin{multline}
  \Action^{R^2} = \int_\BMf d^2\!x e \biggl[ \phi \frac{\susy{R}}{2} -
  4 \tilde{u}_0 \sqrt{\frac{\phi}{3\alpha}}\, \susy{\sigma}^\alpha
  \susy{\sigma}_\alpha - 2i \tilde{u}_0 \phi (\susy{\sigma}
  \gamma^m \epsilon_m\^n \psi_n) - \frac{\alpha}{2} \phi^2 \\
  + \frac{\tilde{u}_0}{4} \sqrt{\frac{\alpha}{3} \phi^3}\, \left( 3
    (\psi^n \psi_n) - \epsilon^{mn} (\psi_n\gamma^3\psi_m) \right)
  \biggr].
\end{multline}
Further elimination of $\phi$ requires the solution of a cubic
equation for $\sqrt{\phi}$ with a complicated explicit solution,
leading to an equally complicated supergravity generalization of the
formulation (\ref{grav}) of this model.

%********************************************************************
\subsubsection{Model of Howe}
\label{sec:Howe}

The supergravity model of Howe \cite{Howe:1979ia}, originally derived
in terms of superfields, is just a special case of our generic model
in the graded PSM approach. Using for the various independent
potentials the particular values
\begin{equation}
  \label{Howe-pot1}
  \tilde{u}_0 = -2, \qquad u = -\phi^2
\end{equation}
we obtain $\detv = 8 Y - \phi^4$ and for the other nonzero potentials
(\cf (\ref{Izq-pot1}), (\ref{Izq-pot2}))
\begin{equation}
  \label{Howe-pot2}
  v = -\half \phi^3, \qquad
  v_2 = -\frac{1}{4}, \qquad
  f_{(s)} = \half \phi.
\end{equation}

%We choose, by normalizing $\dot{\casimir} = 1$,
%\begin{equation}
%  \label{Howe-C}
%  \casimir = Y - \frac{1}{8} \phi^4, \qquad
%  \casimir_2 = -\frac{1}{4} \phi.
%\end{equation}

The Lagrange density for this model in the formulation (\ref{FOG}) is
a special case of (\ref{Izq-S}):
\begin{multline}
  \label{Howe-S}
  \Action^{\mathrm{Howe}} = \int_\BMf \phi d\omega + X^a De_a +
  \chi^\alpha D\psi_\alpha \\ 
  + \half \phi^2 (\psi \gamma^3 \psi) + i X^a (\psi \gamma_a \psi) +
  \frac{i}{2} \phi (\chi \gamma^a e_a \psi) - \half \epsilon
  \left( \phi^3 + \frac{1}{4} \chi^2 \right).
\end{multline}

The local supersymmetry transformations from
(\ref{Izq-symm-X})--(\ref{Izq-symm-ps-f}) are
\begin{equation}
  \label{Howe-symm-X}
  \delta X^a = -\frac{i}{2} \phi (\chi \gamma^a \epsilon), \qquad
  \delta \chi^\alpha = -\phi^2 (\gamma^3 \epsilon)^\alpha - 2i X^b
  (\gamma_b \epsilon)^\alpha,
\end{equation}
and
\begin{align}
  \delta \omega &= -2 \phi (\epsilon \gamma^3 \psi) + \frac{i}{2} (\chi
  \gamma^b \epsilon) e_b, \label{Howe-symm-om} \\
  \delta e_a &= -2i (\epsilon \gamma_a \psi), \label{Howe-symm-e-b} \\
  \delta \psi_\alpha &= -D\epsilon_\alpha + \frac{i}{2} \phi (\gamma^b
  \epsilon)_\alpha e_b. \label{Howe-symm-ps-f}
\end{align}

%The bosonic torsion 2-form can be easily calculated by varying $X^a$
%in (\ref{Howe-S}),
%\begin{equation}
%  \label{Howe-T}
%  T_a = De_a = -i (\psi \gamma_a \psi),
%\end{equation}
%which is usually used to eliminate the Lorentz connection $\omega$ in
%terms of the derivative of the zweibein $e_a$ and the Rarita-Schwinger
%field $\psi_\alpha$. As a consequence of the second order formulation
%the Lagrange multiplier $X^a$ drops out of the Lagrange density
%(\ref{Howe-S}).

Starting from the dilaton action (\ref{Izq-S1}) with (\ref{Howe-pot1})
and (\ref{Howe-pot2}), the remaining difference to the formulation of
Howe is the appearance of the fermionic coordinate $\chi^\alpha$. Due
to the quadratic term of $\chi^\alpha$ in (\ref{Howe-S}) we can use
its own algebraic field equations to eliminate it. Applying the
Hodge-dual yields
\begin{equation}
  \label{Howe-ch}
  \chi_\alpha = 4 \susy{\sigma}_\alpha + 2i \phi (\gamma^n
  \epsilon_n\^m \psi_m)_\alpha.
\end{equation}
Inserting this into the Lagrange density (\ref{Howe-S}) and into the
supersymmetry transformations (\ref{symm-ph}) as well as into
(\ref{Howe-symm-X})--(\ref{Howe-symm-ps-f}) and identifying $\phi$
with the scalar, usually interpreted as auxiliary field $A$, $\phi
\equiv A$, reveals precisely the supergravity model of Howe. That
model, in a notation almost identical to the one used here, is also
contained in \cite{Ertl:1997ib,Ertl:2001PhD}, where a superfield
approach was used.

%*********************************************************************
\subsection{Supergravities with Quadratic Bosonic Torsion}
\label{sec:cIzq}

The algebra (\ref{cIzq-P-bb-alt})--(\ref{cIzq-P-ff-alt}) in
(\ref{gFOG}) leads to
\begin{multline}
  \label{cIzq-S}
  \Action^{\mathrm{QBT}} = \int_\BMf \phi d\omega + X^a De_a +
  \chi^\alpha D\psi_\alpha + \epsilon \left( V + \half X^a X_a Z +
    \half\chi^2 v_2 \right) \\
  + \frac{Z}{4} X^a (\chi\gamma^3\gamma_a\gamma^b e_b \psi) -
  \frac{i\tilde{u}_0 V}{2u} (\chi \gamma^a e_a \psi) \\
  - \frac{i \tilde{u}_0}{2} X^a (\psi \gamma_a \psi) - \frac{1}{2}
  \left( u + \frac{Z}{8} \chi^2 \right) (\psi \gamma^3 \psi),
\end{multline}
with $u(\phi)$ determined from $V(\phi)$ and $Z(\phi)$ according to
(\ref{cIzq-u}) and
\begin{equation}
  v_2 = -\frac{1}{2u} \left( V Z + V' + \frac{\tilde{u}_0^2 V^2}{u^2}
  \right).
\end{equation}

The special interest in models of this type derives from the fact that
because of their equivalence to the dilaton theories with dynamical
dilaton field (\cf \Sec\ref{sec:sdil}) they cover a large class of
physically interesting models. Also, as shown in \Sec\ref{sec:cIzq-P}
these models are connected by a simple conformal transformation to
theories without torsion, discussed in \Sec\ref{sec:Izq}.

Regarding the action (\ref{cIzq-S}) it should be kept in mind that
calculating the prepotential $u(\phi)$ we discovered the condition
$W(\phi) < 0$ (\cf (\ref{cIzq-u})). This excludes certain bosonic
theories from supersymmetrization with real actions, and it may lead
to restrictions on $\phi$, but there is even more information
contained in this inequality: It leads also to a restriction on $Y$.
Indeed, taking into account (\ref{c-sol}) we find
\begin{equation}
  \label{cIzq-resY}
  Y > \casimir(\phi,Y) e^{-Q(\phi)}.
\end{equation}

The local supersymmetry transformations of the action (\ref{cIzq-S})
become (\ref{symm-ph}),
\begin{align}
  \delta X^a &=  -\frac{Z}{4} X^b
  (\chi\gamma_b\gamma^a\gamma^3\epsilon) + \frac{i\tilde{u}_0 V}{2u}
  (\chi\gamma^a\epsilon), \label{cIzq-symm-X-b} \\
  \delta \chi^\alpha &= i \tilde{u}_0 X^b (\gamma_b \epsilon)^\alpha +
  \left( u + \frac{Z}{8} \chi^2 \right) (\gamma^3 \epsilon)^\alpha,
  \label{cIzq-symm-X-f}
\end{align}
and
\begin{align}
  \delta \omega &= \left( -\frac{\tilde{u}_0^2 V}{u} - \frac{Zu}{2} +
    \frac{Z'}{8} \chi^2 \right) (\epsilon \gamma^3 \psi) +
  \frac{Z'}{4} X^a (\chi\gamma^3\gamma_a\gamma^b\epsilon) e_b
  \eqnsplit\hspace{8.5em} - \frac{\tilde{u}_0}{2u} \left( V' +
    \frac{VZ}{2} + \frac{\tilde{u}_0^2 V^2}{u^2} \right)
  (\chi\gamma^b\epsilon) e_b, \label{cIzq-symm-om} \\
  \delta e_a &= i \tilde{u}_0 (\epsilon \gamma_a \psi) + \frac{Z}{4}
  (\chi\gamma^3\gamma_a\gamma^b\epsilon) e_b, \label{cIzq-symm-e-b} \\
  \delta \psi_\alpha &= -D\epsilon_\alpha + \frac{Z}{4} X^a
  (\gamma^3\gamma_a\gamma^b\epsilon)_\alpha e_b - \frac{i\tilde{u}_0
    V}{2u} (\gamma^b\epsilon)_\alpha e_b + \frac{Z}{4} \chi_\alpha
  (\epsilon\gamma^3\psi). \label{cIzq-symm-ps-f}
\end{align}

Finally, we take a closer look at the torsion condition. Variation of
$X^a$ in (\ref{cIzq-S}) yields
\begin{equation}
  \label{cIzq-tor}
  De_a - \frac{i\tilde{u}_0}{2} (\psi\gamma_a\psi) + \frac{Z}{4}
  (\chi\gamma^3\gamma_a\gamma^b e_b\psi) + \epsilon X_a Z = 0.
\end{equation}
For $Z \neq 0$ this can be used to eliminate $X^a$ directly, as
described in \Sec\ref{sec:Legendre} for a generic PSM.  The general
procedure to eliminate instead $\omega_a$ by this equation was
outlined in \Sec\ref{sec:sdil}. There, covariant derivatives were
expressed in terms of $\tilde{\omega}_a$.  For supergravity theories
it is desirable to use SUSY-covariant derivatives instead. The
standard covariant spinor dependent Lorentz connection
$\susy{\omega}_a$ was given in (\ref{Izq-om}).  However, that quantity
does not retain its SUSY-covariance if torsion is dynamical. Eq.\ 
(\ref{cIzq-tor}) provides such a quantity. Taking the Hodge dual,
using (\ref{Ldre-tau}) we find
\begin{align}
  \omega_a &= \susy{\omega}_a + X_a Z, \label{cIzq-om} \\
  \susy{\omega}_a &\equiv \tilde{\omega}_a + \frac{i\tilde{u}_0}{2}
  \epsilon^{mn} (\psi_n\gamma_a\psi_m) + \frac{Z}{4}
  (\chi\gamma^3\gamma_a\gamma^b\epsilon_b\^n\psi_n). \label{cIzq-s-om}
\end{align}
Clearly, $\omega_a$ possesses the desired properties (\cf
(\ref{gPSM-symms})), but it is not the minimal covariant connection.
The last term in (\ref{cIzq-om}) is a function of the target space
coordinates $X^I$ only, thus covariant by itself, which leads to the
conclusion that $\susy{\omega}_a$ is the required quantity.
Unfortunately, no generic prescription to construct $\susy{\omega}_a$
exists however.  The rest of the procedure of \Sec\ref{sec:sdil} for
the calculation of a superdilaton action starting with
(\ref{sdil-phdom}) still remains valid, but with $\susy{\omega}_a$ of
(\ref{cIzq-s-om}) replacing $\tilde{\omega}_a$.

We point out that it is essential to have the spinor field
$\chi^\alpha$ in the multiplet; just as $\phi$ has been identified
with the usual auxiliary field of supergravity in \Sec\ref{sec:Howe},
we observe that general supergravity (with torsion) requires an
additional auxiliary spinor field $\chi^\alpha$.

%*********************************************************************
\subsubsection{Spherically Reduced Gravity (SRG)}
\label{sec:SRG}

The special case (\ref{SRG}) with $d=4$ for the potentials $V$ and $Z$
in (\ref{cIzq-S}) yields
\begin{multline}
  \label{SRG-S}
  \Action^{\mathrm{SRG}} = \int_\BMf \phi d\omega + X^a De_a +
  \chi^\alpha D\psi_\alpha - \epsilon \left( \lambda^2 +
    \frac{1}{4\phi} X^a X_a + \frac{3\lambda}{32\tilde{u}_0
      \phi^{3/2}} \chi^2 \right) \\
  - \frac{1}{8\phi} X^a (\chi\gamma^3\gamma_a\gamma^b e_b \psi) +
  \frac{i\lambda}{4 \sqrt{\phi}} (\chi \gamma^a e_a \psi) \\
  - \frac{i \tilde{u}_0}{2} X^a (\psi \gamma_a \psi) - \frac{1}{2}
  \left( 2\tilde{u}_0 \lambda \sqrt{\phi} - \frac{1}{16\phi} \chi^2
  \right) (\psi \gamma^3 \psi).
\end{multline}
We do not write down the supersymmetry transformations which follow
from (\ref{cIzq-symm-X-b})--(\ref{cIzq-symm-ps-f}).
%Supersymmetry:
%\begin{align}
%  \delta \phi &= \half (\chi \gamma^3 \epsilon), \\
%  \delta X^a &=  \frac{1}{8\phi} X^b
%  (\chi\gamma_b\gamma^a\gamma^3\epsilon) -
%  \frac{i\lambda}{4\sqrt{\phi}} (\chi\gamma^a\epsilon), \\
%  \delta \chi^\alpha &= i \tilde{u}_0 X^b (\gamma_b \epsilon)^\alpha +
%  \left( 2\tilde{u}_0 \lambda \sqrt{\phi} - \frac{1}{16\phi} \chi^2
%  \right) (\gamma^3 \epsilon)^\alpha
%\end{align}
%\begin{align}
%  \delta \omega &= \left( \frac{\tilde{u}_0 \lambda}{\sqrt{\phi}} +
%    \frac{1}{16\phi^2} \chi^2 \right) (\epsilon \gamma^3 \psi) +
%  \frac{1}{8\phi^2} X^a (\chi\gamma^3\gamma_a\gamma^b\epsilon) e_b -
%  \frac{\lambda}{8\phi^{3/2}} (\chi\gamma^b\epsilon) e_b, \\
%  \delta e_a &= i \tilde{u}_0 (\epsilon \gamma_a \psi) -
%  \frac{1}{8\phi} (\chi\gamma^3\gamma_a\gamma^b\epsilon) e_b, \\
%  \delta \psi_\alpha &= -D\epsilon_\alpha - \frac{1}{8\phi} X^a
%  (\gamma^3\gamma_a\gamma^b\epsilon)_\alpha e_b +
%  \frac{i\lambda}{4\sqrt{\phi}} (\gamma^b\epsilon)_\alpha e_b -
%  \frac{1}{8\phi} \chi_\alpha (\epsilon\gamma^3\psi)
%\end{align}
We just note that our transformations $\delta e_a$ and $\delta
\psi_\alpha$ are different from the ones obtained in
\cite{Park:1993sd}. There, the supergravity multiplet is the same as
in the underlying model \cite{Howe:1979ia}, identical to the one of
\Sec\ref{sec:Howe}. The difference is related to the use of an
additional scalar superfield in \cite{Park:1993sd} to construct a
superdilaton action. Such an approach lies outside the scope of the
present paper, where we remain within pure gPSM without additional
fields which, from the point of view of PSM are `matter' fields.

Here, according to the general derivation of \Sec\ref{sec:sdil}, we
arrive at the superdilaton action
%Superdilaton action (in the original sequence):
%\begin{align}
%  \Action &= \Action^{\mathrm{dil}} + \Action^{\mathrm{f}} \\
%  \Action^{\mathrm{f}} &= \int_\BMf d^2\!x e \biggl[ \chi^\alpha
%  \tilde{\sigma}_\alpha + \frac{1}{4\phi} (\partial^n \phi)
%  (\chi\gamma^3\psi_n) - \frac{1}{32\phi} \chi^2 (\psi^n \psi_n) -
%  \frac{3\lambda}{32 \tilde{u}_0 \phi^{3/2}} \chi^2 \eqnsplit -
%  \frac{1}{8\phi} (\partial^n \phi)
%  (\chi\gamma^3\gamma_n\gamma^m\psi_m) - \frac{1}{32\phi} \chi^2
%  (\psi^n\gamma_n\gamma^m\psi_m) - \frac{i\lambda}{4\sqrt{\phi}}
%  (\chi\gamma^3\gamma^m\psi_m) \eqnsplit + i\tilde{u}_0 (\partial^n
%  \phi) (\psi_n\gamma^m\psi_m) + \frac{i\tilde{u}_0}{2}
%  (\chi\gamma^3\psi^n) (\psi_n\gamma^m\psi_m) \eqnsplit + \tilde{u}_0
%  \lambda \sqrt{\phi}\, \epsilon^{mn} (\psi_n\gamma^3\psi_m) -
%  \frac{1}{32} \chi^2 \epsilon^{mn} (\psi_n\gamma^3\psi_m) \biggr].
%\end{align}
\begin{align}
  \Action^\mathrm{SRG} &= \Action^{\mathrm{dil}} +
  \Action^{\mathrm{f}}, \label{SRG-dilS1} \\
  \intertext{with bosonic part (\ref{dil}) and the fermionic extension}
  \Action^{\mathrm{f}} &= \int_\BMf d^2\!x e \biggl[ \chi^\alpha
  \tilde{\sigma}_\alpha + i\tilde{u}_0 \left\{ (\partial^n \phi) +
    \half (\chi\gamma^3\psi^n) \right\} (\psi_n\gamma^m\psi_m)
  \eqnsplit + \frac{1}{8\phi} (\partial^n \phi)
  (\chi\gamma^3\gamma^m\gamma_n\psi_m) - \frac{i\lambda}{4\sqrt{\phi}}
  (\chi\gamma^3\gamma^m\psi_m) + \tilde{u}_0 \lambda \sqrt{\phi}\,
  \epsilon^{mn} (\psi_n\gamma^3\psi_m) \eqnsplit - \frac{1}{32} \chi^2
  \left\{ \frac{1}{\phi} (\psi^n\psi_n) + \frac{1}{\phi}
    (\psi^n\gamma_n\gamma^m\psi_m) + \frac{3\lambda}{\tilde{u}_0
      \phi^{3/2}} + \epsilon^{mn} (\psi_n\gamma^3\psi_m) \right\}
  \biggr].
\end{align}

However, as already noted in the previous section, it may be
convenient to use the SUSY-covariant $\susy{\omega}_a$ (\cf
(\ref{cIzq-s-om})) instead of $\tilde{\omega}_a$, with the result:
\begin{multline}
  \label{SRG-dilS2}
  \Action^{\mathrm{SRG}} = \int_\BMf \phi d\susy{\omega} + \chi^\alpha
  \susy{D}\psi_\alpha + \frac{\epsilon}{4\phi}
  \biggl[ (\partial^n \phi) (\partial_n \phi) + (\partial^n \phi)
    (\chi\gamma^3\psi_n) + \frac{1}{8} \chi^2 (\psi^n \psi_n) \biggr] \\
  - \epsilon \biggl[ \lambda^2 + \frac{3\lambda}{32\tilde{u}_0
      \phi^{3/2}} \chi^2 \biggr]
  + \frac{i\lambda}{4 \sqrt{\phi}} (\chi \gamma^a e_a \psi) -
  \frac{1}{2} \biggl[ 2\tilde{u}_0 \lambda \sqrt{\phi} -
    \frac{1}{16\phi} \chi^2 \biggr] (\psi \gamma^3 \psi).
\end{multline}

%*********************************************************************
\subsubsection{Katanaev-Volovich Model (KV)}
\label{sec:SKV}

The supergravity action (\ref{cIzq-S}) for the algebra of
\Sec\ref{sec:SKV1} reads
\begin{multline}
  \label{SKV-S}
  \Action^{\mathrm{KV}} = \int_\BMf d^2\!x e \biggl[ \phi R + X^a
  \tau_a + \chi^\alpha \sigma_\alpha + \frac{\beta}{2} \phi^2 -
  \Lambda + \frac{\alpha}{2} X^a X_a + \half\chi^2 v_2 \\
  + \frac{\alpha}{4} X^a (\chi \gamma_a \gamma^m \psi_m) -
  \frac{i\tilde{u}_0 \left( \frac{\beta}{2} \phi^2 - \Lambda
  \right)}{2u} (\chi\gamma^m\epsilon_m\^n\psi_n) \\
  + \frac{i\tilde{u}_0}{2} X^a \epsilon^{mn} (\psi_n\gamma_a\psi_m) +
  \half \left( u + \frac{\alpha}{8} \chi^2 \right) \epsilon^{mn}
  (\psi_n\gamma^3\psi_m) \biggr],
\end{multline}
where $v_2$ and $u$ were given in (\ref{SKV1-v2}) and (\ref{SKV1-u}).
It is invariant under the local supersymmetry transformations
(\ref{symm-ph}) and
\begin{align}
  \delta X^a &= -\frac{\alpha}{4} X^b
  (\chi\gamma_b\gamma^a\gamma^3\epsilon) + \frac{i\tilde{u}_0 \left(
      \frac{\beta}{2} \phi^2 - \Lambda \right)}{2u}
  (\chi\gamma^a\epsilon), \\
  \delta \chi^\alpha &= i \tilde{u}_0 X^b (\gamma_b \epsilon)^\alpha +
  \left( u + \frac{\alpha}{8} \chi^2 \right) (\gamma^3
  \epsilon)^\alpha,
\end{align}
in conjunction with the transformations
\begin{align}
  \delta \omega &= \left[ -\frac{\tilde{u}_0^2 \left(
      \frac{\beta}{2} \phi^2 - \Lambda \right)}{u} - \frac{\alpha u}{2}
  \right] (\epsilon \gamma^3 \psi) \eqnsplit - \frac{\tilde{u}_0}{2u}
  \left[ \beta\phi + \frac{\alpha \left( \frac{\beta}{2} \phi^2 -
      \Lambda \right)}{2} + \frac{\tilde{u}_0^2 \left(
      \frac{\beta}{2} \phi^2 - \Lambda \right)^2}{u^2} \right]
  (\chi\gamma^b\epsilon) e_b, \\
  \delta e_a &= i \tilde{u}_0 (\epsilon \gamma_a \psi) +
  \frac{\alpha}{4} (\chi\gamma^3\gamma_a\gamma^b\epsilon) e_b, \\
  \delta \psi_\alpha &= -D\epsilon_\alpha + \frac{\alpha}{4} X^a
  (\gamma^3\gamma_a\gamma^b\epsilon)_\alpha e_b
  -\frac{i\tilde{u}_0 \left( \frac{\beta}{2} \phi^2 - \Lambda
  \right)}{2u}  (\gamma^b\epsilon)_\alpha e_b + \frac{\alpha}{4} \chi_\alpha
  (\epsilon\gamma^3\psi)
\end{align}
of the gauge fields.

Also here the explicit formula for the superdilaton action can be
obtained easily, however, the complicated formulas are not very
illuminating. It turns out that $\delta\omega$ and $\delta\psi_\alpha$
contain powers $u^k$, $k = -3,\ldots,1$. Therefore, singularities
related to the prepotential formulae (\ref{cIzq-u}) also affect these
transformations.

%********************************************************************
\subsection[$\N=(1,0)$ Dilaton Supergravity]%
{\mathversion{bold}$\N=(1,0)$ Dilaton Supergravity}
\label{sec:CPA}

Of special interest among the degenerate algebras of
\Sec\ref{sec:DFS-P} and \ref{sec:CPA-P} is the case where one chiral
component in $\chi^\alpha$ (say $\chi^-$) decouples from the theory.

We adopt the solution (\ref{CPA-P-bb})--(\ref{CPA-F-b}) for the
Poisson algebra accordingly.  To avoid a cross term of the form
$(\chi^- \psi_+)$ appearing in $\Poisson^{a\beta} \psi_\beta e_a =
(\chi F^a e_a \psi)$ we have to set $f_{(s)} = f_{(t)} = 0$. Similarly
$f_{(12)} = 0$ cancels a $(\chi^-\chi^+)$-term in
$\half\Poisson^{\alpha\beta} \psi_\beta \psi_\alpha$. Furthermore we
choose $\tilde{u} = \tilde{u}_0 = \const$:
\begin{multline}
  \label{CPA-S}
  \Action = \int_\BMf \phi d\omega + X^{++} De_{++} + X^{--} De_{--} +
  \chi^+ D\psi_+ + \chi^- D\psi_- \\
  + \epsilon v + \frac{v}{2Y} X^{--} e_{--} (\chi^+ \psi_+ - \chi^-
  \psi_-) + \frac{\tilde{u}_0}{\sqrt{2}} X^{++} \psi_+ \psi_+
\end{multline}
The chiral components $\chi^-$ and $\psi_-$ can be set to zero
consistently. The remaining local supersymmetry has one parameter
$\epsilon_+$ only:
\begin{alignat}{2}
  \delta \phi &= \half (\chi^+ \epsilon_+), \label{CPA-symm-ph} \\
  \delta X^{++} &= 0, &\qquad
  \delta X^{--} &= -\frac{v}{2Y} X^{--} (\chi^+ \epsilon_+),
  \label{CPA-symm-X-b} \\
  \delta \chi^+ &= \sqrt{2} \tilde{u}_0 X^{++} \epsilon_+, &\qquad
  \delta \chi^- &= 0 \label{CPA-symm-X-f},
\end{alignat}
and
\begin{align}
  \delta \omega &= \frac{v'}{2Y} X^{--} e_{--} (\chi^+ \epsilon_+),
  \label{CPA-symm-om} \\
  \delta e_{++} &= -\sqrt{2} \tilde{u}_0 (\epsilon_+ \psi_+) + X_{++}
  \left( \frac{v}{2Y} \right)\spdot X^{--} e_{--} (\chi^+ \epsilon_+),
  \label{CPA-symm-e++} \\
  \delta e_{--} &= \frac{v}{2Y} e_{--} (\chi^+ \epsilon_+) + X_{--}
  \left( \frac{v}{2Y} \right)\spdot X^{--} e_{--} (\chi^+ \epsilon_+),
  \label{CPA-symm-e--} \\
  \delta \psi_+ &= -D\epsilon_+ + \frac{v}{2Y} X^{--} e_{--}, \qquad
  \delta \psi_- = 0. \label{CPA-symm-ps-f}
\end{align}

%*********************************************************************
\section{Solution of the Dilaton Supergravity Model}
\label{sec:sdil-sol}

\newcommand{\chTp}{\tilde{\chi}^{(+)}}
\newcommand{\chTm}{\tilde{\chi}^{(-)}}
\newcommand{\chTpm}{\tilde{\chi}^{(\pm)}}
\newcommand{\chTT}{\Tilde{\Tilde{\chi}}^{(-)}}
\newcommand{\chSm}{\susy{\chi}^{(-)}}

For the dilaton prepotential supergravity model of \Sec\ref{sec:Izq}
\cite{\bibIzq}
the Poisson algebra was derived in \Sec\ref{sec:Izq-P}. The PSM field
equations (\ref{gPSM-eomX}) and (\ref{gPSM-eomA}) simplify
considerably in Casimir-Darboux coordinates which can be found
explicitly here, as in the pure gravity PSM. This is an
improvement as compared to \cite{Strobl:1999zz} where only the
\emph{existence} of such target coordinates was used.

We start with the Poisson tensor (\ref{Izq-P-bb})--(\ref{Izq-P-ff}) in
the coordinate system $X^I = (\phi,X^{++},X^{--},\chi^+,\chi^-)$. The
algebra under consideration has maximal rank $(2|2)$, implying that
there is one bosonic Casimir function $\Casimir$. Rescaling
(\ref{Izq-C}) we choose here
\begin{equation}
  \label{co-C}
  \Casimir = X^{++} X^{--} - \frac{u^2}{2\tilde{u}_0^2} + \half\chi^2
  \frac{u'}{2\tilde{u}_0^2}\,.
\end{equation}
In regions $X^{++} \neq 0$ we use $\Casimir$ instead of $X^{--}$ as
coordinate in target space ($X^{--} \rightarrow \Casimir$). In regions
$X^{--} \neq 0$ $X^{++} \rightarrow \Casimir$ is possible.

Treating the former case explicitly we replace $X^{++} \rightarrow
\lambda = -\ln|X^{++}|$ in each of the two patches $X^{++} > 0$ and
$X^{++} < 0$.  This function is conjugate to the generator of Lorentz
transformation $\phi$ (\cf (\ref{LorentzB}))
\begin{equation}
  \label{PB-la-ph}
  \{\lambda,\phi\} = 1.
\end{equation}
The functions $(\phi,\lambda,\Casimir)$ constitute a Casimir-Darboux
coordinate system for the bosonic sector \cite{\bibPSM}. Now our aim
is to decouple the bosonic sector from the fermionic one. The
coordinates $\chi^\alpha$ constitute a Lorentz spinor (\cf
(\ref{LorentzF})). With the help of $X^{++}$ they can be converted to
Lorentz scalars, \ie $\{\chTpm, \phi\} = 0$,
\begin{equation}
  \label{tr-ScalF}
  \chi^+ \rightarrow \chTp = \frac{1}{\sqrt{|X^{++}|}}
  \chi^+, \qquad
  \chi^- \rightarrow \chTm = \sqrt{|X^{++}|} \chi^-.
\end{equation}
A short calculation for the set of coordinates $(\phi, \lambda,
\Casimir, \chTp, \chTm)$ shows that $\{\lambda, \chTp\} = 0$ but
$\{\lambda, \chTm\} = -\frac{\sigma u'}{\sqrt{2} \tilde{u}_0} \chTp$,
where $\sigma := \sign(X^{++})$. The redefinition
\begin{equation}
  \label{tr-BD-F}
  \chTm \rightarrow \chTT = \chTm +
  \frac{\sigma u}{\sqrt{2}\tilde{u}_0} \chTp
\end{equation}
yields $\{\lambda, \chTT\} = 0$ and makes the algebra block diagonal.
For the fermionic sector
\begin{align}
  \{\chTp,\chTp\} &= \sigma \sqrt{2}
  \tilde{u}_0, \label{PB-F+F+} \\
  \{\chTT,\chTT\} &= \sigma \sqrt{2} \tilde{u}_0 \Casimir,
  \label{PB-F-F-} \\
  \{\chTp,\chTT\} &= 0 \label{PB-F+F-}
\end{align}
is obtained.  We first assume that the Casimir function $\Casimir$
appearing explicitly on the \rhs of (\ref{PB-F-F-}) is invertible.
Then the redefinition
\begin{equation}
  \label{tr-C-F-}
  \chTT \rightarrow \chSm = \frac{1}{\sqrt{|\Casimir|}}
  \chTT
\end{equation}
can be made. That we found the desired Casimir-Darboux system can be
seen from
\begin{equation}
  \label{PB-sF-sF-}
  \{\chSm,\chSm\} = s \sigma \sqrt{2}
  \tilde{u}_0.
\end{equation}
Here $s := \sign(\Casimir)$ denotes the sign of the Casimir function.
In fact we could rescale $\chSm$ and $\chTp$ so as to reduce the
respective right hand sides to $\pm 1$; the signature of the fermionic 
$2 \times 2$ block cannot be changed however. In any case, we call the 
local coordinates  $\bar{X}^I := (\phi, \lambda,
\Casimir, \chTp, \chSm)$ Casimir-Darboux since the respective Poisson
tensor is constant, which is the decisive feature here.  Its
non-vanishing components can be read off from (\ref{PB-la-ph}),
(\ref{PB-F+F+}) and (\ref{PB-sF-sF-}). 

Now it is straightforward to solve the PSM field equations.  Bars are
used to denote the gauge fields $\bar{A}_I = (\bar{A}_\phi,
\bar{A}_\lambda, \bar{A}_C, \bar{A}_{(+)}, \bar{A}_{(-)})$
corresponding to the coordinates $\bar{X}^I$.  The first PSM \eom{}s
(\ref{gPSM-eomX}) then read
\begin{align}
  d\phi - \bar{A}_\lambda &= 0, \label{Izq-eomPh} \\
  d\lambda + \bar{A}_\phi &= 0, \label{Izq-eomLa} \\
  d\Casimir &= 0, \label{Izq-eomC} \\
  d\chTp + \sigma \sqrt{2} \tilde{u}_0 \bar{A}_{(+)} &= 0,
  \label{Izq-eomCh+} \\
  d\chSm + s \sigma \sqrt{2} \tilde{u}_0 \bar{A}_{(-)} &=
  0. \label{Izq-eomCh-}
\end{align}
These equations decompose in two parts (which is true also in the case
of several Casimir functions). In regions where the Poisson tensor is
of constant rank we obtain the statement that any solution of
(\ref{gPSM-eomX}) and (\ref{gPSM-eomA}) lives on symplectic leaves
which is expressed here by the differential equation (\ref{Izq-eomC})
with the one-parameter solution $\Casimir = \casimir_0 = \const$.
Eqs.\ (\ref{Izq-eomPh})--(\ref{Izq-eomCh-}) without (\ref{Izq-eomC})
are to be used to solve for all gauge fields excluding the ones which
correspond to the Casimir functions, thus excluding $\bar{A}_C$ in our
case. Note that this solution is purely algebraic.  The second set of
PSM equations (\ref{gPSM-eomA}) again split in two parts. The
equations $d\bar{A}_\phi = d\bar{A}_\lambda = d\bar{A}_{(+)} =
d\bar{A}_{(-)} = 0$ are identically fulfilled, as can be seen from
(\ref{Izq-eomPh})--(\ref{Izq-eomCh-}). To show this property in the
generic case the first PSM equations (\ref{gPSM-eomX}) in conjunction
with the Jacobi identity of the Poisson tensor have to be used. The
remainder of the second PSM equations are the equations for the gauge
fields corresponding to the Casimir functions. In a case as simple as
ours we find, together with the local solution in terms of an
integration function $F(x)$ (taking values in the commuting
supernumbers),
\begin{equation}
  \label{Izq-A-C}
  d\bar{A}_C = 0 \quad \Rightarrow \quad \bar{A}_C = -dF.
\end{equation}

The explicit solution for the original gauge fields $A_I = (\omega,
e_a, \psi_\alpha)$ is derived from the target space transformation
$A_I = (\partial_I \bar{X}^J) \bar{A}_J$, but in order to compare with
the case $\Casimir = 0$ we introduce an intermediate step and give the
solution in coordinates $\tilde{X}^I = (\phi, \lambda, \Casimir,
\chTp, \chTT)$ first. With $\tilde{A}_I = (\tilde{A}_\phi,
\tilde{A}_\lambda, \tilde{A}_C, \tilde{A}_{(+)}, \tilde{A}_{(-)})$ the
calculation $\tilde{A}_I = (\tilde{\partial}_I \bar{X}^J) \bar{A}_J$
yields
\begin{equation}
  \label{Izq-solA1}
  \tilde{A}_\phi = -d\lambda, \qquad
  \tilde{A}_\lambda = d\phi, \qquad
  \tilde{A}_{(+)} = -\frac{\sigma}{\sqrt{2}\tilde{u}_0} d\chTp,
\end{equation}
and
%\begin{equation}
%  \tilde{A}_C = \bar{A}_C - \frac{1}{2\Casimir} \chSm
%  \bar{A}_{(-)}, \qquad
%  \tilde{A}_{(-)} = \frac{1}{\sqrt{|\Casimir|}} \bar{A}_{(-)},
%\end{equation}
\begin{equation}
  \label{Izq-solA2a}
  \tilde{A}_C = -dF + \frac{\sigma}{2 \sqrt{2}\tilde{u}_0 \Casimir^2}
  \chTT d\chTT, \qquad
  \tilde{A}_{(-)} = -\frac{\sigma}{\sqrt{2}\tilde{u}_0 \Casimir}
  d\chTT.
\end{equation}

This has to be compared with the case $\Casimir = 0$. Obviously, the
fermionic sector is no longer of full rank, and $\chTT$ is an
additional, fermionic Casimir function as seen from (\ref{PB-F-F-})
and also from the corresponding field equation
\begin{equation}
  \label{Izq-eomChBD}
  d\chTT = 0.
\end{equation}
Thus, the $\tilde{X}^I$ are Casimir-Darboux coordinates on the
subspace $\Casimir = 0$. The PSM \eom{}s in barred coordinates
continue to be of the form (\ref{Izq-eomPh})--(\ref{Izq-eomCh+}).
Therefore, the solution (\ref{Izq-solA1}) remains unchanged, but
(\ref{Izq-solA2a}) has to be replaced by the solution of $d\tilde{A}_C
= 0$ and $d\tilde{A}_{(-)} = 0$. In terms of the bosonic function
$F(x)$ and an additional fermionic function $\rho(x)$ the solution is
\begin{equation}
  \label{Izq-solA2b}
  \tilde{A}_C = -dF, \qquad
  \tilde{A}_{(-)} = -d\rho.
\end{equation}

Collecting all formulas the explicit solution for the original gauge
fields $A_I = (\omega, e_a, \psi_\alpha)$ calculated with $A_I =
(\partial_I \tilde{X}^J) \tilde{A}_J$ reads (\cf \App\ref{sec:spinors}
for the definition of $++$ and $--$ components of Lorentz vectors)
\begin{align}
  \omega &= \frac{d X^{++}}{X^{++}} + V \tilde{A}_C +
  \frac{\sigma u'}{\sqrt{2}\tilde{u}_0} \chTp \tilde{A}_{(-)},
  \label{Izq-solOm} \\
  e_{++} &= -\frac{d\phi}{X^{++}} + X^{--}
  \tilde{A}_C \eqnsplit + \frac{1}{2 X^{++}} \left[
    \frac{\sigma}{\sqrt{2}\tilde{u}_0} \chTp d\chTp + \left( \chTm -
      \frac{\sigma u}{\sqrt{2}\tilde{u}_0} \chTp \right) \tilde{A}_{(-)}
  \right], \label{Izq-solE++} \\
  e_{--} &= X^{++} \tilde{A}_C, \label{Izq-solE--} \\
  \psi_+ &= -\frac{u'}{2\tilde{u}_0^2} \chi^-
  \tilde{A}_C - \frac{\sigma}{\sqrt{2}\tilde{u}_0}
  \frac{1}{\sqrt{|X^{++}|}} \left( d\chTp - u \tilde{A}_{(-)} \right),
  \label{Izq-solPs+} \\
  \psi_- &= \frac{u'}{2\tilde{u}_0^2} \chi^+
  \tilde{A}_C + \sqrt{|X^{++}|} \tilde{A}_{(-)}. \label{Izq-solPs-}
\end{align}
$\tilde{A}_C$ and $\tilde{A}_{(-)}$ are given by (\ref{Izq-solA2a})
for $\Casimir \neq 0$ and by (\ref{Izq-solA2b}) for $\Casimir = 0$.
%The explicit solution for the original gauge fields $A_I = (\omega,
%e_a, \psi_\alpha)$ is derived from $A_I = (\rpartial_I \bar{X}^J)
%\bar{A}_J$ with the result
%\begin{align}
%  \omega &= -d\lambda - V dF - \frac{s\sigma}{\sqrt{2}\tilde{u}_0}
%  (\partial_\phi \chSm) d\chSm, \\
%  e_{++} &= -\frac{d\phi}{X^{++}} - X^{--} dF + \frac{\sigma}{2
%    \sqrt{2}\tilde{u}_0 X^{++}}\, \chTp d\chTp
%  - \frac{s\sigma}{\sqrt{2}\tilde{u}_0}
%  (\partial_{++} \chSm) d\chSm, \\
%  e_{--} &= -X^{++} dF + \frac{s\sigma X^{++}}{2 \sqrt{2}\tilde{u}_0
%    \Casimir}\, \chSm d\chSm, \\
%  \psi_+ &= \frac{u'}{2\tilde{u}_0^2}\, \chi^- dF -
%  \frac{\sigma}{\sqrt{2}\tilde{u}_0 \sqrt{|X^{++}|}}\,
%  d\chTp - \frac{s u}{2\tilde{u}_0^2 \sqrt{|X^{++}|}
%    \sqrt{|\casimir|}}\, d\chSm, \\
%  \psi_- &= -\frac{u'}{2\tilde{u}_0^2}\, \chi^+ dF - \frac{s\sigma
%    \sqrt{|X^{++}|}}{\sqrt{2}\tilde{u}_0 \sqrt{|\casimir|}}\,
%  d\chSm.
%\end{align}

For $\Casimir \neq 0$ our solution depends on the free function $F$
and the coordinate functions $(\phi, X^{++}, X^{--}, \chi^+, \chi^-)$
which, however, are constrained by $\Casimir = \casimir_0 = \const$
according to (\ref{co-C}). For $\Casimir = 0$ the free functions are
$F$ and $\rho$. The coordinate functions $(\phi, X^{++}, X^{--},
\chi^+, \chi^-)$ here are restricted by $\Casimir = 0$ in (\ref{co-C})
and by $\chTT = \const$.

This solution holds for $X^{++} \neq 0$; an analogous set of relations
can be derived exchanging the role of $X^{++}$ and $X^{--}$.

The solution (\cf (\ref{Izq-solE++}) and (\ref{Izq-solE--})) is free
from coordinate singularities in the line element, exhibiting a sort
of `super Eddington-Finkelstein' form. For special choices of the
potentials $v(\phi)$ or the related prepotential $u(\phi)$ we refer to
\Tab\ref{tab:IzqMdls}.

This provides also the solution for the models with quadratic bosonic
torsion of \Sec\ref{sec:cIzq} by a further change of variables
(\ref{conf-tr-X}) with parameter (\ref{cIzq-tr}). Its explicit form is
calculated using (\ref{conf-tr-A}).

Up to here we did not use the gauge freedom. Actually, in supergravity
theories this is generically not that easy, since the fermionic part
of the symmetries are known only in their \emph{infinitesimal} form
(the bosonic part corresponds on a global level to diffeomorphisms and
local Lorentz transformations). This changes in the present context
for the case that Casimir-Darboux coordinates are available. Indeed,
for a constant Poisson tensor the otherwise field dependent, nonlinear
symmetries (\ref{gPSM-symms}) can be integrated easily: Within the
range of applicability of the target coordinates, $\bar{X}^I$ may be
shifted by some arbitrary function (except for the Casimir function
$\Casimir$, which, however, was found to be constant over $\BMf$), and
$\bar{A}_C$ may be redefined by the addition of some exact part. The
only restrictions to these symmetries come from nondegeneracy of the
metric (thus \eg $\tilde{A}_C$ should not be put to zero, \cf
(\ref{Izq-solE--})). In particular we are thus allowed to put both
$\chTp$ and $\chSm$ to zero in the present patch, and thus, if one
follows back the transformations introduced, also the original fields
$\chi^\pm$. Next, in the patches with $X^{++} \neq 0$, we may fix the
local Lorentz invariance by $X^{++} := 1$ and the diffeomorphism
invariance by choosing $\phi$ and $F$ as local coordinates on the
spacetime manifold $\BMf$. The resulting gauge fixed solution agrees
with the one found in the original bosonic theory, \cf \eg
\cite{Klosch:1996fi}. This is in agreement with the general
considerations of \cite{Strobl:1999zz}, here however made explicit.

A final remark concerns the discussion following (\ref{Izq-u}): As
noted there, for some choices of the bosonic potential $v$ the
potential $u$ becomes complex valued if the range of $\phi$ is not
restricted appropriately. It is straightforward to convince oneself
that the above formulas are still valid in the case of complex valued
potentials $u$ (\ie complex valued Poisson tensors). Just in
intermediary steps, such as due to (\ref{tr-BD-F}), one used complex
valued fields (with some reality constraints). The final gauge fixed
solution, however, is not affected by this, being real as it should
be.

%*********************************************************************
\section{Summary and Outlook}
\label{sec:summary}

The extension of the concept of Poisson Sigma Models (PSM) to the
graded case \cite{\bibNLSGT,Strobl:1999zz} has been explored
in some detail for the application in general two-dimensional
supergravity theories, when a dilaton field is present. Adding one
($N=1$) or more ($N>1$) pairs of Majorana fields representing
respectively a target space (spinor) variable $\chi^\alpha$ and a
related `gravitino' $\psi_m\^\alpha$, automatically leads to a
supergravity with local supersymmetry closing on-shell. Our approach
yields the minimal supermultiplets, avoiding the imposition and
evaluation of constraints which is necessary in the superfield
formalism. Instead we have to solve Jacobi identities, which the
(degenerate) Poisson structure $\Poisson^{AB}$ of a PSM must obey. In
our present paper we have performed this task for the full $N=1$
problem. The solution for the algebras turns out to be quite different
according to the rank (defined in \Sec\ref{sec:remJac}) of the
fermionic extension, but could be reduced essentially to an algebraic
problem---despite the fact that the Jacobi identities represent a set
of nonlinear first order differential equations in terms of the target
space coordinates.

In this argument the Casimir functions are found to play a key role.
If the fermionic extension is of full rank that function of the
corresponding bosonic PSM simply generalizes to a quantity $\Casimir$,
taking values in the (commuting) supernumbers, because a quadratic
contribution of $\chi^\alpha$ is included (\cf (\ref{Casimir})). If
the extension is not of full rank, beside that commuting $\Casimir$
also anticommuting Casimir functions of the form (\ref{DFS-Cf+}) and
(\ref{DFS-Cf-}) appear.

In certain cases, but not in general, the use of target space
diffeomorphisms (\cf \Sec\ref{sec:diffeo}) was found to be a useful
tool for the construction of the specific algebras and ensuing
supergravity models.  The study of `stabilisators', target space
transformations which leave an initially given bosonic algebra
invariant, also clarified the large arbitrariness (dependence of the
solution on arbitrary functions) found for the Poisson superalgebras
and the respective supergravity actions.

Because of this we have found it advisable to study explicit
specialized algebras and supergravity theories of increasing
complexity (\Sec\ref{sec:Poisson} and \ref{sec:models}). Our examples
are chosen in such a way that the extension of known bosonic 2d models
of gravity, like the Jackiw-Teitelboim model \cite{\bibJT}, the
dilaton black-hole \cite{\bibDil}, spherically symmetric gravity, the
Katanaev-Volovich model \cite{\bibKV}, $R^2$-gravity and others could
be covered (\cf \Sec\ref{sec:models}).  The arbitrariness referred to
above has the consequence that in all cases examples of several
possible extensions can be given. For a generic supergravity, obtained
in this manner, obstructions for the allowed values of the bosonic
target space coordinates emerge. Certain extensions are even found to
be not viable within real extensions of the bosonic algebra. We
identified two sources of these problems: the division by a certain
determinant (\cf (\ref{detv})) in the course of the (algebraic)
solution of the Jacobi identities, and the appearance of a
`prepotential' which may be nontrivially related to the potential in
the original PSM.  The hope for the existence of an eventual criterion
for a reduction of the inherent arbitrariness, following from the
requirement that such obstructions should be absent, unfortunately did
not materialize: \eg for the physically interesting case of an
extension of spherically reduced Einstein gravity no less than four
different obstruction-free supergravities are among the examples
discussed here, and there exist infinitely more.

The PSM approach for 2d gravities contains a preferred formulation of
gravity as a `first order' (in derivatives) action (\ref{FOG}) in the
bosonic, as well as in the supergravity case (\Sec\ref{sec:action}).

In this formulation the target space coordinates $X^I = (\phi, X^a,
\chi^\alpha)$ of the gPSM are seen to coincide with the momenta in a
Hamiltonian action. A Hamiltonian analysis is not pursued in our
present paper. Instead we discuss the possibility to eliminate $X^I$
in part or completely.

The elimination of $X^a$ is possible in the action of a generic
supergravity PSM together with a torsion dependent part of the spin
connection. We show that in this way the most general superdilaton
theory with usual bosonic part (\ref{dil}) and minimal content of
fermionic fields in its extension (the Majorana spinor $\chi^\alpha$
as partner of the dilaton field $\phi$, and the 1-form `gravitino'
$\psi_\alpha$) is produced.

By contrast, the elimination of the dilaton field $\phi$ and/or the
related spinor $\chi^\alpha$ can only be achieved in particular cases.
Therefore, these fields should be regarded as substantial ingredients
when extending a bosonic 2d gravity action of the form (\ref{grav}),
depending on curvature and torsion.

The supergravity models whose bosonic part is torsion-free ($Z=0$ in
(\ref{FOG}) with (\ref{vdil}), or in (\ref{dil})) have been studied
before \cite{\bibIzq}. Specializing the potential $v(\phi)$ \emph{and}
the extension appropriately, one arrives at the supersymmetric
extension of the $R^2$-model ($v = -\frac{\alpha}{2} \phi^2$) and the
model of Howe ($v = -\half\phi^3$) \cite{Howe:1979ia} originally
derived in terms of superfields (\cf also \cite{Ertl:1997ib}).  In the
latter case the `auxiliary field' $A$ is found to coincide with the
dilaton field $\phi$. It must be emphasized, though, that in the PSM
approach all these models are obtained by introducing a cancellation
mechanism for singularities and ensuing obstructions (for actions and
solutions in the real numbers).

When the bosonic model already contains torsion in the PSM form
(\ref{FOG}) or when, equivalently, $Z \neq 0$ in (\ref{dil}) an
extension of a conformal transformation to a target space
diffeomorphism in the gPSM allows an appropriate generalization of the
models with $Z=0$. Our discussion of spherically symmetric gravity
(\ref{SRG}) and of the Katanaev-Volovich model (\ref{KV}) show basic
differences. Whereas no new obstruction appears for the former
`physical' theory ($\phi > 0$), as already required by (\ref{SRG}),
the latter model develops a problem with real actions, except when the
parameters $\alpha$, $\beta$ and $\Lambda$ are chosen in a very
specific manner.

We also present a field theoretic model for a gPSM with rank $(2|1)$,
when only one component of the target space spinor $\chi^\alpha$ is
involved.  Its supersymmetry only contains one anticommuting function
so that this class of models can be interpreted as $\N = (1,0)$
supergravity.

Finally (\Sec\ref{sec:sdil-sol}) we make the general considerations of
\cite{Strobl:1999zz} more explicit by giving the full (analytic)
solution to the class of models summarized in the two preceding
paragraphs. It turns out to be sufficient to discuss the case $Z=0$,
because $Z \neq 0$ can be obtained by conformal transformation. Our
formulation in terms of Casimir-Darboux coordinates (including the
fermionic extension) allows the integration of the infinitesimal
supersymmetry transformation to finite ones. Within the range of
applicability for the target space coordinates $X^I$ this permits a
gauging of the target space spinors to zero. In this sense
supergravities (without matter) are `trivial'.  However, as stressed
in the introduction, such arguments break down when (supersymmetric)
matter is coupled to the model. This leads us to an outlook on
possible further applications.

Clearly starting from any of the models described here, its
supertransformations could be used---at least in a trial-and-error
manner as in the original $d=4$ supergravity \cite{\bibSUGRA}---to
extend the corresponding bosonic action \cite{Izquierdo:1998hg}.

The even simpler introduction of matter in the form of a scalar
`testparticle' in gravity explicitly (or implicitly) is a necessary
prerequisite for defining the global manifold geometrically to its
geodesics (including null directions). We believe that (properly
defined) spinning testparticle would be that instrument for 2d
supergravity.  `Trivial' supergravity should be without influence on
its (`super')-geodesics. This should work in the same way as
coordinate singularities are not felt in bosonic gravity.

Another line of investigation concerns the reduction of $d \geq 4$
supergravities to a $d=2$ effective superdilaton theory. In this way
it should be possible to perhaps nail down the large arbitrariness of
superdilaton models, when---as in our present work---this problem is
regarded from a strictly $d=2$ point of view. It can be verified in
different ways \cite{Ertl:2001PhD} that the introduction of Killing
spinors within such an approach inevitably leads to complex fermionic
(Dirac) components. Thus 2d (dilatonic) supergravities with $\N \geq
2$ must be considered. As explained in \Sec\ref{sec:gPSM} also then
the gPSM approach seems to be the method of choice. Of course, the
increase in the number of fields, together with the restrictions of
the additional $SO(\N)$ symmetry will provide an even more complicated
structure.  Already for $\N=1$ we had to rely to a large extend on
computer-aided techniques.

Preliminary computations show that the `minimal' supergravity actions,
provided by the PSM approach, also seem to be most appropriate for a
Hamiltonian analysis leading eventually to a quantum 2d supergravity,
extending the analogous result for a purely bosonic case
\cite{Schaller:1994es,\bibQGr,Strobl:1999Habil}. The role of the
obstruction for real supersymmetric extensions, encountered for some
of the models within this paper, has to be reconsidered carefully in
this context.

%*********************************************************************
\section*{Acknowledgements}

The authors thank H.~Balasin, D.~Grumiller and M.~Volkov for
discussions. The project has been supported by project P~12.815-TPH
and, in its final stage by project P~13.126-PHY of the FWF
(\"O{}sterreichischer Fonds zur F\"o{}rderung der wissenschaftlichen
Forschung). One of the authors (M.\,E.) is grateful to A.~Wipf for
hospitality during a research visit at the
Friedrich-Schiller-Universit\"at Jena.

%*********************************************************************
\begin{appendix}

%********************************************************************
\section{Notation and Identities}
\label{sec:notation}

%********************************************************************
\subsection{Forms and Vectors}
\label{sec:forms}

Let $x^m$ be local coordinates on a manifold $\BMf$. We define the
components of a $p$-form $\Phi$ according to
\begin{equation}
  \label{form}
  \Phi = \frac{1}{p!} dx^{m_p} \wedge \cdots \wedge dx^{m_1}
  \Phi_{m_1\cdots m_p},
\end{equation}
and the exterior derivative 
\begin{equation}
  \label{form-d}
  d\Phi = \frac{1}{p!} dx^{m_p} \wedge \cdots \wedge dx^{m_1} \wedge
  dx^n \rpartial_n \Phi_{m_1\cdots m_p}.
\end{equation}
As a consequence $d$ acts from the \emph{right}, \ie for a $q$-form
$\Psi$ and a $p$-form $\Phi$ the Leibniz rule is
\begin{equation}
  \label{form-Leibnitz}
  d(\Psi \wedge \Phi) = \Psi \wedge d\Phi + (-1)^{p} d\Psi
  \wedge \Phi.
\end{equation}
This convention is advantageous for the extension to spinors and
superspace where we assume similar summations of indices as in
(\ref{form}), (\ref{form-d}).

The relation between the right partial derivative $\rpartial_I$ and
the left partial derivative $\lpartial_I$ for the graded case becomes
\begin{equation}
  \label{lr-partial}
  \rpartial_I f = f \lpartial_I (-1)^{I(f+1)},
\end{equation}
where in the exponent $I$ and $f$ are $1$ for anticommuting quantities
and $0$ otherwise.

We use the characters $a,b,c\ldots$ to denote Lorentz indices taking
values $(0,1)$. In $d=2$ our Minkowski metric is
\begin{equation}
  \label{lor-metric}
  \eta_{ab} = \eta^{ab} = \mtrx{1}{0}{0}{-1},
\end{equation}
and for the antisymmetric $\epsilon$-tensor we set $\epsilon_{ab} =
\epsilon(a,b)$ and consistently $\epsilon^{ab} = -\epsilon(a,b)$,
where $\epsilon(0,1) \equiv 1$ is the $\epsilon$-symbol, so that
\begin{equation}
  \label{lor-ep}
  \epsilon_{ab} = -\epsilon^{ab} = \mtrx{0}{1}{-1}{0}.
\end{equation}
It obeys $\epsilon_a\^b \epsilon_b\^c = \delta_a\^c$ and
$\epsilon_{ab} \epsilon^{cd} = \delta_a\^d \delta_b\^c - \delta_a\^c
\delta_b\^d$, and $\epsilon_a\^b$ is the generator of Lorentz
transformations in $d=2$. The transition to world indices with the
help of the vielbein $e_a\^m$ and its inverse $e_m\^a$ yields
\begin{equation}
  \label{ep-tensor}
  \epsilon_{mn} = e\, \epsilon(m,n), \qquad
  \epsilon^{mn} = - \frac{1}{e}\, \epsilon (m,n),
\end{equation}
where $e = \det(e_m\^a)$. We use $\partial_a = e_a\^m \partial_m$ to
denote the moving frame and $e^a = dx^m e_m\^a$ for the 1-forms. Thus,
the quantity $e_a$ used in the PSM context is the 1-form with index
lowered, $e_a = e^b \eta_{ba}$. In terms of the metric $g_{mn} =
e_n\^b e_m\^a \eta_{ab}$ and its determinant $g = \det(g_{mn})$ we
have $e = \sqrt{-g}$. The induced volume form
\begin{equation}
  \label{volume-form}
  \epsilon = \half e^a \wedge e^b \epsilon_{ba} = e^1 \wedge e^0
  = e\, dx^1 \wedge dx^0
\end{equation}
enables us to derive the useful relation $dx^m \wedge dx^n =
\epsilon\, \epsilon^{mn}$, and to define the Hodge dual according to
\begin{equation}
  \label{HodgeDual}
  \star 1 = \epsilon, \qquad
  \star dx^m = dx^n \epsilon_n\^m, \qquad
  \star\epsilon = 1. \qquad
\end{equation}
It is a linear map, \ie $\star(\Phi f) = (\star\Phi) f$ for functions
$f$ and forms $\Phi$, and bijective $\star\star = \id$.

%*********************************************************************
\subsection{Spinors}
\label{sec:spinors}

The Dirac matrices obey
\begin{equation}
  \gamma^a \gamma^b + \gamma^b \gamma^a = 2 \eta^{ab}.
\end{equation}
We use the explicit representation
\begin{equation}
  \label{ga-matrices}
  \gamma^0\_\a\^\b = \mtrx{0}{1}{1}{0}, \qquad
  \gamma^1\_\a\^\b = \mtrx{0}{1}{-1}{0},
\end{equation}
\begin{equation}
  \gamma^3 = \gamma^1 \gamma^0 = \mtrx{1}{0}{0}{-1}, \qquad
  (\gamma^3)^2 = \1.
\end{equation}
Further useful identities are
\begin{equation}
  \gamma^a \gamma^b = \eta^{ab} \1 + \epsilon^{ab} \gamma^3,
\end{equation}
\begin{equation}
  \gamma^a \gamma^3 + \gamma^3 \gamma^a = 0, \qquad
  \gamma^a \gamma^3 = \gamma^b \epsilon_b\^a, \qquad
  \gamma^a \gamma_b \gamma_a = 0,
\end{equation}
and the Fierz identity
\begin{equation}
  \label{Fierz}
    2 \d_\a\^\g \d_\b\^\d = \d_\a\^\d \d_\b\^\g
  +\gamma^3\_\a\^\d \gamma^3\_\b\^\g + \gamma^a\_\a\^\d
  \gamma_{a\b}\^\g
\end{equation}
which implies the completeness relation
\begin{equation}
  \label{ga-compl}
  \Gamma_\alpha\^\beta = \half \Gamma_\gamma\^\gamma\,
  \delta_\alpha\^\beta + \half (\Gamma\gamma_a)_\gamma\^\gamma\,
  \gamma^a\_\alpha\^\beta + \half (\Gamma\gamma^3)_\gamma\^\gamma\,
  \gamma^3\_\alpha\^\beta.
\end{equation}

For a spinor
\begin{equation}
  \chi_\alpha = \clmn{\chi_+}{\chi_-}
\end{equation}
the Dirac conjugation $\bar\chi^\alpha = \chi^\dagger\^{\dot\alpha}
A_{\dot\alpha}\^\alpha$ depends on the matrix $A$, which obeys
\begin{equation}
  \label{DiracAProp}
  A \gamma^a A^{-1} = (\gamma^a)^\dagger, \qquad
  A^\dagger = A.
\end{equation}
We make the usual choice
\begin{equation}
  \label{DiracA}
  A = \mtrx{0}{1}{1}{0} = \gamma^0.
\end{equation}

The charge conjugation of a spinor using complex conjugation is
\begin{equation}
  \label{ChargeConj}
  \chi^c = B \chi^*
\end{equation}
\begin{equation}
  \label{ChargeBProp}
  B^{-1} \gamma^a B = -(\gamma^a)^*, \qquad B B^* = \1.
\end{equation}
For our choice of $\gamma^a$ (\ref{ga-matrices})
\begin{equation}
  \label{ChargeB}
  B = \mtrx{-1}{0}{0}{1}.
\end{equation}
Alternatively, one can define the charge conjugated spinor with the
help of the Dirac conjugation matrix (\ref{DiracA}),
\begin{equation}
  \chi^c = (\bar\chi C)^T = C^T A^T \chi^*, \qquad
  \chi^c\_\alpha = \bar\chi^\beta C_{\beta\alpha},
\end{equation}
\begin{equation}
  \label{ChargeCProp}
  C^{-1} \gamma^a C = - (\gamma^a)^T, \qquad C^T = -C,
\end{equation}
\begin{equation}
  \label{ChargeC}
  C = (C_{\beta\alpha}) = \mtrx{0}{1}{-1}{0}.
\end{equation}

By means of
\begin{equation}
  \label{spin-metric}
  \epsilon_{\a\b} = \epsilon^{\a\b} = \mtrx{0}{1}{-1}{0}
\end{equation}
indices of Majorana spinors $\chi^c = \chi$, in components $\chi_+ =
-(\chi_+)^*, \chi_- = (\chi_-)^*$, can be raised and lowered as
$\chi^\alpha = \epsilon^{\alpha\beta} \chi_\beta$ and $\chi_\alpha =
\chi^\beta \epsilon_{\beta\alpha}$. In components we get
\begin{equation}
  \label{spin-down}
  \chi^+ = \chi_-, \qquad \chi^- = -\chi_+.
\end{equation}
This yields $\varphi^\alpha \chi_\alpha = -\varphi_\alpha \chi^\alpha
= \chi^\alpha \varphi_\alpha = \varphi^- \chi^+ - \varphi^+ \chi^-$
for two anticommuting Majorana spinors $\varphi$ and $\chi$. For
bilinear forms we use the shorthand
\begin{equation}
  (\varphi\chi) = \varphi^\alpha \chi_\alpha, \qquad
  (\varphi\gamma^a\chi) = \varphi^\alpha \gamma^a\_\alpha\^\beta \chi_\beta,
  \qquad
  (\varphi\gamma^3\chi) = \varphi^\alpha \gamma^3\_\alpha\^\beta \chi_\beta.
\end{equation}
A useful property is
\begin{equation}
  \label{spin-metric-prop}
  \epsilon_{\alpha\beta} \epsilon^{\gamma\delta} =
  \delta_\alpha\^\gamma \delta_\beta\^\delta - \delta_\alpha\^\delta
  \delta_\beta\^\gamma.
\end{equation}
The Fierz identity (\ref{Fierz}) yields
\begin{equation}
  \chi_\alpha \varphi^\beta = -\half (\varphi \chi) \delta_\alpha\^\beta -
  \half (\varphi \gamma_a \chi) \gamma^a\_\alpha\^\beta - \half (\varphi
  \gamma^3 \chi) \gamma^3\_\alpha\^\beta.
\end{equation}

Among the spinor matrices $(\gamma^a)^{\alpha\beta}$ and
$(\gamma^3)^{\alpha\beta}$ are symmetric in $\alpha \leftrightarrow
\beta$, whereas $\epsilon^{\alpha\beta}$ is antisymmetric.

The chiral projectors
\begin{equation}
  \label{ChiralP}
  P_{\pm} = \half (\1 \pm \gamma^3)
\end{equation}
project $\chi_+$ and $\chi_-$.

Light cone components of a vector $v^a$ are defined by
\begin{equation}
  \label{LC-comp}
  \clmn{v^{++}}{v^{--}} = \frac{i}{\sqrt{2}} \clmn{v^0 + v^1}{-v^0 +
    v^1}.
\end{equation}
Then the indices $(++,--)$ exactly coincide with components of the
2-spinor $v^{\alpha\beta}$ which can be related to a Lorentz vector
$v^a$ by
\begin{equation}
  \label{spin-vector}
  v^{\alpha\beta} := \frac{i}{\sqrt{2}} v^a \gamma_a\^{\alpha\beta},
\end{equation}
where $\gamma^a$ is given by (\ref{ga-matrices}). The pre-factor (up
to a sign) is necessary for (\ref{spin-vector}) to be consistent with
the metric in light-cone coordinates $\eta_{++--} = \eta_{--++} = 1$.
This metric can be extended to a full metric in 2-spinor space
\begin{equation}
  \eta_{\alpha\beta}\^{\delta\gamma} = -\half \gamma^a\_{\alpha\beta}
  \gamma_a\^{\delta\gamma} - \half \gamma^3\_{\alpha\beta}
  \gamma^3\^{\delta\gamma} - \half \epsilon_{\alpha\beta}
  \epsilon^{\delta\gamma},
\end{equation}
where the Fierz identity (\ref{Fierz}) yields
\begin{equation}
  \eta_{\alpha\beta}\^{\delta\gamma} = \delta_\alpha\^\gamma
  \delta_\beta\^\delta.
\end{equation}

%*********************************************************************
%\bibliographystyle{utphys}
%\bibliography{paper}

%***** copy of paper.bbl *********************************************

\providecommand{\href}[2]{#2}\begingroup\raggedright\endgroup

\end{appendix}

\end{document}